\documentclass[fleqn,usenatbib]{mnras}
\usepackage{newtxtext,newtxmath}
\usepackage{comment}
\usepackage{multirow}
\usepackage{mathtools}
\usepackage[T1]{fontenc}
\usepackage{ae,aecompl}

\usepackage{graphicx}	
\usepackage{amsmath}	
\usepackage[ruled,vlined]{algorithm2e}
\usepackage{algpseudocode}
\usepackage{enumitem} 
\usepackage{xfrac} 

\setlist[itemize]{leftmargin=*}

\newcommand{\norm}[1]{\|#1\|}

\newcommand{\ud}{\mathrm{d}}

\newcommand{\vect}[1]{\boldsymbol{#1}}

\newcommand{\derfrac}[2]{\frac{\ud #1}{\ud #2}}

\newcommand{\msol}[1]{{#1}\:\mathrm{M_\odot}}

\newcommand{\rhosol}[1]{{#1}\:\mathrm{M_\odot\:pc^{-3}}}
\newcommand{\nsol}[1]{{#1}\:\mathrm{\:pc^{-3}}}

\newcommand{\gadget}{\texttt{GADGET-4}}
\newcommand{\sevn}{\texttt{SEVN}}

\newcommand{\bifrost}{\texttt{BIFROST}}

\newcommand{\mcluster}{\texttt{McLuster}}
\newcommand{\myosotis}{\texttt{MYOSOTIS}}
\newcommand{\catone}{\texttt{IMBH-LM}}
\newcommand{\cattwo}{\texttt{IMBH-JAM}}
\newcommand{\catthree}{\texttt{IMBH-M}}

\title[IMBH growth in assembling star clusters]{{FROST-CLUSTERS -- I. Hierarchical star cluster assembly boosts intermediate-mass black hole formation}}
\author[A. Rantala et al.]{Antti Rantala$^{1}$\thanks{E-mail: anttiran@mpa-garching.mpg.de}, Thorsten Naab$^{1}$, Natalia Lah\'en$^{1}$\\
$^{1}$Max-Planck-Institut f\"ur Astrophysik, Karl-Schwarzschild-Str. 1, 
D-85748, Garching, Germany\\
}
\date{Accepted XXX. Received YYY; in original form ZZZ}
\pubyear{2023}

\begin{document}
\label{firstpage}
\pagerange{\pageref{firstpage}--\pageref{lastpage}}
\maketitle

\begin{abstract}
Observations and high-resolution hydrodynamical simulations indicate that massive star clusters assemble hierarchically from sub-clusters with a universal power-law cluster mass function. We study the consequences of such assembly for the formation of intermediate-mass black holes (IMBHs) at low metallicities ($Z=0.01\;Z_\mathrm{\odot}$) with our updated N-body code \bifrost{} based on the hierarchical fourth-order forward integrator. \bifrost{} integrates few-body systems using secular and regularized techniques including post-Newtonian equations of motion up to order PN3.5 and gravitational-wave recoil kicks for BHs. Single stellar evolution is treated using the fast population synthesis code \sevn. We evolve three cluster assembly regions with $N_\mathrm{tot} = 1.70$--$2.35 \times 10^6$ stars following a realistic IMF in $\sim$1000 sub-clusters for $t=50$ Myr. IMBHs with masses up to $m_\bullet \sim \msol{2200}$ form rapidly mainly via the collapse of very massive stars (VMSs) assembled through repeated collisions of massive stars followed by growth through tidal disruption events and BH mergers. No IMBHs originate from the stars in the initially most massive clusters. We explain this by suppression of hard massive star binary formation at high velocity dispersions and the competition between core collapse and massive star life-times. Later the IMBHs form subsystems resulting in gravitational-wave BH-BH, IMBH-BH and IMBH-IMBH mergers with a $m_\bullet\sim\msol{1000}$ gravitational-wave detection being the observable prediction. Our simulations indicate that the hierarchical formation of massive star clusters in metal poor environments naturally results in formation of potential seeds for supermassive black holes.
\end{abstract}

\begin{keywords}
gravitation -- celestial mechanics -- methods: numerical -- galaxies: star clusters: general -- stars: black holes
\end{keywords}


\section{Introduction}

The runaway stellar collision scenario in dense, massive star clusters has been proposed to explain the emergence of massive black hole (MBH) seeds at high redshifts \citep{Gold1965, Spitzer1967, Spitzer1971, Peebles1972, Begelman1978}. If the MBH seeds form but later fail to further grow in mass, they should be in principle observable in the present-day Universe as intermediate-mass black holes (IMBHs\footnote{In this work we divide the black hole (BH) population at $Z=0.01\:Z_\mathrm{\odot}$ as follows: stellar-mass BHs with $m_\bullet \lesssim \msol{63}$, stellar-mass pulsational pair-instability mass-gap BHs in the range of $\msol{63} \lesssim m_\bullet \lesssim \msol{100}$, IMBHs with $\msol{100} \lesssim m_\bullet \lesssim \msol{10^5}$ including the mass gap IMBHs at $\msol{100} \lesssim m_\bullet \lesssim \msol{290}$, and finally MBHs above $m_\bullet \gtrsim \msol{10^5}$. We use the terms IMBH and MBH seed interchangeably.}).

The runaway collision scenario for MBH seed and IMBH formation has been widely studied using numerical simulations (see e.g. the reviews of \citealt{Mezcua2017} and \citealt{Greene2020} and references therein). The so-called slow formation channel \citep{Miller2002} has a stellar mass BH with $m_\bullet>\msol{50}$ growing in a globular cluster up to masses of $m_\bullet\sim \msol{10^3}$ in 10 Gyr, mainly by merging with lower-mass BHs. The bottleneck for the slow collision channel is the initial formation of a stellar mass BH with $m_\bullet>\msol{50}$. Due to the physics of pair instability and pulsational pair-instability supernovae, hereafter (P)PISN, it is expected that no BHs with masses in the range $\msol{50}\lesssim m_\bullet \lesssim \msol{130}$ should from through single stellar evolution at solar metallicity \citep{Fowler1964,Woosley2007,Woosley2017}. At lower metallicities this (P)PISN mass gap widens and moves into somewhat higher BH masses.

The so-called fast runaway collision scenario relies on a very massive star (VMS\footnote{We refer any stars more massive than $m_\star>\msol{150}$ as VMSs in this work.}) growing via repeated mergers to sufficiently high masses \citep{Lee1987,Quinlan1990,Bonnell1998} to produce an IMBH remnant above the (P)PISN mass gap range. The IMBH may further grow by accreting stellar material in tidal disruption events (TDEs) and by merging with other BHs. Numerical simulations of the fast scenario (e.g. \citealt{PortegiesZwart1999, PortegiesZwart2002, Gurkan2004, PortegiesZwart2004, Freitag2006a, Freitag2006b, Rizzuto2021, Rizzuto2022, ArcaSedda2023a, ArcaSedda2023b, ArcaSedda2023c,GonzalesPrieto2024arXiv}) show that the stellar collisions can result in a formation of a VMS or a supermassive star ($m_\star>\msol{10^3}$, e.g. \citealt{Denissenkov2014,Gieles2018}) with a mass from hundreds to potentially tens of thousands of solar masses. In a true runaway collision sequence, each collision further increases the collision rate. Rapid growth is however possible even without fulfilling this criterion. Typically only $1$--$2$ stars per cluster can grow via collisions \citep{Baumgardt2011}. In the global instability variant of the fast runaway collision scenario, stellar collisions become relevant if the collision time-scale is shorter than the characteristic age of the system \citep{Escala2021,Vergara2023}. 

There are other ingredients in the runaway VMS scenario in addition to the star cluster masses, densities and metallicities. Primordial mass segregation may play a role in the collision dynamics but does not necessarily increase collision rate \citep{Ardi2008}. An external potential may delay the cluster evolution and the onset of collisions by increasing the velocities of the stars \citep{Reinoso2020}. Gas may be very important in the growth of VMSs \citep{Clarke2008, Baumgardt2011, Moeckel2011, Roupas2015}, while the initial mass function (IMF) of the cluster \citep{Goswami2012} and the primordial binary fraction also likely play a role \citep{Gurkan2006} especially due to single-binary interactions \citep{Fregeau2004,Gaburov2008}. And early variant of the VMS growth channel may have occurred in the first clusters \citep{Boekholt2018,Reinoso2018} of Pop-III stars \citep{Schaerer2002}.

The fast and the slow collisional IMBH formation scenarios are not mutually exclusive. Even if the stellar collisions do not result in the formation of an IMBH via the fast VMS formation and collapse channel, repeated tidal disruption events or BH-BH collisions may still result in a formation of mass gap BHs or IMBHs \citep{Morscher2015,Rodriguez2016,Askar2017,Banerjee2018,Fragione2018,Antonini2019,Rodriguez2019,Belczynski2020,ArcaSedda2021,Banerjee2021,Rizzuto2021,Mapelli2021,Rizzuto2022,Rizzuto2023,ArcaSedda2023b}. IMBH (or MBH seed) formation via the fast collisional runaway channel most likely occurs in young, massive, star clusters \citep{PortegiesZwart2010} with low metallicity. First, low metallicity environments are favored due to weaker stellar winds of massive stars resulting in more massive IMBH remnants (e.g. \citealt{Mapelli2016, DiCarlo2021}). There is still uncertainty about the wind mass-loss rates of massive stars. High wind mass-loss rates (especially at high metallicities) might suppress the VMS IMBH formation channel or inhibit it altogether (e.g. \citealt{Belkus2007,Yungelson2008,Glebbeek2009,Pauldrach2012}).

Next, the massive star clusters need to be dense to have high enough collision rates. The collision rates of massive stars are strongly enhanced by the mass segregation and core collapse of the star clusters. Due to the negative heat capacity of gravitating systems, the centers of star clusters may undergo a catastrophic core collapse \citep{LyndenBell1968,LyndenBell1980,Sugimoto1983,Bettwieser1984,Heggie1993,Spurzem1996,Heggie2003,Kamlah2022} which is eventually halted by the dynamical formation of binary stars that act as an additional energy source at the cluster core. For clusters with a stellar mass function, mass segregation proceeds on a dynamical friction time-scale $t_\mathrm{df}$ of the most massive stars, and the core collapse time-scale is only a small fraction of the two-body relaxation time-scale of the cluster. The core bounce becomes ambiguous in low-mass clusters when the mass of the most massive star exceeds $0.1\%$ of the cluster mass \citep{Fujii2014}. The mass segregation and core collapse of star clusters increase the stellar collision rate by increasing the number density of massive stars at their centers. Only $0.1\%$ of the total cluster mass may collapse into the very center \citep{Gurkan2004}, which is far below the typical definitions for the core size, density and mass \citep{Casertano1985}. 

Dense, low-metallicity star clusters have been studied observationally both in the local Universe and at high redshift. Massive star clusters are characterized by their flat constant-density cores with core radii $r_\mathrm{c}$ and core densities $\rho_\mathrm{c}$ while the outer parts are described by power-laws optionally including a tidal cut-off \citep{Plummer1911,King1962,Casertano1985,Elson1987}. In the local Universe, the Magellanic clouds offer the best-resolved view of the internal structure of young low-metallicity star clusters \citep{Hunter2003}.  The core-radii and half-mass radii are often less than a parsec \citep{Hill2006, Werchan2011, Sun2017, Gatto2021}. Extra-galactic surveys, that offer a larger sample size but limited resolution, have also reported extremely compact young star clusters \citep{Brown2021}, although with a significant scatter in the mass-size relation. The Tarantula nebula is an extreme local example of the complex hierarchical structure of young star cluster-forming regions: sub-clusters range from 30 Myr old clusters in the outer regions \citep{Cignoni2016} to the 1-2 Myr old core \citep{Crowther2010} to still ongoing star formation \citep{Kalari2018}. The dense core region R136 \citep{Selman2013} contains several young stars more massive than $m_\star\gtrsim \msol{120}$ \citep{Massey1998} while even higher stellar masses of $m_\star\sim\msol{160}$--$\msol{320}$ have been suggested \citep{Crowther2010}. Outside the cluster, massive runaway star candidates can be observed \citep{Bestenlehner2011, Gvaramadze2011}. For further details about the massive stars in the region see \cite{Vink2015} and references therein.

Low-mass galaxies that host proto-globular clusters, observed with the aid of gravitational lensing \citep{Livermore2017, Vanzella2019}, present the densest star forming environments in the high-redshift Universe \citep{Vanzella2023}. The recent James Webb Space Telescope (JWST) observations have revealed clumpy, clustered star formation regions at redshifts beyond $z>6$, such as the Cosmic Grapes galaxy \citep{Fujimoto2024arXiv}, enriched by the young, massive stars \citep{Topping2024}. The JWST observations of the Cosmic Gems arc \citep{Adamo2024arxiv} have revealed five massive star clusters at $z=10.2$ with masses of $M_\star\sim\msol{10^6}$ and lensing-corrected sizes of $\sim1$ pc, corresponding to high stellar surface densities of $\Sigma_\star\sim\msol{10^5}\:\mathrm{pc}^{-2}$. The galaxy GN-z11 observed by the JWST at $z=10.6$ with super-solar N/O abundance at low metallicity \citep{Bunker2023} might be an indication of early stages of globular cluster formation, or even a chemical fingerprint of supermassive stars \citep{Charbonnel2023, MarquesChaves2024}.

The evolution of massive star clusters \citep{Wang2016} and IMBH formation and growth within them \citep{Rizzuto2023,ArcaSedda2023a} is typically studied in isolated simulation setups. There are a number of cosmological setups focusing on the formation of proto-globular clusters and their IMBH growth \citep{Katz2015,Yuya2017,Ma2020}. However, the cosmological setups typically lack the necessary mass or spatial resolution to follow the collisional stellar dynamics at the level of the isolated setups. 

Observations of star cluster formation regions \citep{Zhang2001, Lada2003, Larsen2004, Bastian2005, Grasha2017, Menon2021} suggest that massive star cluster formation resembles a complex hierarchical assembly rather than a simple monolithic collapse, and gas and gas expulsion likely has a role in shaping the young star clusters \citep{Dale2015,Krause2016}. Hydrodynamical simulations of star cluster formation in low-metallicity star-burst environments support this view \citep{Lahen2020}. Star cluster formation from fractal or otherwise structured initial conditions has been studied in the literature (e.g. \citealt{Aarseth1972, Smith2011, Fujii2012, Howard2018, Grudic2018, Sills2018,  Torniamenti2022,Farias2024}), though definitely not as widely as the evolution of simple, isolated spherical systems, partially due to the fact that complex, structured initial conditions evolve relatively rapidly into spherical configurations. Mass segregation and core collapse may occur earlier in these setups, especially if the initial sub-clusters are sub-virial \citep{Allison2009,Yu2011} or already mass segregated \citep{McMillan2007,Moeckel2009}. It has been pointed out by \cite{Fujii2013} that stellar collisions may occur earlier in ensemble star cluster formation in a filamentary setup as the sub-clusters have shorter relaxation and mass segregation time-scales compared to the final assembled star cluster.

Observationally, IMBHs still remain elusive \citep{Greene2020}. Extrapolating the well-established the $M_\bullet$--$\sigma$ relation (e.g. \citealt{Ferrarese2000,Gebhardt2000}) for massive galaxies into systems with lower masses and velocity dispersions, it has been proposed that dwarf galaxies, especially those with an active low-luminosity nucleus, might host IMBHs \citep{Greene2007, Dong2012,Chilingarian2018,Greene2020,Mezcua2018,Reines2022}. Several ultra- or hyper-luminous X-ray emitters (ULXs, HLXs) in nearby galaxies have been interpreted as candidates for accreting IMBHs with masses ranging from $m_\bullet \sim \msol{200}$ to $m_\bullet \sim \msol{10^5}$ (e.g. \citealt{Colbert1999, Kaaret2001, Matsumoto2001, Strohmayer2003, Patruno2006, Farrell2009, Mezcua2013, Pasham2014, Mezcua2015, Mezcua2017, Kim2020}). ALMA observations of gas dynamics in dwarf elliptical NGC 404 \citep{Davis2020} have revealed a BH with a mass of $m_\bullet\sim\msol{5.5\times10^5}$ consistent with both a high-mass IMBH or a low-mass MBH. In addition, there is stellar-dynamical evidence of dark objects in the centers of some globular clusters (GCs) and stripped dwarf galaxies in the IMBH mass range, consistent with concentrations of stellar-mass black holes, or possibly IMBHs (e.g. \citealt{Gebhardt2002, Ibata2009, Noyola2010, vanderMarel2010, Lutzgendorf2011, Jalali2012, Lanzoni2013, Kamann2016, Baumgardt2017, Pechetti2022, DellaCroce2024}). However, a number of studies do not find such stellar-dynamical evidence of IMBHs \citep{Baumgardt2003, Baumgardt2019, Mann2019, HenaultBrunet2020}, and characteristic IMBH accretion signatures have not been observed in the clusters in question \citep{Tremou2018}. Still, the most definite evidence of IMBHs to date remains the gravitational-wave (GW) transient GW190521 \citep{Abbott2020}, a merger between two stellar-mass black holes with masses of $m_\bullet \sim\msol{66}$ and $m_\bullet \sim\msol{85}$ resulting in a formation of an IMBH with $m_\bullet \sim \msol{150}$. In addition to gravitational waves, electromagnetic transients such as tidal disruption events may help to uncover the elusive IMBH population (e.g. \citealt{Lin2018,Fragione2018,Fragione2018b}).

The fact that there are no universally accepted solid detections of IMBHs in globular clusters is not necessarily a problem for the IMBH formation models. After their formation, IMBHs are initially located at the centers of their host clusters. In numerical simulations, BHs sink into the center of the star cluster to form a subsystem of BHs which is an ideal place for both Newtonian and relativistic BH-BH and IMBH-BH interactions \citep{Breen2013,Kremer2020a,GonzalezPrieto2022,Maliszewski2022}. Even Newtonian three-body interactions can be sufficiently strong to eject (IM)BHs from their star clusters, and gravitational-wave recoil kicks after BH-BH mergers with recoil velocities up to several times $1000$ km/s can easily unbind the merger remnant from its host cluster. Merging binaries with misaligned spins or highly spinning BH components are more likely to receive large kicks \citep{Poisson2014}. As the effective spin parameter of a multi-generation BH merger product approaches $\sim0.7$ of the maximal BH spin, it is likely that the BH growing by mergers will other BHs will eventually be ejected from its host star cluster, especially if the cluster mass and thus escape velocity is low. It is thus expected that the BH-BH channel can efficiently grow IMBHs in the environments with the highest escape velocities, such as nuclear star clusters (e.g. \citealt{Fragione2020,Fragione2022,Fragione2022b}). Even though IMBHs would be ejected from their host clusters through relativistic recoil kicks, their gravitational-wave signal in the IMBH mass range should be observable in the near future with either ground-based or space-borne gravitational-wave detectors (e.g. \citealt{Jani2020}).

In this first study of our FROST-CLUSTERS project, we examine the formation of IMBHs during the hierarchical assembly of massive star clusters. IMBH formation in massive cluster star assembly setups (final cluster masses $M_\mathrm{\star}>\msol{5\times10^5}$ and total particle numbers $N_\mathrm{tot}=2.35\times10^6$) with a full IMF from $\msol{0.08}$ to $\msol{150}$ including post-Newtonian equations of motion for the BHs and gravitational-wave recoil kicks for the BH-BH mergers. For simplicity, the first study focuses on single-star setups. We will examine the effect of binary fraction on our results in a forthcoming study (FROST-CLUSTERS II).

Our setup relies on two main building blocks. First, the individual sub-clusters follow an universal young star cluster mass function with a power-law slope of $\alpha=-2$. This is supported both by observations \citep{Elmegreen1996,Zhang1999,Adamo2020} and numerical simulations \citep{Lahen2020}. Second, we assume a shallow power-law mass-size relation slope $\alpha\sim0.18$ (e.g. \citealt{Brown2021}) for our sub-cluster population. The chosen normalization results in small sub-pc birth radii for the cluster population, motivated by the recent JWST observations of young, massive and dense star clusters at $z>10$ \citep{Adamo2024arxiv} as well as massive star cluster formation in state-of-the art star-by-star low-metallicity hydrodynamical simulations of star-bursting dwarf galaxies \citep{Lahen2020}. 

IMBHs from $m_\bullet \sim \msol{100}$ to $m_\bullet \sim\msol{2200}$ form during the first $t=50$ Myr in our simulations. Most interestingly, we find that the initially most massive star clusters do not form IMBHs from their stars, but rather inherit IMBHs formed in lower-mass star clusters through star cluster mergers. We study the reason for the suppressed IMBH formation in isolated, massive star clusters, in a series of $100$ isolated setups containing up to million stars per cluster.

The article is structured as follows. After the introduction we review basic pathways to stellar collisions in dense star clusters in section 2. Next, we introduce our updated simulation code \bifrost{} in section 3 as well as in the Appendix, and the initial conditions for our simulations. We present the results concerning the IMBH formation in the hierarchical star cluster assembly simulations, detailing the complex growth histories of the IMBHs, in section 4, and in isolated star clusters in section 5. After discussing the implications and caveats of our results in section 6, we summarize our results and conclude in section 7.


\section{Stellar collisions in dense star clusters}

\subsection{Stellar collision rate: the cross section argument}\label{section: rate-crosssection}

Estimates for collision rates of stars can be made based on cross-section arguments \citep{Spitzer1987}. Assuming an equilibrium stellar system with a number density $n$ and velocity dispersion $\sigma$, consisting of stars with masses $m_\star$ and radii $r_\star$, the collision rate $t_\mathrm{coll}^\mathrm{-1}$ of stars can be estimated as 
\begin{equation}\label{eq: coll-cross}
\frac{1}{t_\mathrm{coll}} = 4 \sqrt{\pi} n \sigma r_\mathrm{\star}^2 \left( 1 + \frac{G m_\star}{\sigma^2 r_\mathrm{\star}}\right),
\end{equation}
following \cite{Binney2008}. If gravitational focusing is unimportant, i.e. the second term in the parenthesis is small, the collision rate scales with the velocity dispersion $\sigma$. In a typical star cluster, this is not the case as
\begin{equation}
\frac{G m_\star}{\sigma^2 r_\mathrm{\star}} \sim 1.9\times10^3 \left( \frac{m_\star}{1\:\mathrm{M_\odot}}\right) \left( \frac{r_\star}{1\:\mathrm{R_\odot}}\right)^{-1} \left( \frac{\sigma}{10\:\mathrm{km/s}}\right)^{-2} \gg 1.
\end{equation}
Instead, gravitational focusing is important and the collision rate is inversely proportional to $\sigma$, and the collision rate decreases in high velocity dispersion environments. The collision rates predicted by the simple cross section argument of Eq. \eqref{eq: coll-cross} are typically low. For example, a dense star cluster following a \cite{Plummer1911} density profile with a total mass of $M_\star = \msol{2\times10^5}$ and half-mass radius of $r_\mathrm{h}=0.25$ pc consisting of equal-mass main-sequence stars of $m=\msol{10}$ has a collision rate of only $t_\mathrm{coll}^\mathrm{-1}\sim10^{-3}$ Myr$^{-1}$.

Numerical simulations of evolving star clusters have shown that the collision rate can be orders of magnitude higher compared to the simple cross section arguments. This is due to mass segregation and cluster core collapse which increase the central stellar density $\rho_\mathrm{c}$. In addition, after mass segregation and core collapse, massive stars can form binaries and more complex hierarchies in the cluster core, enhancing the collision rate (e.g. \citealt{PortegiesZwart1999}).

\subsection{Mass segregation and core collapse}

In star clusters consisting of equal-mass stars, core collapse cannot be driven by mass segregation. Instead, the core collapse proceeds on a time-scale $t_\mathrm{cc}$ by a factor of $15$--$20$ \citep{Spitzer1975,Cohn1980,Takahashi1995,Makino1996} longer than the common two-body relaxation time-scale $t_\mathrm{rlx}$ \citep{Spitzer1987,Giersz1994,Aarseth2003,Heggie2003}. The expression for the two-body relaxation time-scale can be written as
\begin{equation}\label{eq: rlx}
    t_\mathrm{rlx} = \frac{0.138 N}{\ln{(\gamma N)}} \left( \frac{r_\mathrm{h}^3}{G M_\star} \right)^{1/2}.
\end{equation}
Here $M_\star$ is the total stellar mass, $r_\mathrm{h}$ is the half-mass radius of the cluster and $N=M_\star/\tilde{m}$ the total number of stars with $\tilde{m}$ being the mean stellar mass. The parameter $\gamma$ has a value of $\gamma \sim 0.11$ for single-mass systems \citep{Giersz1994}.

In multi-mass systems, massive stars will reach the center of the cluster due to energy equipartition on a time-scale of the order of the segregation time-scale $t_\mathrm{seg}$ \citep{Spitzer1971, PortegiesZwart2004}, defined as
\begin{equation}\label{eq: tseg}
    t_\mathrm{seg} = \frac{M_\mathrm{\star}}{m_\mathrm{max}} \frac{0.138}{\ln{(\gamma M_\mathrm{\star}/m_\mathrm{max})}} \left( \frac{r_\mathrm{h}^3}{G M_\mathrm{\star}} \right)^{1/2},
\end{equation}
in which $m_\mathrm{max}$ is the mass of the most massive individual star in the cluster. Comparing the expression for $t_\mathrm{seg}$ to the definition of the two-body relaxation time-scale in Eq. \ref{eq: rlx}, we see that the total number of particles $N=M_\star/\tilde{m}$ is replaced by the effective number of particles $M_\star/m_\mathrm{max}$ in the expression for $t_\mathrm{seg}$. 

For our dense example star cluster from section \ref{section: rate-crosssection} with $M_\star = \msol{2\times10^5}$ and $r_\mathrm{h}=0.25$, this time populated from a \cite{Kroupa2001} IMF with $m_\mathrm{max}=\msol{150}$, $t_\mathrm{rlx}\sim18.5$ Myr and $t_\mathrm{seg}\sim0.15$ Myr. For realistic IMFs the core collapse time-scale accelerated by mass segregation is $t_\mathrm{cc} \sim 0.2\:t_\mathrm{rlx}$ \citep{PortegiesZwart2004}, a considerably shorter time-scale than for single-mass systems for which $t_\mathrm{cc} \sim 15$--$20\:t_\mathrm{rlx}$. For our example cluster, this yields $t_\mathrm{cc}\sim3.7$ Myr.

Assuming a power-law mass-size relation $r_\mathrm{h} \propto M_\star^\alpha$ for the star clusters, the central density $\rho_\mathrm{c}$, the two-body relaxation $t_\mathrm{rlx}$ and mass segregation times scale as
\begin{equation}
\begin{split}
    \rho_\mathrm{c} &\propto M_\star^\mathrm{1-3\alpha}\\
    t_\mathrm{rlx} &\propto \frac{M_\mathrm{\star}^{\frac{1}{2} ({3\alpha+1})}}{\ln{(\gamma M_\mathrm{\star}/\tilde{m})}}\\
    t_\mathrm{seg} &\propto \frac{M_\mathrm{\star}^{\frac{1}{2} ({3\alpha+1})}}{\ln{(\gamma M_\mathrm{\star}/m_\mathrm{max})}}.
\end{split}
\end{equation}
The central density formula assumes the common \cite{Plummer1911} profile. Cluster populations with $\alpha<1/3$ have their central density $\rho_\mathrm{c}$ increasing when the mass of the clusters $M_\star$ increases. For any mass-size relation with a power-law slope larger than $\alpha>-1/3$, more massive star clusters have longer two-body relaxation time-scales $t_\mathrm{rlx}$ as well as longer mass segregation time-scales $t_\mathrm{seg}$. The observed mass-size relation slopes $\alpha$ for young star clusters in the local Universe are in general positive but fairly shallow, with $\alpha\sim 0.13$ \citep{Marks2012} for the three-dimensional $r_\mathrm{h}$ and $\sim0.1$--$0.3$ for the projected effective radii \citep{Larsen2004,Brown2021}. Thus, more massive star clusters have higher central densities $\rho_\mathrm{c}$ but longer mass segregation time-scales $t_\mathrm{seg}$ and thus longer core collapse time-scales $t_\mathrm{cc}$ than their less massive counterparts in the local, young star cluster population. 

More detailed core collapse time-scale estimates have been proposed in the literature based on the sinking time-scale for massive stars due to dynamical friction, the core relaxation time-scale, or the formation time-scale for hard binaries which eventually halt the core collapse (e.g. \citealt{PortegiesZwart2002,Gurkan2004,Goswami2012,Fujii2014}). The mass segregation time-scale $t_\mathrm{seg}$ is typically short, less than $1$-$2$ Myr for dense ($r_\mathrm{h}\lesssim1$ pc) star clusters. However, mass segregation and core collapse can only enhance the collision rate of massive stars if the stars are still alive when reaching the core \citep{PortegiesZwart2004}. The short life-times of massive stars, of the order of a few Myr, limit the collision rate: for example, a star with an initial mass of $m_\star = \msol{150}$ at a low metallicity of $Z=0.01 Z_\odot$ has a life-time of only $t=2.82$ Myr. \footnote{The estimates for stellar life-times in this work originate from the \texttt{PARSEC} stellar tracks (e.g. \citealt{Bressan2012}) used by the fast stellar population synthesis code \sevn{} \citep{Iorio2023}.}

\subsection{Binary formation, binary-single interactions and collisions}

If there are no primordial binaries in the star cluster, first binaries form during the core collapse from encounters of three initially unbound stars via the three-body binary formation channel (hereafter 3BBF). Assuming a system consisting of identical stars of mass $m_\star$ with a number density of $n$ and velocity dispersion of $\sigma$, it can be shown \citep{Goodman1993,Binney2008} that the rate $\bar{C}$\footnote{Note that textbook of \cite{Binney2008} erroneously has $n^2$, not $n^3$, in their formula 7.11 discussing the \cite{Goodman1993} calculation.} in which binaries form in triple encounters is 
\begin{equation}\label{eq: goodman}
\bar{C} \approx \frac{G^5 m_\star^5 n^3}{\sigma^9}.
\end{equation}
Thus, for a system consisting of equal-mass bodies, the 3BBF binary formation rate $\bar{C}$ steeply increases with increasing stellar mass $m_\star$ and number density $n$, and strongly decreases with increasing velocity dispersion $\sigma$.

However, the \cite{Goodman1993} binary formation rate estimate of Eq. \eqref{eq: goodman} assumes equal-mass stars, and it is unclear to what extent their 3BBF rates are applicable to realistic multi-mass encounters. Nevertheless, it has been commonly assumed that two most massive bodies are the most likely to pair. Very recently, \cite{Atallah2024arxiv} showed that this assumption does not actually hold. Instead, from the three bodies ($m_\mathrm{1}$, $m_\mathrm{2}$, $m_\mathrm{3}$), the two most massive ($m_\mathrm{1}$, $m_\mathrm{2}$) are the least likely to become bound. Wide binaries\footnote{Wide and hard binaries are characterized by their long-term survival in repeated encounters with the cluster stars. Wide binaries will eventually dissolve which hard binaries can further shrink \citep{Heggie1975}. The threshold is approximately the binary orbital velocity $v_\mathrm{orb}\sim\sigma$. For example, a binary with $m_\mathrm{b}=20\:\mathrm{M_\odot}$ and $a=10^{-3}$ pc is hard in star clusters with $\sigma\lesssim9.3$ km/s.}, which are a considerably more likely outcome than hard binaries, and stellar collisions are more common than hard binary formation. The binary eccentricity distribution is always thermal or steeper, i.e. $f(e)\propto e^\beta$ with $\beta\gtrsim1$.

Binaries formed through the 3BBF channel subsequently interact with single stars and other binaries in the star cluster via fly-bys and exchanges, in which the incoming third body replaces a binary member. The probability for exchanges strongly increases with the increasing mass of the third body \citep{Valtonen2006}, so massive stars prominently end up in binary systems through the exchange channel. In fly-bys, binaries just exchange energy and angular momentum with the single cluster members. On average, hard binaries will further harden, and wide binaries will widen in dynamical interactions. This fundamental result is know as Heggie's law or the Heggie-Hills law \citep{Heggie1975,Hills1975}.

The increased interaction cross-section of the binary stars compared to single stars enhances the stellar collision rate in the star clusters. It has been demonstrated that the runaway collision sequence in dense star clusters typically begins with such a single-binary interaction including massive stars \citep{Gaburov2008}.

\subsection{Can the clusters retain their (IM)BHs?}\label{section: gwkick}

\begin{figure}
\includegraphics[width=\columnwidth]{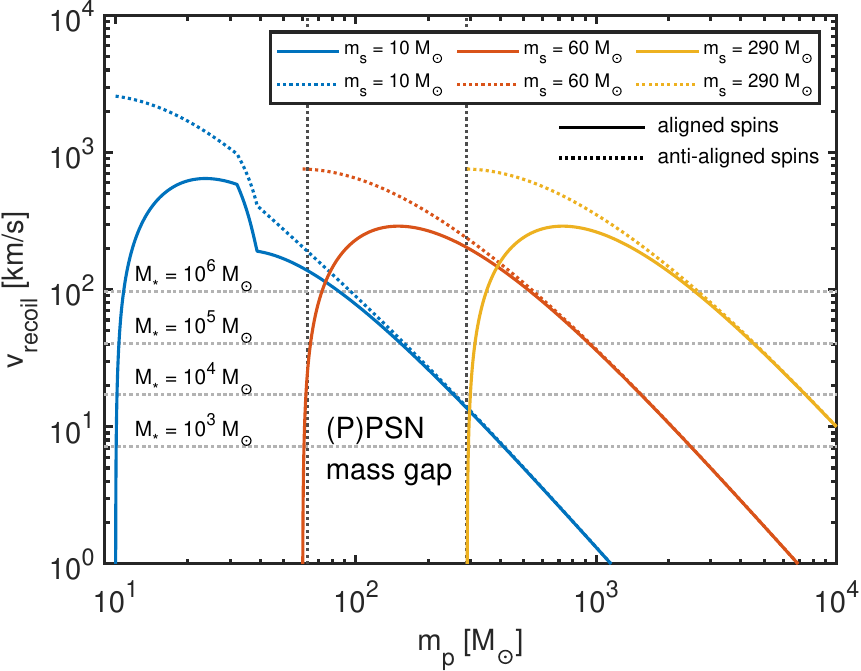}
\caption{The gravitational wave recoil kick velocities $v_\mathrm{recoil}$ \citep{Zlochower2015} for merging BH binaries as a function of the primary mass $m_\mathrm{p}$ for three secondary masses $m_\mathrm{s} = \msol{10}$ (in blue), $m_\mathrm{s} =\msol{60}$ (in orange) and $m_\mathrm{s} =\msol{290}$ (in yellow). The primary mass range is $m_\mathrm{s}\leq m_\mathrm{p} \leq \msol{10^4}$. We show the two extreme cases of aligned (solid lines) and anti-aligned spins (dashed lines). The spin magnitudes originate from the Geneva model \citep{BelczynskiKlencki2020}. Escape velocities for star clusters with masses up to $M_\star = \msol{10^6}$ are shown in horizontal gray lines. The cluster densities are similar as in numerical simulations in this work. In general, only equal-mass mergers or mergers with considerable mass ratios ($m_\mathrm{s}/m_\mathrm{p} \lesssim0.04$--$0.1$) lead to weak kicks after which the remnant BH can remain in the cluster. Increasingly misaligned spins cause stronger recoil kicks.}
\label{fig: gwrecoil}
\end{figure}

Even if IMBHs would form in young, massive star clusters, they might not be indefinitely retained in their host clusters. Assuming the common \cite{Plummer1911} model, the escape velocity $v_\mathrm{esc}$ from the center of the cluster is
\begin{equation}
    v_\mathrm{esc} \approx 33.4\: \mathrm{km/s} \left( \frac{M_\mathrm{\star}}{10^5\:\mathrm{M_\odot}} \right)^{1/2} \left( \frac{r_\mathrm{h}}{1.0\:\mathrm{pc}} \right)^{-1/2},
\end{equation}
only of the order of a few tens of km/s to $\sim 100$ km/s for star clusters in the mass range of $M_\star=\msol{10^5}$--$\msol{10^6}$. Even Newtonian three-body and multiple interactions are occasionally strong enough to unbind BHs and stars from their host clusters, especially in low escape velocity environments of lower-mass star clusters and in clusters containing multiple IMBHs (\citealt{HolleyBockelmann2008, Fragione2018c, Maliszewski2022, ArcaSedda2023b}, Souvaitzis et al., in prep.).

Relativistic gravitational-wave recoil kicks \citep{Bekenstein1973,Campanelli2007a,Campanelli2007b,Zlochower2011} caused by anisotropic emission of gravitational radiation during BH-BH mergers can easily eject the BH merger products from their stellar environments. The recoil kick velocity is low for merging equal-mass BH binaries ($m_\mathrm{p}=m_\mathrm{s}$) and binaries with a considerable mass ratio, i.e. $m_\mathrm{s}/m_\mathrm{p}\ll1$. Spinning BHs receive in general stronger recoil kicks compared to non-spinning BHs, especially if the BH spin directions are anti-aligned. The maximum recoil kick velocity is of the order of $v_\mathrm{recoil}\sim4000$ km/s for spinning BHs \citep{Campanelli2007b}, and $v_\mathrm{recoil}\sim175$ km/s for non-spinning ones \citep{Gonzales2007}.

We illustrate typical gravitational-wave recoil kick velocities compared to escape velocities from dense star clusters in Fig. \ref{fig: gwrecoil}. The recoil velocities are calculated using the fitting formulas provided by \cite{Zlochower2015}. We use the same formulas in \bifrost{} as briefly described in \cite{Rantala2023a}. We use two spin direction options (aligned and anti-aligned spins), and three different secondary BH masses of $m_\bullet=\msol{10}$ (common low-mass BHs), $m_\bullet=\msol{60}$ (BHs just below the (P)PISN mass gap) and $m_\bullet=\msol{290}$ (IMBHs just above the (P)PISN mass gap). The primary BH mass $m_\mathrm{p}$ ranges in each model from $m_\mathrm{s}$ up to $\msol{10^4}$ in the IMBH mass range. As in \bifrost, we assume the Geneva model for BH spins \citep{BelczynskiKlencki2020} as in their Eq. (3) with $Z=0.0004=0.02\:Z_\odot$. 

From Fig. \ref{fig: gwrecoil} it is evident that it is difficult to retain a merger remnant originating from a BH-BH binary that includes a BH below the mass gap ($m_\mathrm{p}\lesssim \msol{63}$), unless the merging binary is equal-mass or nearly so, and has almost aligned spins. Primary IMBHs above the mass gap ($m_\mathrm{p}\gtrsim\msol{290}$) can merge with low-mass BHs ($m_\mathrm{s}=\msol{10}$) with any relative spin orientation and be retained already in star clusters with  masses $M_\star\sim \msol{10^3}$--$\msol{10^4}$. However, the primary IMBH masses above $m_\bullet\gtrsim\msol{500}$ and $m_\bullet\gtrsim\msol{1000}$ are required to retain the merger product in the star clusters with masses $M_\star=\msol{10^6}$ and $M_\star=\msol{10^5}$, respectively, after a merger with any stellar-mass BH below the mass gap. IMBH primaries with $m_\mathrm{p}\gtrsim\msol{2000}$ are difficult to remove from any star cluster in mergers with stellar-mass BHs. For further discussion about merging IMBHs with stellar mass BHs see e.g. \cite{ArcaSedda2021b}. Finally, IMBH-IMBH mergers may remove even massive primary IMBHs from their host clusters, provided that the IMBH formation channel results in multiple IMBHs in a single star cluster. A primary mass of $m_\mathrm{p}\sim\msol{4500}$ is required to retain the remnant after a merger with an above-mass-gap secondary IMBH with $m_\bullet = \msol{290}$ in a star cluster with $M_\star = \msol{10^5}$.

\section{Numerical methods and initial conditions}

\subsection{Numerical methods: the updated \bifrost{} code}

We use an updated version of the \bifrost{} code \citep{Rantala2021,Rantala2023a} for the simulations of this study. In summary, \bifrost{} is a novel GPU-accelerated collisional direct-summation N-body code based on the hierarchical fourth-order forward symplectic integrator. Few-body subsystems such as binary stars, close fly-bys, hierarchical multiples and small clusters around black holes are treated using both secular and regularized integration techniques, including post-Newtonian equations of motion for BHs up to order PN1.0 in secular techniques and PN3.5 in regularized integration. \bifrost{} has already been used to study a number of stellar-dynamical topics involving BHs, IMBHs and MBHs, such as IMBH growth through repeated TDEs \citep{Rizzuto2023}, the dynamics of parsec-scale stellar disks in galactic nuclei and milliparsec-scale S-cluster analogues around MBHs \citep{Rantala2024a}, and escapers from merging older star clusters with and without pre-existing IMBHs (Souvaitzis et al., in prep.).

The main differences of the updated \bifrost{} code version used in this study compared to the original code of \cite{Rantala2023a} are the following. First, a key element of the fourth-order forward integration, the calculation of the term involving the acceleration gradient, is simplified (appendix \ref{section: hhsfsi-omelyan}). Next, the secular triple integration and the standard algorithmically regularized few-body integration (LogH) are replaced by an implementation of the slow-down algorithmic regularization (SDAR) integrator (appendix \ref{section: sdar}). Finally, we present the coupling of the fast stellar population synthesis code \sevn{} \citep{Iorio2023} into \bifrost{} allowing stellar evolution in the N-body simulations (appendix \ref{section: sevn}). In this study, we focus on single stellar evolution in the initial stellar mass range of $\msol{2.2}\lesssim m_\star \lesssim \msol{600}$. Stars in dynamically formed binaries evolve as single stars, ignoring any binary stellar evolution physics. Setups with primordial binary stars and full binary stellar evolution will be explored in a forthcoming study. Each of the three code updates is described in detail in the appendix.

\subsection{Initial conditions: individual star cluster models}

The individual star clusters in our initial models follow the \cite{Plummer1911} density profile
\begin{equation}
    \rho(r) = \frac{3 M_\mathrm{\star}}{4 \pi a^3} \left(1+\frac{r^2}{a^2} \right)^{-5/2}
\end{equation}
and the corresponding potential. Here $M_\mathrm{\star}$ is the total stellar mass of the model and $a$ is the scale radius related to the three-dimensional half-mass radius $r_\mathrm{h}$ of the model as $r_\mathrm{h} = (2^{2/3}-1)^{-1/2}a \approx 1.3a$. The projected effective radius $R_\mathrm{e}$ is simply $R_\mathrm{e}=a$. The spherically symmetric positions and isotropic velocities of the individual stars are sampled from the distribution function of the Plummer density-potential pair.

Simulation studies of massive star cluster evolution or collisional IMBH formation do not typically assume an initial mass-size relation for their sets of star cluster models, but rather select $M_\star$ (or $N$), and the central $\rho_\mathrm{c}$ or the half-mass density $\rho_\mathrm{h}$ to examine the cluster evolution or IMBH formation at different chosen cluster masses and densities. Our approach in this study is to choose an observationally motivated mass-size relation which all of our star cluster models will initially follow. 

The observed effective radii $R_\mathrm{e}$ of star clusters follow a power-law star cluster mass-size relation
\begin{equation}\label{eq: brown-gnedin}
    \frac{R_\mathrm{e}}{\mathrm{pc}} = R_\mathrm{4}\left( \frac{M_\mathrm{\star}}{10^4M_\odot}\right)^\beta
\end{equation}
for which \cite{Brown2021} provide values of $\beta=0.180\pm0.028$ and $R_\mathrm{4}=2.365\pm0.106$ for star cluster of ages between $1$ Myr and $10$ Myr in the LEGUS sample \citep{Adamo2017,Cook2019}. However, it has been suggested that the birth radii of star clusters may be smaller, and clusters increase in size during later evolution. Studying young clusters in their embedded phase, \cite{Marks2012} find that the half-mass radius scales as the function of the embedded cluster mass $M_\mathrm{ecl}$ as
\begin{equation}\label{eq: marks-kroupa}
    \frac{r_\mathrm{h}}{\mathrm{pc}} = 0.10_\mathrm{-0.04}^\mathrm{+0.07}\times\left( \frac{M_\mathrm{ecl}}{M_\odot}\right)^\mathrm{0.13\pm0.04}.
\end{equation}
Thus, a size increase by a factor of up to $\sim 10$ is required from the embedded phase to the young massive cluster phase. The sizes of star clusters increase due to internal collisional dynamics, external tides and stellar evolution (e.g. \citealt{ArcaSedda2023a}), with gas expulsion suggested as an additional physical process contributing to the size growth \citep{Banerjee2017}.

We adopt small initial birth radii for our star cluster model population. The relation between the stellar mass $M_\star$ to the half-mass radii $r_\mathrm{h}$ of our clusters is
\begin{equation}\label{eq: adopted-mass-size-relation}
    \frac{r_\mathrm{h}}{\mathrm{pc}} = \frac{f_\mathrm{h} R_\mathrm{4}}{1.3} \left( \frac{M_\mathrm{\star}}{10^4M_\odot} \right)^\beta
\end{equation}
in which $f_\mathrm{h}$ is a free parameter and the constant $r_\mathrm{h}/R_\mathrm{e} \sim 1.3$ arises from the properties of the Plummer model. While using the values of $\beta$ and $R_\mathrm{4}$ and their uncertainties from \cite{Brown2021} in Eq. \eqref{eq: brown-gnedin}, we set $f_\mathrm{c}=1/8$ so that at $M_\star=\msol{10^5}$ the half-mass radius of the star cluster ($r_\mathrm{h} = 0.45$ pc) corresponds to the size from the relation of \cite{Marks2012} in Eq. \eqref{eq: marks-kroupa}.

We illustrate the particle numbers and half-mass densities of a number of direct N-body \citep{Wang2015,DiCarlo2021,Banerjee2021,Rastello2021,Rizzuto2021,Kamlah2022,ArcaSedda2023a} and cluster Monte Carlo \citep{Askar2017,Rodriguez2016,Rodriguez2019,Kremer2020a,Maliszewski2022} simulations following Fig. 1 of \cite{ArcaSedda2023a} in Fig. \ref{fig: nbody_literature}. We show both our isolated star cluster models, and the population of star clusters for the hierarchical cluster assembly simulations. The cluster assembly region simulations consist of individual star clusters with particle numbers in the range $1.7\times10^2\lesssim N \lesssim 4.0\times 10^5$ particles, up to $N_\mathrm{tot}\sim2.35\times10^6$ particles per simulation. The isolated simulations we run have up to $N=1.0 \times 10^6$ particles. Our adopted mass-size relation in Eq. \eqref{eq: adopted-mass-size-relation} results in an initial star cluster population consistent with other recent simulation studies of massive star cluster evolution and IMBH formation.

\begin{figure}
\includegraphics[width=\columnwidth]{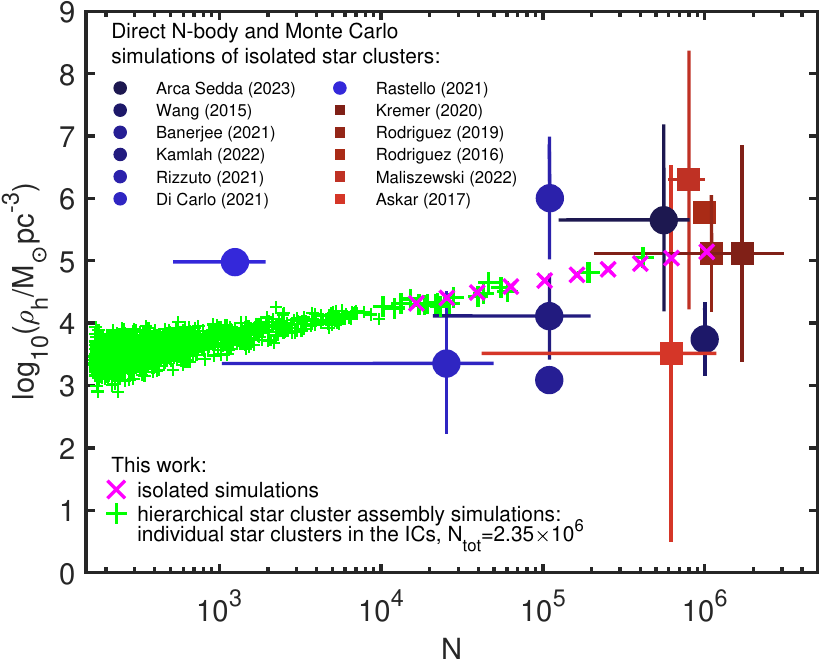}
\caption{A comparison of the initial conditions of recent direct N-body (blue filled circles) and Monte Carlo (red filled squares) studies of star clusters following Fig. 1 of \citet{ArcaSedda2023a}. The figure shows the half-mass density $\rho_\mathrm{h}$ of the star clusters as a function of their particle number $N$. The references to the individual studies are listed in the main text. Our isolated star clusters (up to $N=10^6$) are indicated in purple cross symbols while the initial star clusters in the hierarchical cluster assembly simulations (in total $N_\mathrm{tot}=2.35\times10^6$ particles) are shown in green. The densities of our star cluster models are consistent with the other recent star cluster studies in the literature. 
}
\label{fig: nbody_literature}
\end{figure}

We sample the masses of individual stars from the standard piece-wise power-law stellar initial mass function of \cite{Kroupa2001} as
\begin{equation}\label{eq: imf-powerlaw}
    \derfrac{N_\mathrm{\star}}{m_\mathrm{\star}} \propto m_\star^{-\alpha}
\end{equation}
with $\alpha=1.3$ when $0.08M_\odot \leq m_\star \leq 0.50M_\odot$ and $\alpha=2.3$ for $0.50M_\odot< m_\star \leq m_\mathrm{max}$. The maximum mass of an individual star $m_\mathrm{max}$ in a single star cluster depends on the mass $M_\mathrm{\star}$ of the cluster in question \citep{Weidner2006}. We adopt a simple piece-wise power-law model
\begin{equation}\label{eq: Mcl-mmax-relation}
    \log_\mathrm{10}\left(\frac{m_\mathrm{max}}{M_\odot}\right) = a\times\log_\mathrm{10}\left(\frac{M_\mathrm{\star}}{M_\odot}\right) + b.
\end{equation}
We choose the constant coefficients $a$ and $b$ presented in Table \ref{table: mmax} in such a manner that the simple model reproduces the main features of the cluster mass -- maximum stellar mass relation of \cite{Yan2023}. The maximum individual stellar masses for star clusters with masses $M_\mathrm{\star}=10^2M_\odot$, $M_\mathrm{\star}=10^3M_\odot$ and $M_\mathrm{\star}=10^4M_\odot$ are approximately $m_\mathrm{max} \sim 11 M_\odot$, $m_\mathrm{max} \sim25 M_\odot$ and $m_\mathrm{max} \sim85 M_\odot$, respectively. Furthermore, we adopt an overall maximum initial stellar mass of $m_\mathrm{max} = 150M_\odot$ for the most massive star clusters. The maximum mass of a single star in a star cluster is not typically strictly enforced when generating star cluster initial conditions for simulations. Instead, massive stars are simply less probable to occur in low-mass clusters. In our case, our hierarchical star cluster assembly initial setup contains a large number of low-mass clusters, and we wish to avoid small star clusters with very massive individual stars.

\begin{table}
\centering
\begin{tabular}{l c c}
\hline
Cluster mass range & coeff. a & coeff. b\\
\hline
$\log_\mathrm{10}(M_\mathrm{\star}/M_\odot) \leq 1.5$ & $+0.51$ & $-0.17$\\
$1.5 < \log_\mathrm{10}(M_\mathrm{\star}/M_\odot) \leq 2.0$ & $+0.95$ & $-0.84$\\
$2.0 < \log_\mathrm{10}(M_\mathrm{\star}/M_\odot) \leq 2.6$ & $+0.26$ & $+0.52$\\
$2.6 < \log_\mathrm{10}(M_\mathrm{\star}/M_\odot)$ & $+0.53$ & $-0.23$\\
\hline
\end{tabular}
\caption{The constant coefficients $a$ and $b$ for the simple model for the maximum mass of an individual star in a star cluster with a mass of $M_\mathrm{\star}$ for Eq. \eqref{eq: Mcl-mmax-relation}.}
\label{table: mmax}
\end{table}

We set the stellar metallicity to a constant value of $Z=0.0002=0.01\;Z_\odot$ to emulate the conditions of the low-metallicity star cluster environment at high redshifts. The initial stellar radii are obtained by using the stellar evolutionary tracks of \sevn. Each star begins the simulation at the beginning of its \sevn{} main sequence stellar evolutionary phase, i.e. initially \texttt{phase\_star=1}. All stars in the initial conditions are single stars. We will explore the effect of primordial binaries on our setup in a forthcoming study.

\subsection{Initial conditions: star cluster assembly regions}

The masses $M_\star$ of the star clusters are obtained from the power-law cluster mass function
\begin{equation}
\derfrac{N_\mathrm{cl}}{M_\mathrm{\star}} \propto M_\mathrm{\star}^{\alpha}
\end{equation}
for which both observations (e.g. \citealt{Elmegreen1996,Zhang1999,Adamo2020}) and solar-mass resolution hydrodynamical simulations \citep{Lahen2020} suggest a slope of $\alpha\sim-2$. We sample star cluster masses from the cluster mass function until the desired total stellar particle number $N_\mathrm{tot}$ is reached. For the hierarchical cluster assembly simulations of this study $N_\mathrm{tot}$ is in the range $1.7\times10^6\leq N_\mathrm{tot} \leq2.35\times10^6$. We show the particle numbers $N$ and the half-mass densities $\rho_\mathrm{h}$ of the individual star clusters in Fig. \ref{fig: nbody_literature}.

After the star cluster masses $M_\mathrm{\star}$ have been obtained, we set the initial center-of-mass positions and velocities of the clusters. Our setup is motivated by the structure of star cluster formation regions in the solar-mass resolution hydrodynamical simulations of star-bursting dwarf galaxies of \cite{Lahen2020}. We choose a representative star formation region with dimensions of $47$ pc, $77$ pc and $84$ pc from the simulation volume which contains a large number of young ($t\leq 2$ Myr) star clusters (for an illustration, see the top panel in Fig. 2 of \citealt{Lahen2019}). The total stellar mass in young stars within the region is $M_\star\sim4.8\times10^5 M_\odot$ while most massive individual gravitationally bound cluster has a mass of $M_\mathrm{\star} \sim1.5\times10^5 M_\odot$. Centering on the most massive cluster, the radial cumulative mass function is close to a power-law distribution $M_\mathrm{tot}(<r)\propto r^\gamma$ with $\gamma \sim 3$, indicating a nearly constant density of the star cluster region. The region is in a state of collapse with a mean radial infall velocity of $\langle v_\mathrm{r} \rangle \approx -7$ km/s. In addition to the collapse, the star clusters have random motions up to $\norm{\vect{v}-\langle \vect{v}_\mathrm{r} \rangle} \sim 14$ km/s. Not all clusters are gravitationally bound to the bulk of the system. Half of the stellar mass of the final proto-globular cluster is still yet to form and the sub-clusters are embedded in a filamentary gaseous structure. 

We begin the initial setup construction by placing our most massive sampled Plummer model at the origin at zero velocity. For simplicity, we choose a spherically symmetric distribution of cluster positions even though the spatial distribution of the representative star cluster region in the hydrodynamical simulation of \cite{Lahen2020}
is a somewhat flattened configuration. The assumption allows us to sample the star cluster positions within a homogeneous sphere. For the cluster maximum separation from the origin we set $r_\mathrm{max} = 50$ pc. For the star cluster velocities we use an isotropic random component up to $3.5$ km/s combined with a radial component of $v_\mathrm{r} = -3.5$ km/s for each cluster. The radial cluster infall velocities are by a factor of $\sim 2$ smaller than in the representative region of \cite{Lahen2020} to compensate for the fact that half of the stellar mass of the region in the hydrodynamical simulation is still to be formed. The chosen random velocity component ensures that most of the stellar mass in the region will end up in a single cluster or its diffuse envelope.

\begin{table*}
\begin{center}
\begin{tabular}{c c c c c c c c}
\hline
Simulation & $N_\mathrm{cl}$ & $N_\mathrm{tot}$ & $N_\mathrm{min}$ & $N_\mathrm{max}$ & $M_\mathrm{tot}$ & $M_\mathrm{min}$ & $M_\mathrm{max}$\\ 
 &  &  &  &  & $[M_\odot]$ & $[M_\odot]$ & $[M_\odot]$\\
\hline
H1 & $1.10\times10^3$ & $1.70\times10^6$ & $170$ & $3.45\times10^5$ & $9.42\times10^5$ & $63.1$ & $2.02\times10^5$\\
H2, H3 & $1.43\times10^3$ & $2.35\times10^6$ & $204$ & $4.07\times10^5$ & $1.29\times10^6$ & $73.2$ & $2.30\times10^5$\\
\hline
\end{tabular}
\caption{The three hierarchical star cluster assembly simulation setups H1, H2 and H3. The table lists the total number of star clusters ($N_\mathrm{cl}$), the total number of stars ($N_\mathrm{tot}$), the minimum and the maximum numbers of particles in a single cluster ($N_\mathrm{min}$ and $N_\mathrm{max}$), the total stellar mass ($M_\mathrm{tot}$), the minimum and the maximum masses of a single cluster ($M_\mathrm{min}$ and $M_\mathrm{max}$). The setups H2 and H3 are different random realizations from the same list of cluster masses with different particle masses, positions and velocities within the clusters as well as different initial cluster positions and velocities.}
\label{table: H1-H2-H3}
\end{center}
\end{table*}

\begin{table}
\begin{tabular}{l l l}
\hline
\bifrost{} parameter & symbol & value\\
\hline
integration interval /& $\epsilon_\mathrm{max}$ & $5\times10^{-3}$ Myr \\
maximum time-step &  & \\
forward integrator time-step factor & $\eta_\mathrm{ff}$, $\eta_\mathrm{fb}$, $\eta_\mathrm{\nabla}$ & $0.2$\\
subsystem neighbor radius & $r_\mathrm{rgb,\star}$ & $0.25$ mpc\\
SDAR GBS tolerance  & $\eta_\mathrm{GBS}$ & $5\times10^{-8}$\\
SDAR GBS end-time tolerance & $\eta_\mathrm{endtime}$ & $10^{-2}$\\
regularization highest PN order &  & PN3.5\\
\hline
\end{tabular}
\caption{The user-given \bifrost{} parameters used in the simulations of this study.}
\label{table: bifrost_params}
\end{table}

We construct in total three hierarchical cluster assembly region setups which we label H1, H2 and H3. The properties of the cluster assembly setups are listed in Table \ref{table: H1-H2-H3}. The setups H2 and H3 have the same cluster particle numbers and cluster masses, but are otherwise different random realizations with differing initial cluster positions and velocities as well as different randomly drawn particle masses, positions and velocities for stars within the clusters. Each cluster formation region is evolved using our \bifrost{} code until $t=50$ Myr with user-given accuracy parameters listed in Table \ref{table: bifrost_params}. The simulations were run between November 2023 and March 2024 using the FREYA cluster and the supercomputer RAVEN hosted by the Max Planck Computing and Data Facility (MPCDF), in Garching, Germany using $2$--$4$ computing nodes with either two Nvidia Tesla P100-PCIE-16GB or four Nvidia Tesla A100-PCIE-40GB GPUs and $40$-$72$ CPUs per node.

\section{IMBHs from star cluster assembly}

\begin{table*}
\begin{center}
\begin{tabular}{c c c c c c c c c}
\hline
Cluster & IMBH & $M_\mathrm{\star}$ & $\rho_\mathrm{c,init}$ & $t_\mathrm{seg}$ & $t_\mathrm{cc}$ & $t_\mathrm{acc}$ & $t_\mathrm{cc}<t_\mathrm{acc}?$ & $M_\mathrm{\bullet}(t_\mathrm{acc})$\\
 & & $\mathrm{[M_\odot]}$ & $[\rhosol{10^6}]$ & $\mathrm{[Myr]}$ & $\mathrm{[Myr]}$ & $\mathrm{[Myr]}$ & & $\mathrm{[M_\odot]}$\\
\hline
H1-A &  & $2.0\times10^5$ & 1.77 & 0.32 & 7.60 & & &\\
H1-B & BH1-B & $4.4\times10^4$ & 0.88 & 0.10 & 1.95 & 6.5& $\checkmark$ & $1194$\\
H1-C & BH1-C & $1.1\times10^5$ & 1.34 & 0.20 & 4.43 &3.1& - & $536$\\
H1-D & BH1-D & $5.9\times10^4$ & 1.01 & 0.12 & 2.53 &2.3& - & $348$\\
\hline
H2-A & & $2.3\times10^5$ & 1.89 & 0.35 & 8.63 & & &\\
H2-B & BH2-B & $3.2\times10^4$ & 0.76 & 0.08 & 1.47 & 4.8 & $\checkmark$ & $1256$\\
H2-C & BH2-C & $2.6\times10^4$ & 0.69 & 0.08 & 1.23 &- & & $350$\\
\hline
H3-A & & $2.3\times10^5$ & 1.89 & 0.35 & 8.63 & & &\\
H3-B & BH3-B & $1.1\times10^5$ & 1.34 & 0.20 & 4.43 & 2.2 & - & $1748$\\
H3-C & BH3-C & $1.5\times10^4$ & 0.54 & 0.06 & 0.76 & 8.7 & $\checkmark$ & $347$\\
H3-D & BH3-D & $2.3\times10^4$ & 0.65 & 0.07 & 1.10 & 3.7 & $\checkmark$ & $338$\\
H3-E & BH3-E & $1.4\times10^4$ & 0.52 & 0.06 & 0.71 & 6.9 & $\checkmark$ & $289$\\
H3-F & BH3-F & $2.6\times10^4$ & 0.69 & 0.08 & 1.23 & 7.6 & $\checkmark$ & $105$\\
\hline
\end{tabular}
\caption{The initial stellar masses $M_\mathrm{\star}$ of the star clusters which produce an IMBH during the cluster assembly simulations. The most massive, central initial cluster is indicated with a label A. The estimates for the mass segregation and core collapse time-scales ($t_\mathrm{seg}$, $t_\mathrm{cc}$) are obtained using Eq. \eqref{eq: rlx} and \eqref{eq: tseg}.} The accretion time $t_\mathrm{acc}$ indicates when the star cluster merges with the central cluster, and $M_\mathrm{\bullet}(t_\mathrm{acc})$ describes the mass of the most massive IMBH at this time. If $t_\mathrm{acc}$ is small enough, IMBHs have had no time to form yet. In this case the mass of the IMBH descending from the most massive star at the time of the accretion is given.
\label{table: mcluster}
\end{center}
\end{table*}

\begin{figure*}
\centering
\includegraphics[width=0.75\textwidth]{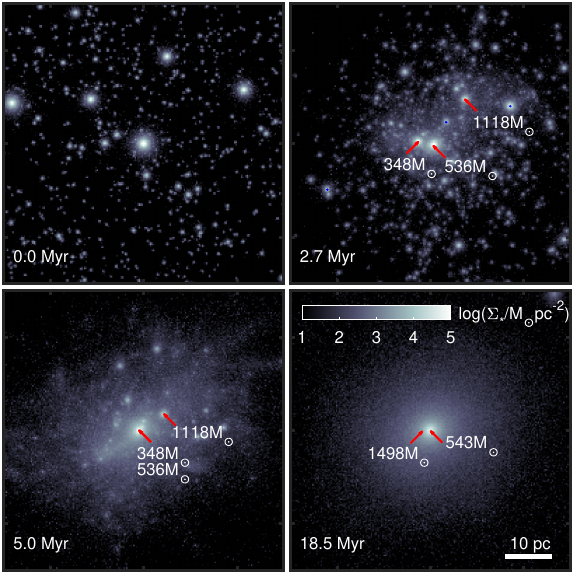}
\caption{The evolution of the star cluster assembly regions into a single, massive star cluster in the simulation H1. The gray and white colors indicate the projected stellar density of the clusters. Initially, the star clusters populate a region with a radius of $50$ pc (top left panel). Most of the in-falling star clusters interact and merge with the central star cluster between $\sim2$ Myr and $\sim10$ Myr (top right and bottom left panels). Intermediate-mass black holes (indicated with red arrows) from between $\sim 2$ Myr and $\sim 4$ Myr as collapsed remnants of VMSs formed in sequences of repeated collisions of massive stars. VMSs in the (P)PISN mass gap leaving either a small BH or no remnant are indicated in the top-right panel as small, blue dots in dense star clusters. After $\gtrsim10$ Myr the region settles into a single, smooth, spherical star cluster with the remaining sub-clusters either escaping the region or being disrupted by the main cluster. The bottom right panel shows the cluster at $18.5$ Myr when an IMBH with $m_\bullet = 1498\;\mathrm{M_\odot}$ escapes the cluster due to a gravitational-wave recoil kick of $v_\mathrm{recoil} = 206$ km/s.}
\label{fig: largescale}
\end{figure*}

\begin{figure*}
\centering
\includegraphics[width=0.75\textwidth]{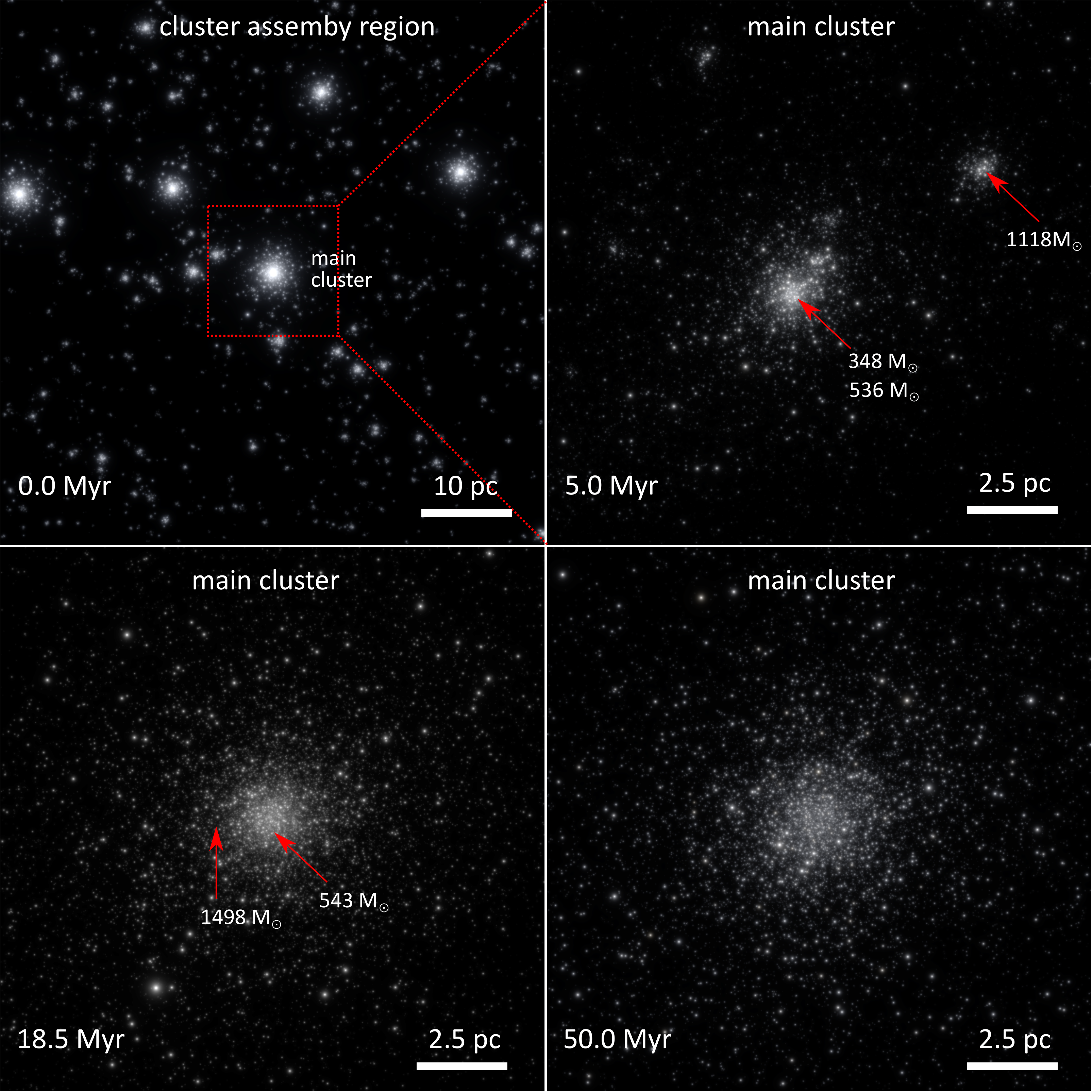}
\caption{A simulated BVR filter view of the massive stars ($m_\star>2\;\mathrm{M_\odot}$) in the star cluster H1-A during its evolution generated using the \myosotis{} visualization code \citep{Khorrami2019}. The top left panel displays the initial cluster assembly region as in Fig. \ref{fig: largescale} while the remaining three panels focus on the central cluster at $t=5$ Myr (top right), $t=18.5$ Myr (bottom left) and $t=50.0$ Myr (bottom right). At $t=5.0$ Myr, the cluster assembly is ongoing and substructure such as the disrupting remains of the star cluster H1-B is clearly visible in the image near and around the central cluster. At later times the stellar density of the main cluster decreases due to the combined effects of collisional gravitational dynamics and stellar evolution. In the simulation H1, no IMBHs inhabit the star cluster at the end of the simulation due to gravitational-wave recoil kicks exceeding the cluster escape velocity in the aftermath of BH-IMBH and IMBH-IMBH mergers.}
\label{fig: smallscale}
\end{figure*}

\subsection{Star cluster assembly}

The collapse of a star cluster assembly region H1 into a single, massive star cluster is illustrated in Fig. \ref{fig: largescale}. Initially, the in-falling star cluster population resides in a volume with $r_\mathrm{max} = 50$ pc in radius with the most massive star cluster at the center. The massive star clusters begin to strongly interact and merge with the central cluster after $t\gtrsim 2$ Myr. Before merging, clusters are tidally stripped in close encounters with the main cluster, in some cases multiple times. Most, though not all, in-falling star clusters eventually merge into the main cluster with most star cluster mergers occurring before $t\lesssim 10$ Myr. As massive stars have short life-times ($t = 2.82$ Myr for $m_\star=\msol{150}$ at $Z=0.01Z_\odot$), massive stars may already have reached the end of their lifetimes when their host cluster merges into the main star cluster.

The initial substructure around the main cluster is gradually erased as unbound material escapes from the vicinity of the cluster while the remaining bound remnants of the in-falling clusters are tidally disrupted. After $t\gtrsim 10$ Myr, the cluster assembly region has settled into a single massive, spherical and smooth star cluster. We use a modified version of the publicly available \myosotis{} code \citep{Khorrami2019} to generate mock BVR images of the simulated star clusters. The evolution of the star cluster region H1 and its initially most massive cluster H1-A is illustrated in mock observations in Fig. \ref{fig: smallscale}. The visualizations only involve stars with masses above $m_\star \gtrsim2.2\:\mathrm{M_\odot}$ as this is the minimum stellar mass in stellar tracks used in this study. IMBH positions and masses are also indicated in the illustration.

\begin{figure}
\includegraphics[width=\columnwidth]{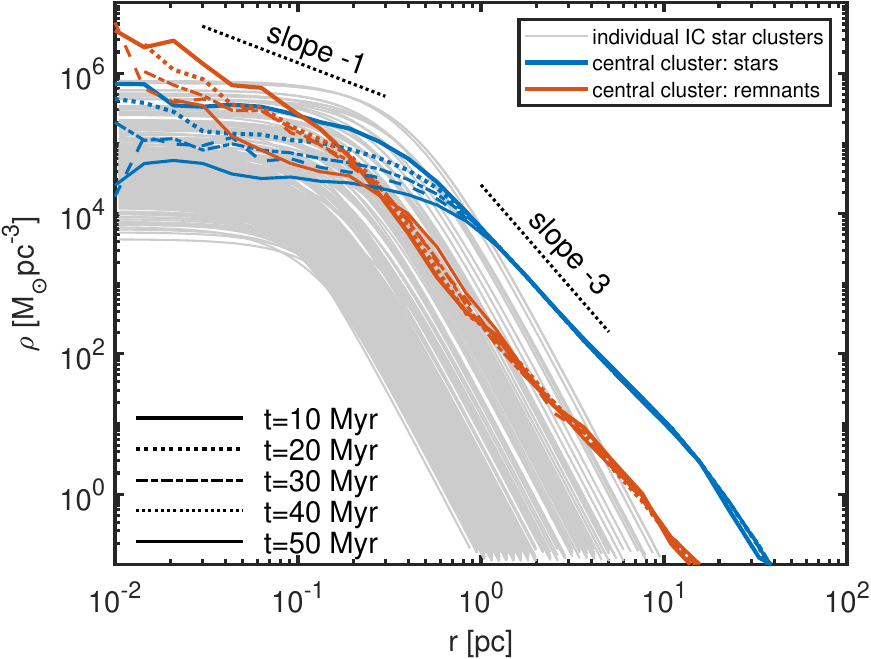}
\caption{The density profiles of the initial individual star cluster models (gray lines) as well as the stars (blue lines) and the compact remnants (orange lines) in the central cluster from $t=10$ Myr to $t=50$ Myr in intervals of $10$ Myr. At $t=10$ Myr, the density of the central $\lesssim0.10$ pc is dominated by compact remnants, mostly BHs, while at $t=50$ Myr the BHs dominate the region within the central $0.22$ pc. Both the central stellar and compact remnant densities decrease as a function of time while the outer parts beyond $1.0$ pc (stars) and $0.3$ pc (remnants) remain relatively unchanged. While the stellar profiles are almost flat at the center, the compact remnants have a more cuspy central profile.}
\label{fig: density-star-remnant}
\end{figure}

We present the density profiles of stars and compact remnants (mostly stellar-mass BHs) in the central cluster in Fig. \ref{fig: density-star-remnant} between $t=10$ Myr and $t=50$ Myr in intervals of $10$ Myr. IMBHs are excluded from the density profile calculation. Already at $t=10$ Myr the profiles are smooth with little indication of remaining substructure. The central stellar profile shows an almost flat core as in the initial Plummer profile while the BHs have a cuspy central structure with a power-law slope close to $-1$. The central stellar density at $t=10$ Myr is $\rho_\mathrm{c,\star}\sim\rhosol{7\times10^5}$ while the BH density is
somewhat higher, $\rho_\mathrm{c,\bullet}\sim\rhosol{3\times10^5}$--$\rhosol{5\times10^6}$. BHs are the dominant density component within the central regions of the cluster, $r\lesssim0.10$ pc. At later times the central densities decrease due to stellar evolution (stellar mass loss, BH formation) and dynamical interactions for both stars and BHs. At $t=50$ Myr, the central stellar density is $\rho_\mathrm{c,\star}\sim\rhosol{5\times10^4}$ while the BH density is $\rho_\mathrm{c,\bullet}\sim\rhosol{4\times10^5}$, and BHs are the dominating density component within the central part ($r\lesssim0.22$ pc) of the cluster. Both the stellar and BH density profiles remain unchanged in the outer parts within the first $t=50$ Myr of the simulations. The outer density profiles slopes are close to a power-law slope of $-3$ indicated in Fig. \ref{fig: density-star-remnant}, considerably less steep than the initial Plummer outer parts, which have a power-law slope of $-5$.

\subsection{Merger trees of star clusters, their VMSs and IMBHs}\label{section: mergertrees-intro}

We have modified the \texttt{FOF} (friends-of-friends) and \texttt{subfind} structure finder routines \citep{Springel2001,Dolag2009} of the \gadget{} simulation code \citep{Springel2021} to identify the individual star clusters and their merger histories from \bifrost{} output during the during the hierarchical assembly simulations. As the snapshot format of \bifrost{} closely resembles the \gadget{} HDF5 format, this is very straightforward. Given the \texttt{subfind} catalogs constructed from the \bifrost{} snapshots (interval $\Delta t_\mathrm{output}=10^{-2}$ Myr), we construct the star cluster merger trees and identify massive stars and IMBHs associated with each cluster. The IMBH progenitor merger trees can be constructed from the \bifrost{} snapshots and merger data output alone, supplemented by the information about cluster membership of the IMBH progenitors from the \texttt{subfind} data.

We present a detailed description of the formation and assembly history of the IMBHs and their host star clusters in the simulations H1, H2 and H3 in the following sections \ref{section: sim-H1}, \ref{section: sim-H2} and \ref{section: sim-H3}. In each simulation, at least two IMBHs form in the mass range from $m_\bullet \sim 105\;M_\odot$ up to $m_\bullet \sim 2200\:\mathrm{M_\odot}$.

Although the formation histories of the IMBHs in our simulations differ in details, a number of common characteristics can be readily observed. First, IMBHs form almost always from a collapsing VMS built in a sequence of mergers of massive stars. The sequence of stellar mergers eventually leading to a formation of an IMBH through direct collapse always begins from a star more massive than $m_\mathrm{\star}\sim 80\:\mathrm{M_\odot}$. The total number of mergers in a merger tree of a single IMBH can be up to $N_\mathrm{merg}\sim100$. Most of the stellar mass in the growing VMSs is acquired in mergers with stars with masses $10\:\mathrm{M_\odot} \lesssim m_\star \lesssim 150\:\mathrm{M_\odot}$. The branches other than the main branch of the merger tree before the IMBH formation are almost always short, i.e. the stars merging with the growing VMS have typically never merged before with other stars. There are only two mergers occurring in our simulation sample in which the colliding stars have both masses over $m_\star>\msol{150}$. Next, most of the IMBH mass originates from the VMS merger phase, and IMBHs grow in mass by less than $25\%$--$35\%$ by disrupting main sequence stars and evolved stars, and by merging with stellar-mass BHs and other IMBHs. More massive IMBHs acquire more mass by disrupting stars and merging with other BHs compared to less massive IMBHs. Finally, and most interestingly, the main progenitor stars of the IMBHs do not originate from the initially most massive star cluster. Instead, the main cluster accretes IMBHs and evolved VMSs formed in the lower-mass star clusters as they merge into the central cluster. This somewhat unexpected phenomenon is examined in section \ref{section: isolated-simulations}.

In simulations H1 and H3 more than a single IMBH ends up in the main star cluster. The interaction histories of the IMBHs in the main cluster are complex and rich in dynamical phenomena. These involve IMBH binaries, IMBH binary member exchanges, hierarchical triple IMBH configurations, strong triple interactions, gravitational-wave driven mergers of BH-IMBH and IMBH-IMBH binaries as well as IMBHs escaping the star clusters after strong gravitational-wave recoil kicks.

The initial masses $M_\star$ and central densities $\rho_\mathrm{c,init}$ of the star clusters hosting IMBH progenitors are presented in Table \ref{table: mcluster}. The table also presents the times $t_\mathrm{acc}$ when the clusters merge with the main central star cluster, and the IMBH mass at that time. If the IMBH has not formed yet at $t=t_\mathrm{acc}$, we provide the mass of the IMBH descending from the most massive VMS of the cluster at the time of its collapse. The table also shows the estimates for the mass segregation and core collapse time-scales ($t_\mathrm{seg}$, $t_\mathrm{cc}$) for the clusters based on Eq. \eqref{eq: rlx} and Eq. \eqref{eq: tseg} with $t_\mathrm{cc}\sim0.2\;t_\mathrm{rlx}$ \citep{PortegiesZwart2004}. Based on the estimate, mass segregation occurs rapidly in all models with $t_\mathrm{seg}<t_\mathrm{acc}$. However, not all clusters that form IMBHs are core collapsed when they merge with the central cluster. Thus, core collapse seems not to be the absolute bottleneck for collisional VMS and IMBH formation in the simulations of this study. Dense, mass segregated clusters can support repeated stellar collisions already at times $t_\mathrm{seg}<t<t_\mathrm{cc}$.

\subsubsection{Simulation H1}\label{section: sim-H1}

\begin{figure*}
\includegraphics[width=0.85\textwidth]{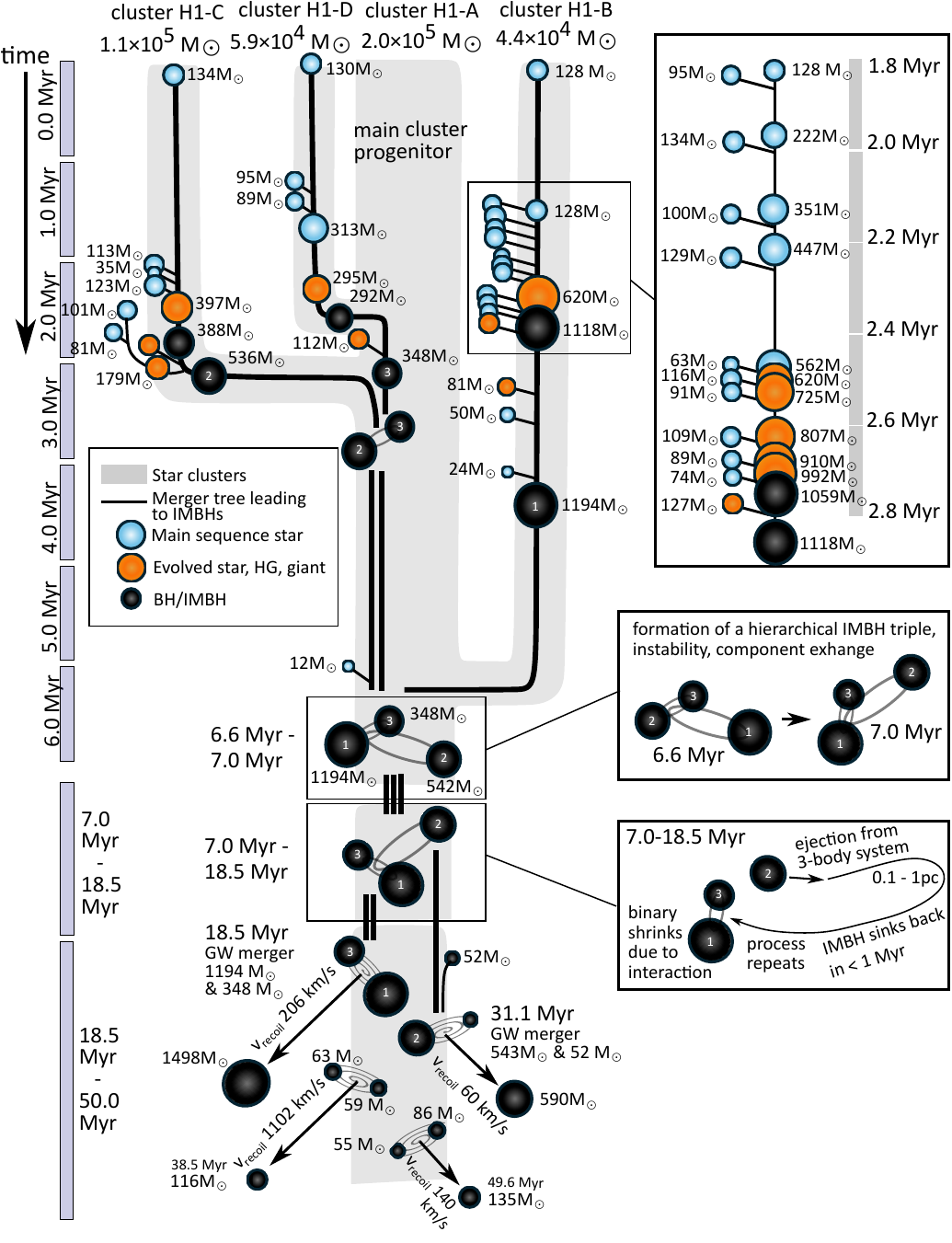}
\caption{The merger trees of stars leading to IMBHs (black symbols) in the simulation H1. IMBHs form through collapse of VMSs which have been built up in a series of collisions of massive main sequence (blue symbols) and giant (orange symbols) stars. No IMBHs originate from stars initially at the most massive, central star cluster H1-A. Three IMBHs with masses ranging from $m_\bullet\sim\msol{350}$ to $m_\bullet\sim\msol{1500}$ form in star clusters H1-B--H1-D less massive than the main cluster and are accreted into the main cluster between $t=3.0$ Myr and $t=6.5$ Myr. In the main cluster, the IMBHs undergo a series of IMBH binary and triple phases between until $t=18.5$, when IMBHs with masses of $m_\bullet=\msol{348}$ and $m_\bullet=\msol{1194}$ merge. The resulting IMBH with a mass of $m_\bullet=\msol{1498}$ receives a gravitational-wave recoil kick of $v_\mathrm{recoil}=206$ km/s, unbinding it from the star cluster. The final IMBH of the cluster is also ejected at later times after a merger with a stellar mass BH, as well as the remnants of two low-mass IMBHs formed though mergers of two stellar-mass BHs. Thus, no IMBHs remain in the cluster H1-A at the end of the simulation at $t=50$ Myr. The complex merger histories of the IMBH progenitors are described in more detail in the text.}
\label{fig: mergertree-23A}
\end{figure*}

The merger trees of stars leading to IMBHs and their host star clusters of the simulation H1 are presented in Fig. \ref{fig: mergertree-23A}. The mass of the central star cluster in the run is initially $M_\star = \msol{2.0\times10^5}$. Only four stars originating from the main cluster H1-A merge with stars more massive than $m_\star\gtrsim\msol{10}$, and the mass of these merger remnant stars never exceeds $m_\star=\msol{150}$. Thus, the most massive stars from the cluster H1-A end up in the (P)PISN mass gap, leaving behind no massive remnant at the end of their lives.

The IMBHs in the simulation H1 are formed in three smaller star clusters labeled H1-B (IMBH label BH1-B), H1-C (BH1-C) and H1-D (BH1-D) with decreasing IMBH mass. The initial stellar masses of the clusters are $M_\star=\msol{4.4\times10^4}$, $M_\star=\msol{1.1\times10^5}$ and $M_\star=\msol{5.9\times10^4}$, respectively. The progenitor of the IMBH BH1-B in the star cluster H1-B is a massive star with an initial mass of $m_\star=\msol{128}$. After the mass segregation phase, stellar collisions between the main progenitor and other massive stars begin at the center of the cluster at $t\gtrsim1.8$ Myr. Before $t=2.5$ Myr, the star has already merged with five massive main sequence (MS) stars with $\msol{63} \lesssim m_\star \lesssim \msol{134}$. At this point the star reaches the giant phase with $m_\mathrm{VMS} = \msol{620}$. In the giant phase, the star further merges with five massive MS stars. When the star reaches the end of its life at $t=2.73$ Myr, the mass of the resulting IMBH BH1-B is $m_\bullet = \msol{1049}$. Later, the IMBH tidally disrupts a massive giant and two massive MS stars. The mass of the IMBH BH1-B is $m_\bullet=\msol{1194}$ at $t=6.5$ Myr when the star cluster H1-B merges with the central cluster.

The second-most-massive IMBH of the simulation (BH1-C) originates from a progenitor of $m_\star=\msol{134}$ in a somewhat more massive star cluster ($M_\mathrm{\star} = \msol{1.1\times10^5}$). When the star begins its giant phase at $t=2.58$ Myr it has a mass of $m_\star=\msol{397}$, having merged with one MS star with $m_\star = \msol{35}$ and two MS stars with $m_\star>\msol{100}$. In the giant phase there are no mergers with massive stars. The star collapses into the IMBH BH1-C ($m_\bullet=\msol{388}$) at $t=2.81$ Myr. Before $t=3.0$ Myr, the IMBH tidally disrupts two massive giants ($m_\star=\msol{119}$ and $m_\star=\msol{179}$). The latter giant is one of the two cases in which the star merging with the main progenitor is itself a stellar merger remnant with $m_\star>\msol{150}$. This star originates from an  earlier merger of two massive MS stars with $m_\star = \msol{101}$ and $m_\star=\msol{81}$. After these tidal disruption and accretion events the mass of the BH1-C is $m_\bullet=\msol{536}$. At this point the host star cluster merges with the main central cluster.

The original progenitor of the lowest-mass IMBH of the simulation (BH1-D) has a mass of $m_\star = \msol{130}$ in a star cluster H1-D with a mass of $M_\star=\msol{5.9\times10^4}$. The star has its first mergers earlier than the two more massive IMBH progenitors, reaching a mass of $m_\star = \msol{313}$ already by $t=1.6$ Myr after merging with two massive MS stars of $m_\star\sim\msol{90}$. The star evolves into the giant phase and later collapses into an IMBH with $m_\bullet=\msol{292}$. The host cluster now merges with the main star cluster of the simulation, after which BH1-D disrupts a massive giant reaching the mass of $m_\star=\msol{348}$ at $t=2.90$ Myr. 

After the star cluster H1-C merges with the main cluster, the IMBHs BH1-C and BH1-D form a bound IMBH binary at the center of the main star cluster. The binary remains at the center of the star cluster for about $3$ Myr until the most massive IMBH of the simulation, BH1-B, is accreted into the main star cluster at $t=6.5$ Myr. At this point, the IMBHs form a hierarchical triple system with the inner binary consisting of IMBHs BH1-C ($m_\bullet=\msol{542}$) and BH1-D ($m_\bullet=\msol{348}$) with the third body being BH1-B ($m_\bullet=\msol{1194}$). The hierarchical IMBH triple is not stable \citep{Mardling2001}, and an exchange event occurs soon at $t=7.0$ Myr when BH1-B replaces BH1-C in the inner binary. Next, an evolutionary phase lasting from $t=7.0$ Myr until $t=18.5$ Myr begins, illustrated in the bottom-right box in Fig. \ref{fig: mergertree-23A}. In this phase, the outer IMBH ($m_\bullet=\msol{542}$) is repeatedly ejected into a wide orbit (typical apocenters at $r_\mathrm{a} \sim 0.1$ pc -- $1.0$ pc) in the star cluster by the inner binary in three-body interactions. In the interaction, the inner binary shrinks towards smaller separations. The outer third body sinks back into the center of the star cluster on a time-scale of $t_\mathrm{sink}\lesssim 1$ -- $2$ Myr due to dynamical friction. The cycle consisting of interaction, ejection and sinking phases repeats multiple times. This cluster-aided pump-like process transfers the orbital energy of the inner binary to the cluster stars until the inner binary finally reaches the regime in which gravitational-wave emission becomes dynamically important to the system. The IMBHs BH1-B ($m_\bullet=\msol{1194}$) and BH1-D ($m_\bullet=\msol{348}$) merge into an IMBH with $m_\bullet = \msol{1498}$ at $t=18.5$ Myr. At the time of the merger, the remnant receives a gravitational-wave recoil kick of $v_\mathrm{recoil} = 206$ km/s, which is enough to unbind the IMBH from the central star cluster. The third IMBH, BH1-C sinks into the center of the massive star cluster and remains there until $t=31.1$ Myr at which point it merges with a stellar-mass BH ($m_\bullet=\msol{52}$). Even though the gravitational-wave recoil kick is relatively weak, $v_\mathrm{recoil} = 60$ km/s, it is enough to marginally unbind the merger remnant from the star cluster. At later times, two low-mass IMBHs ($m_\bullet=\msol{116}$ and $m_\bullet=\msol{135}$) form in mergers of stellar-mass BHs in the cluster, but both are instantly ejected from the cluster with recoil kick velocities of  $v_\mathrm{recoil} = 1102$ km/s and $v_\mathrm{recoil} = 140$ km/s, respectively. At the end of the simulation at $t=50$ Myr, the star cluster H1-A contains no IMBHs and only evolving stars and interacting stellar-mass BHs remain.

\subsubsection{Simulation H2}\label{section: sim-H2}

\begin{figure}
\includegraphics[width=\columnwidth]{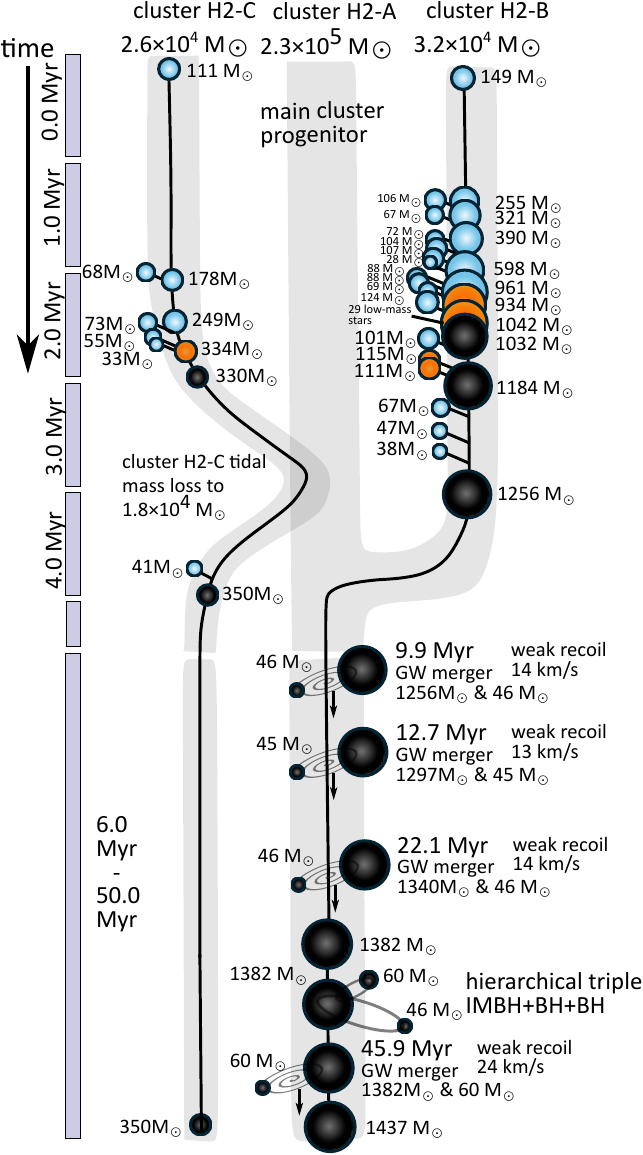}
\caption{The merger trees of IMBHs and their host star clusters in the simulation H2. As in simulation H1, the initially most massive star cluster does not produce an IMBH, but two IMBHs with masses of $m_\bullet=\msol{350}$ and $m_\bullet=\msol{1256}$ form in lower-mass star clusters. The host star cluster H2-C of the lower-mass IMBH has a close encounter with the main cluster at $t\sim4$ Myr and is tidally stripped, but does not merge with the main cluster. The more massive IMBH with its host cluster H2-B is accreted by the main cluster, where it repeatedly merges with stellar-mass black holes. The resulting gravitational-wave recoil kicks are weak, $v_\mathrm{recoil}<25$ km/s, and the IMBH of mass $m_\bullet=\msol{1437}$ is retained in the main cluster at the end of the simulation at $t=50$ Myr. }
\label{fig: mergertree-23B}
\end{figure}

The IMBH and star cluster merger trees of the simulation H2 are presented in Fig. \ref{fig: mergertree-23B}. As in the case of the simulation H1, the initially most massive central star cluster H2-A ($M_\mathrm{\star}=\msol{2.3\times10^5}$) does not itself form an IMBH, but rather accretes an IMBH formed in the less massive cluster H2-B ($M_\mathrm{\star}=\msol{3.2\times10^4}$) with its IMBH BH2-B at $t=4.8$ Myr when $m_\bullet=\msol{1256}$. A third star cluster H2-C forms an IMBH BH2-C with a mass of $m_\bullet=\msol{350}$, but its host star cluster is never accreted into the main cluster, although the two clusters have a close encounter around $t\sim3.6$ Myr in which the smaller star cluster is tidally stripped and loses $\sim30\%$ of its stellar mass.

The progenitor of the BH2-B is a star with $m_\star=\msol{149}$ in the initial conditions. In the MS phase, this star merges with $10$ other MS stars with masses ranging from $m_\star=\msol{28}$ to $m_\star=\msol{124}$ between $t=1.42$ Myr and $t=2.09$ Myr. When the star reaches the giant phase at $t=2.29$ Myr, it has a mass of $m_\mathrm{VMS}=\msol{934}$. In the giant phase, the star only merges with a single massive MS star, but accretes $29$ low-mass stars before collapsing into an IMBH with $m_\bullet=\msol{1132}$ at $t=2.52$ Myr. Before the IMBH reaches the main star cluster at $t=4.8$ Myr, it disrupts three massive MS stars. In the main cluster, the IMBH BH2-B repeatedly merges with stellar-mass black holes, but the resulting gravitational-wave recoil kicks are weak ($v_\mathrm{recoil}\lesssim24$ km/s) and unable to unbind the IMBH from its host cluster. The final mass of the IMBH BH2-B is $m_\bullet=\msol{1437}$ at $t=50$ Myr.

The progenitor to the BH2-C is a star with $m_\star=\msol{111}$ which has a relatively simple merger history. Before the beginning of its giant phase at $t=2.76$ Myr, the star merges with four MS stars with masses between $m_\star=\msol{33}$ and $m_\star=\msol{73}$, reaching a mass of $m_\star=\msol{334}$. The resulting IMBH forms at $t=2.97$ Myr with a mass of $m_\bullet=\msol{330}$. The IMBH later accretes a single MS star, and its final mass is $m_\bullet=\msol{350}$ at the end of the simulation. At this point, the IMBH is located at the center of the star cluster ($M_\mathrm{\star}=\msol{1.4\times10^4}$) in which it originally formed.

\subsubsection{Simulation H3}\label{section: sim-H3}

\begin{figure*}
\includegraphics[width=0.9\textwidth]{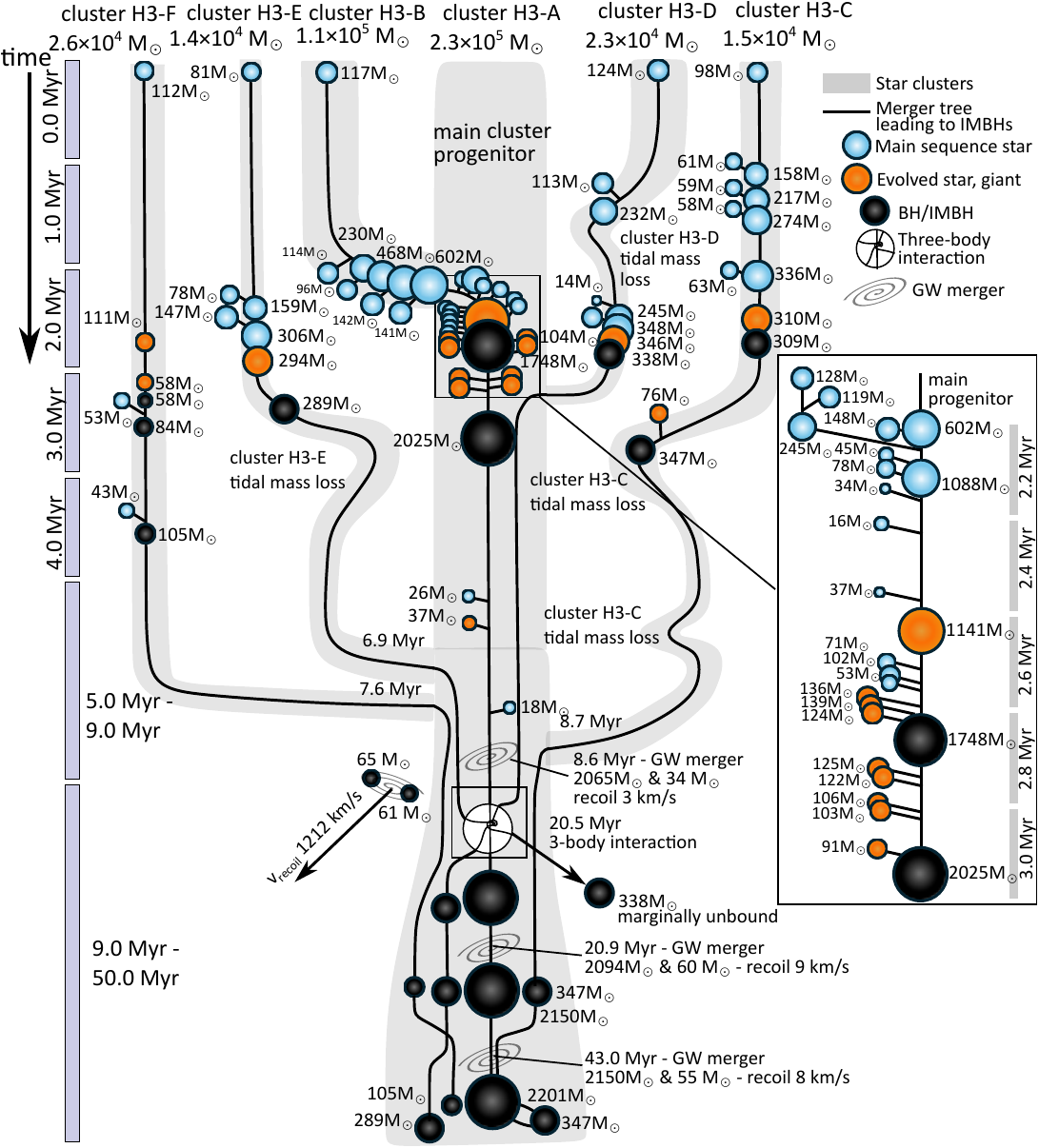}
\caption{The merger trees of IMBH progenitors and their host star clusters in the simulation H3. As in the other two hierarchical assembly simulations, the stars from the initially most massive star clusters are not the progenitors of the IMBHs. The IMBH BH3-B with $m_\bullet=\msol{2201}$ formed in a large number of collision with MS and giant stars mainly in only $0.9$ Myr between $t=2.0$ Myr and $t=2.9$ Myr is the most massive IMBH formed in the simulations of this study. By $t=8.7$ Myr, five IMBHs with eventual final masses $m_\bullet=\msol{2201}$, $m_\bullet=\msol{347}$, $m_\bullet=\msol{338}$, $m_\bullet=\msol{289}$ and $m_\bullet=\msol{105}$ of have ended up in the main cluster. At $t=20.5$ Myr, a strong three-body interaction among the IMBHs ejects the IMBH with $m_\bullet=\msol{338}$ from the cluster. While a gravitational-wave recoil kick ($v_\mathrm{recoil}=1212$ km/s) after a merger of two stellar-mass BHs ejects a new low-mass remnant IMBH ($m_\bullet=\msol{105}$) from the cluster, mergers between the most massive IMBH and stellar-mass BHs only result in weak recoil kicks. At $t=30$ Myr, four IMBHs populate the main star cluster, the most massive cluster at the center and the other three in the outskirts ($r\lesssim10$ pc) of the cluster. At later times, the two most massive IMBHs form a binary, and by the end of the simulation $t=50$ Myr the two other remaining IMBHs ($m_\bullet=\msol{105}$ and $m_\bullet=\msol{289}$) have sunk into separations less than $r\lesssim2$ pc from the central binary.}
\label{fig: mergertree-23C}
\end{figure*}

The IMBH and star cluster merger trees of the simulation H3 are presented in Fig. \ref{fig: mergertree-23C}. The simulation H3 has the most complex IMBH formation histories of the three simulations of this study, as in total five star clusters produce an IMBH. As in the case of the simulations H1 and H2, the most massive star cluster H3-A ($M_\mathrm{\star}=\msol{2.3\times10^5}$) in the simulation does not itself produce an IMBH, but rather accretes IMBHs or their VMS progenitors of IMBHs from lower-mass star clusters ($\msol{1.4\times10^4} \lesssim M_\mathrm{\star} \lesssim \msol{1.1\times10^5}$).

The progenitor of the most massive IMBH of this study (BH3-B) is a star with a mass of $m_\star=\msol{117}$. Its host star cluster H3-B ($M_\mathrm{\star}=\msol{1.1\times10^5}$) merges with the central star cluster already at $t=2.20$ Myr, at which point the star has reached a mass of $m_\mathrm{VMS}\sim\msol{600}$ after merging with four massive MS stars with masses exceeding $m_\star=\msol{96}$, but is still in the MS phase. Before reaching the giant phase at $t=2.62$ Myr, the VMS further merges with seven MS stars, from which one star is a VMS ($m_\star=\msol{245}$) formed earlier in a merger of two massive MS stars with masses of $m_\star=\msol{129}$ and $m_\star=\msol{119}$. At the beginning of the giant phase, the BH3-B progenitor VMS has a mass of $m_\mathrm{VMS}=\msol{1141}$. In the giant phase, the star merges with three more MS stars and three massive giants before collapsing into an IMBH with a mass of $m_\bullet=\msol{1748}$ at $t=2.84$ Myr. The IMBH BH3-B grows beyond $m_\bullet = \msol{2000}$ by disrupting mainly massive giant stars.

The progenitors of IMBHs from star clusters H3-C, H3-D and H3-E have similar merger and evolutionary histories. The initial masses of their host star clusters are $M_\star=\msol{1.5\times10^4}$, $M_\star=\msol{2.3\times10^4}$ and $M_\star=\msol{1.4\times10^4}$, respectively. Most mergers of the IMBH progenitors occur already in the MS phase with relatively few mergers in the giant and IMBH phases. The masses of the IMBHs are $m_\bullet=\msol{347}$ (BH3-C), $m_\bullet=\msol{338}$ (BH3-D), and $m_\bullet=\msol{289}$ (BH3-E) when their host clusters merge with the central star cluster at $t=8.7$ Myr, $t=3.2$ Myr and $t=6.9$ Myr, respectively. The lowest-mass IMBH formed through collisions is BH3-F which originates from a star cluster with $M_\mathrm{\star} = \msol{2.6\times{10^4}}$. The star evolves in isolation until losing its envelope in the giant phase in a collision and leaves a stellar-mass remnant of $m_\bullet=\msol{58}$. However, the stellar-mass BH disrupts two massive MS stars, and reaches $m_\bullet=\msol{105}$ before being accreted into the main cluster at $t=7.6$ Myr. 

Between $t=8.7$ Myr and $t=20.5$ Myr, the main star cluster H3-A hosts in total five IMBHs. In addition, a pair of stellar-mass BHs with masses of $m_\bullet=\msol{65}$ and $m_\bullet=\msol{61}$ merges in the cluster, leaving behind a low-mass IMBH with $m_\bullet = \msol{106}$. However, the remnant receives a strong gravitational-wave recoil kick of $v_\mathrm{recoil} = 1212$ km/s, immediately unbinding the remnant from the star cluster. At $t=20.5$ Myr, BH3-D ($m_\bullet = \msol{338})$ is ejected from the star cluster after a strong three-body encounter with IMBHs BH3-B ($m_\bullet=\msol{2094}$) and BH3-E ($m_\bullet=\msol{289}$). The most massive IMBH of the cluster also merges with two stellar-mass BHs, but the resulting gravitational-wave recoil kicks are weak ($v_\mathrm{recoil}\lesssim10$ km/s) and thus unable to move the IMBH away from the cluster. At $t=30$ Myr the star cluster hosts four IMBHs: BH3-B ($m_\bullet=\msol{2150}$) at the center and three IMBHs (BH3-C, BH3-E and BH3-F) orbiting at the outskirts of the cluster ($r\lesssim10$ pc). The masses of the three IMBHs are at this point $m_\bullet=\msol{347}$, $m_\bullet=\msol{289}$ and $m_\bullet=\msol{105}$, respectively. At later times, the most massive IMBH BH3-B merges with another stellar-mass BH, reaching its final mass of $m_\bullet=\msol{2201}$. The two most massive IMBHs (BH3-B and BH3-C) form a binary, and by the end of the simulation the two other remaining IMBHs in the star cluster have sunk into separations less than $r\lesssim2$ pc from the central binary.

\subsection{Three categories of early IMBH growth histories}

\begin{figure}
\includegraphics[width=\columnwidth]{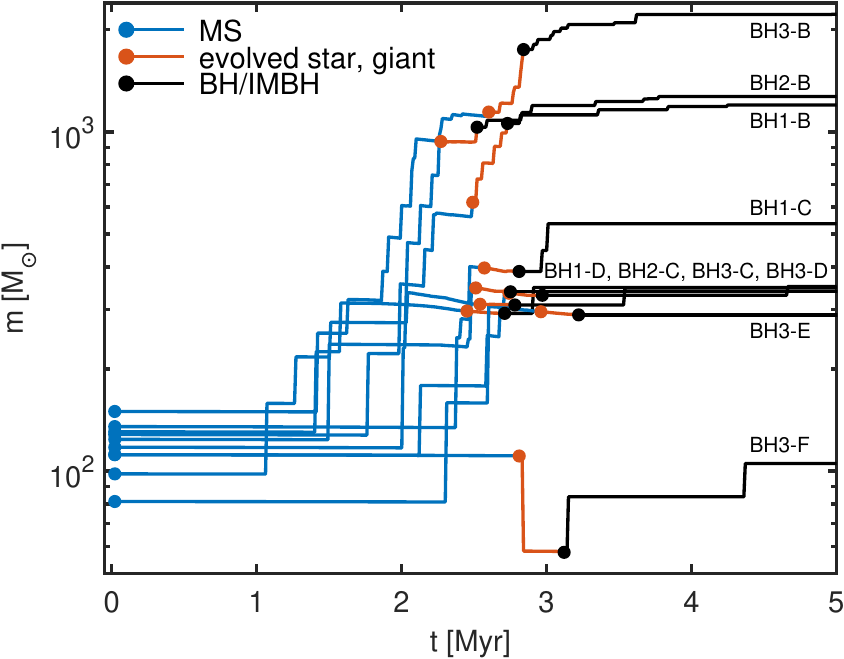}
\caption{The mass growth histories of the ten IMBHs that reached $m_\bullet \gtrsim \msol{100}$ by $t=5$ Myr in the simulations H1, H2 and H3. There are three approximate categories as elaborated in Table \ref{table: categories}: the low-mass IMBHs with $m_\bullet\sim \msol{100}$ originating below the mass gap, the numerous just-above-the-mass-gap IMBHs with masses $m_\bullet\sim\msol{300}$--$\msol{550}$, and finally the more massive ($m_\bullet\gtrsim \msol{1200}$) IMBHs experiencing growth also after the MS phase.}
\label{fig: imbh-mass-time}
\end{figure}

\begin{table*}
\begin{center}
\begin{tabular}{l l l l}
\hline
category & \texttt{imbh-lowmass} & \texttt{imbh-just-above-massgap} & \texttt{imbh-massive}\\
 & \texttt{(IMBH-LM)} & \texttt{(IMBH-JAM)} & \texttt{(IMBH-M)}\\
\hline
summary & initial BH mass below the mass gap, & IMBH mass just above the mass gap, & MS end mass well above the mass gap\\
& growth by TDEs or BH mergers & no substantial later growth & substantial growth in the giant and IMBH phases\\
\hline
MS end mass & $m_\star\lesssim \msol{150}$ & $m_\star\sim\msol{300}$--$\msol{400}$ & $m_\star\gtrsim \msol{600}$ \\
\hline
growth after MS & TDEs and/or BH mergers required & typically no substantial growth & substantial growth by up to a factor of $\sim2$\\
until $t=50$ Myr & & &\\
\hline
IMBH mass at & $m_\bullet\sim\msol{100}$--$\msol{200}$ & $m_\bullet\sim\msol{300}$--$\msol{550}$ & $m_\bullet\gtrsim \msol{1200}$ \\
$t=5$ Myr & & &\\
\hline
IMBHs of & BH3-F & BH3-E, BH1-D, BH2-C, & BH1-B, BH2-B, BH3-B, BH1-C (?)\\ 
this study & & BH3-C, BH3-D, BH1-C (?)& \\ 
\hline
\end{tabular}
\caption{The three categories of IMBHs formed in the hierarchical star cluster assembly simulations and their basic formation pathways. The IMBHs formed in the recent direct-summation N-body simulation studies of massive, isolated star clusters of \citet{Rizzuto2021,Rizzuto2022} and \citet{ArcaSedda2023b} would correspond the categories \catone{} and \cattwo{} of this classification.}
\label{table: categories}
\end{center}
\end{table*}

We present the mass growth histories of stars which became IMBHs ($m_\bullet>\msol{100}$) during the first $t=5$ Myr in the hierarchical clustering simulations H1--H3 in Fig. \ref{fig: imbh-mass-time}. There are in total ten such IMBHs in the three simulations, as detailed in sections \ref{section: sim-H1}--\ref{section: sim-H3}. Although the stars growing by repeated stellar mergers rapidly enter the VMS phase, their merger histories do not classify as a true collisional runaways as the merger rate does not monotonically increase after each stellar merger. The ten growth histories can be qualitatively divided into three categories: low-mass IMBHs in the mass gap (\catone), IMBHs just above the mass gap (\cattwo), and massive IMBHs well above the mass gap (\catthree). We summarize these three categories in Table \ref{table: categories}.

In the lowest-mass category \catone, the IMBH mass threshold of $m_\bullet=\msol{100}$ is reached from below. The only particular example of the category during the first $t=5$ Myr is the IMBH BH3-F. The progenitor star of BH3-F experiences sudden mass loss in a common envelope event before its core collapses into a stellar-mass BH, but later disrupts two massive stars, reaching the IMBH mass range from below.

In the second category \cattwo, the growing IMBH progenitor stars have grown by mergers into masses of $m_\mathrm{\star}\sim \msol{290}$--$\msol{400}$ already in the main sequence. The stars do not grow in mass in their giant phases, and collapse into IMBHs above the mass gap. IMBHs BH3-E, BH1-D, BH2-C, BH3-C and BH-D with masses $m_\bullet=\msol{289}$--$\msol{350}$ fall into this category. The final (the most massive) IMBH of the category \catthree, BH1-C, already shows some characteristics of the third and the most massive category \catthree as its mass increases by almost $\sim\msol{150}$ via disrupting stars, reaching a mass of $m_\bullet=\msol{536}$ by $t=5$ Myr.

In the high-mass category \catthree, the VMSs have reached a mass of $m_\mathrm{VMS} \gtrsim \msol{620}$ by the end of their MS life-time. The category is characterized by high initial mass and substantial mass growth in the giant and early IMBH phases, up to a factor of $\sim 2$ from their mass at the end of the MS. The category contains three IMBHs: BH1-B, BH2-B and BH3-B, each being the most massive object in its simulation with IMBH masses ranging from $m_\bullet\sim\msol{1200}$ to $m_\bullet\sim\msol{2000}$ at $t=5$ Myr. This highlights the fact that even though the growth histories of the most massive objects do not technically classify as runaway growth histories, there is a large difference between the most massive and the second-most-massive objects formed in the clusters.

\subsection{No IMBHs form in isolated comparison simulations}

\begin{table}
\begin{center}
\begin{tabular}{c c c c c c}
Cluster & $N$ & $M_\star$ & $a$ & $\gamma_\mathrm{EFF}$ & $\rho_\mathrm{c,\star}$\\
 & & $\mathrm{[M_\odot]}$ & [pc] & & [$\mathrm{M_\odot}$ pc$^{-3}$]\\
\hline
C3 & $10^6$ & $5.9\times10^5$ & $0.3$ & $2.25$ & $9.5\times10^5$\\
C50 & $10^6$ & $5.9\times10^5$ & $1.0$ & $2.25$ & $4.1\times10^4$\\
\hline
\end{tabular}
\caption{The EFF model parameters $M_\star$ (total mass), $a$ (scale radius) and $\gamma_\mathrm{EFF}$ (projected outer power-law slope) with the resulting central stellar density $\rho_\mathrm{c,\star}$ for the isolated comparison simulations C3 and C50. We set the outer cut-off radius to $r_\mathrm{max} = 100$ pc.}
\label{table: EFF}
\end{center}
\end{table}

We now examine whether IMBHs form in isolated massive star clusters with masses and central densities comparable to the final, hierarchically assembled clusters with two comparison simulations C3 and C50. We focus on the final, central star cluster H1-A from the simulation H1. We select two times, $t=3$ Myr at which point the central stellar density is the highest, and the end of the simulation $t=50$ Myr. The stellar surface density profiles of the central clusters at these times are well described by the Elson-Freeman-Fall (EFF) profile \citep{Elson1987} with the outer slope of $\gamma_\mathrm{EFF}=2.25$. This corresponds to a three-dimensional density profile slope of $3.25$. We generate the two EFF initial models C3 and C50 using the \mcluster{} code \citep{Kupper2011} which has the EFF profile readily available. We use the Nuker \citep{Lauer1995} template of \mcluster{} with $\gamma_\mathrm{Nuker}=0.0$, $\beta_\mathrm{Nuker}=2.25$ and $\alpha_\mathrm{Nuker}=2.0$. The EFF model parameters for the two comparison simulation setups are listed in Table \ref{table: EFF}. The central stellar densities $\rho_\mathrm{c,\star} = \rhosol{9.5\times10^5}$ (C3) and $\rho_\mathrm{c,\star} = \rhosol{4.1\times10^4}$ (C50) in the two comparison setups are similar than in Fig. \ref{fig: density-star-remnant} for the cluster H3-A. The stellar IMF is the same as in the hierarchical assembly simulations. Even though the initial density profiles of the comparison models correspond to the cluster H1-A at $t=3$ Myr and $t=50$ Myr, respectively, each star in the comparison simulation initial conditions is initialized to the beginning of its MS phase. 

We run the two comparison simulations with \bifrost{} for either $t=20$ Myr (C3) or $t=30$ Myr (C50) at which point the massive stars required for VMS growth have reached the end of their lives. We find that neither run forms IMBHs. The maximum stellar mass reached through stellar collisions in the simulation C3 is $m_\star=\msol{231}$, which falls into the (P)PISN mass gap and the massive star leaves no massive remnant. The maximum BH mass reached in C3 is $m_\bullet=\msol{89}$. In the simulation C50 due to relatively low central density, only five stellar mergers during the simulation, and only one of which involves a massive star with $m_\star=\msol{116}$ accreting a low-mass star. Thus, the BHs formed in the simulation all have masses below the mass gap with maximum BH mass being  $m_\bullet=\msol{63}$.

Based on the hierarchical star cluster assembly simulations H1, H2, H3 and the isolated comparison simulations C3 and C50, it seems that key to IMBH formation in the simulated massive star clusters is the hierarchical formation process itself. Massive, isolated comparison clusters alone with monolithic origins could not produce IMBHs, even at corresponding initial central densities.

\section{IMBHs from isolated star clusters}\label{section: isolated-simulations}


\begin{table*}
\begin{center}
\begin{tabular}{l l c c c c c r r}
\hline
Cluster & $N_\mathrm{seed}$ & $N$ & $M_\mathrm{\star}$ & $\rho_\mathrm{c,init}$ & $t_\mathrm{seg}$ & $t_\mathrm{cc}$ & $\max(m_\star)$ & $\max(m_\bullet)$ \\
& & & $\mathrm{[M_\odot]}$ & $\rhosol{10^6}$ & Myr & Myr & $\mathrm{[M_\odot]}$ & $\mathrm{[M_\odot]}$\\
\hline
I1 & $10$ & $1.7\times10^4$ & $9.4\times10^3$& 0.43 & 0.05 & 0.41 & ${197}$ & ${72}$ \\
I2 & $10$ & $2.6\times10^4$ & $1.5\times10^4$& 0.54 & 0.05 & 0.62 & ${290}$ & ${87}$ \\
I3 & $10$ & $3.9\times10^4$ & $2.3\times10^4$& 0.65 & 0.06 & 0.91 & ${419}$ & ${378}$ \\
I4 & $10$ & $6.3\times10^4$ & $3.7\times10^4$& 0.81 & 0.07 & 1.38 & ${932}$ & ${964}$ \\
I5 & $10$ & $1.0\times10^5$ & $5.9\times10^4$& 1.00 & 0.10 & 2.09 & ${982}$ & ${1065}$ \\
I6 & $10$ & $1.6\times10^5$ & $9.4\times10^4$& 1.25 & 0.14 & 3.17 & ${1429}$ & ${1510}$ \\
I7 & $10$ & $2.5\times10^5$ & $1.5\times10^5$& 1.55 & 0.21 & 4.83 & ${1494}$ & ${1672}$ \\
I8 & $20$ & $4.0\times10^5$ & $2.3\times10^5$& 1.89 & 0.29 & 7.11 & ${362}$ & ${91}$ \\
I9 & $5$ & $6.2\times10^5$ & $3.7\times10^5$& 2.34 & 0.43 & 10.95 & ${199}$ & ${81}$ \\
I10 & $5$ & $1.0\times10^6$ & $5.9\times10^5$& 2.91 & 0.63 & 16.76 & ${150}$ & ${63}$ \\

\hline
\end{tabular}
\caption{The ten isolated star cluster simulation sets I1--I10, in total $100$ isolated simulations with their initial particle numbers $N$, initial masses $M_\star$ and the masses for the most massive star ($\max{m_\star}$) and (IM)BH $(\max{m_\bullet}$) formed through collisions in the simulation sets. We also show the initial central densities of the cluster models $\rho_\mathrm{c,init}$ as well as estimates for the mass segregation and core collapse time-scales ($t_\mathrm{seg}$, $t_\mathrm{cc}$) of the models as in Table \ref{table: mcluster}.}
\label{table: isolated}
\end{center}
\end{table*}

\begin{figure*}
\includegraphics[width=\textwidth]{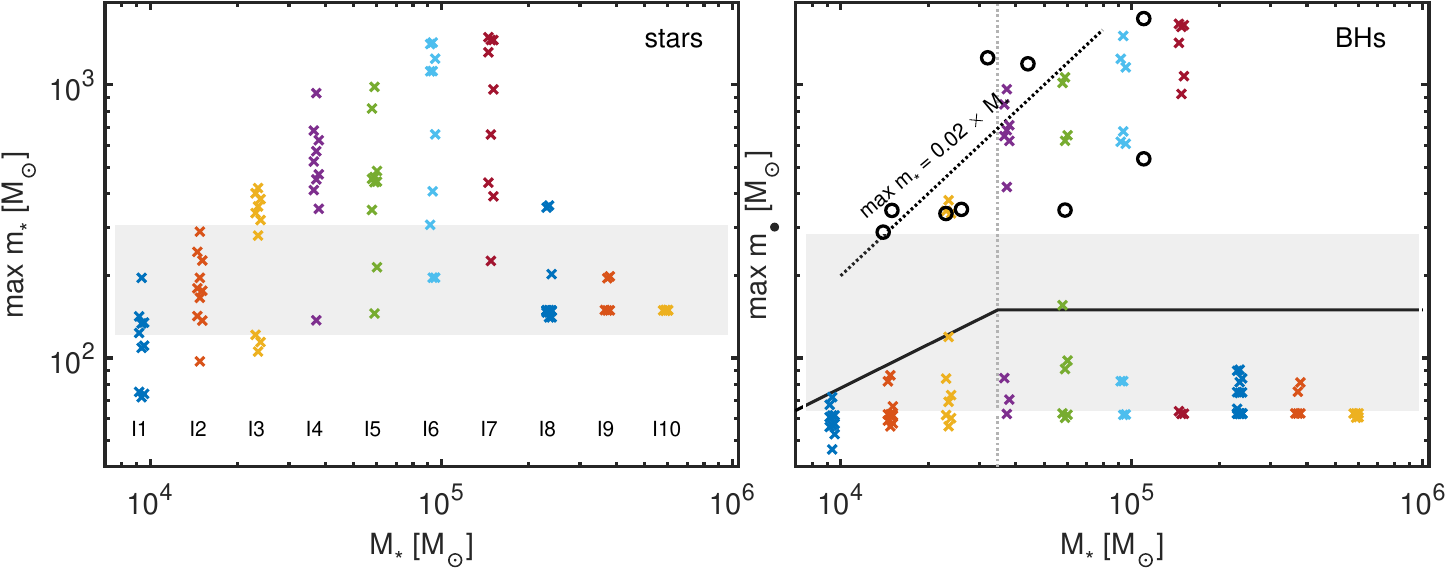}
\caption{The maximum stellar (left panel) and the BH (right panel) masses $\max{(m_\star)}$ and $\max{(m_\bullet)}$ reached in the isolated star clusters during the first $t=5$ Myr of the simulations. The solid black line indicates the expectation without mergers from the \citet{Kroupa2001} initial mass function a maximum mass of $m_\star=\msol{150}$ including the suppression of high-mass stars in low-mass clusters \citep{Yan2023}. The maximum final individual stellar mass grows linearly as $\max{(m_\star)} \approx 0.02 \times M_\star$ up to star cluster masses of $M_\star \sim \msol{10^5}$, after which the VMS growth in more massive clusters is strongly suppressed. The maximum BH mass in the isolated runs (colored crosses) follows the trends of the maximum stellar mass, processed by stellar evolution. Below cluster masses $M_\mathrm{\star}\lesssim\msol{2\times10^4}$, no BHs with masses above the (P)PISN mass gap (in gray shading) form. The maximum reached IMBH masses are $m_\bullet\sim\msol{1630}$. At cluster masses higher than $M_\star\gtrsim\msol{2 \times10^5}$ there are no IMBHs above the (P)PISN mass, but BHs in the mass gap below the IMBH limit $m_\bullet=\msol{100}$ still form. Note that the maximum IMBH mass can exceed the maximum VMS mass as the IMBHs still grow via disrupting stars in the remnant phase. For comparison, the IMBH (or their progenitor) masses from the hierarchical runs at the time of their accretion into the main cluster (as in Table \ref{table: mcluster}) are marked with black, open circles. The IMBHs formed in the in-falling sub-clusters can be more massive than their counterparts in isolated star clusters of the same mass and initial density.}
\label{fig: isolated-maxmass}
\end{figure*}

\subsection{Maximum VMS and BH masses in isolated simulations}

In the second part of the study we examine, given a density profile shape and a mass-size relation, which star clusters can form IMBHs. The motivation for the study is twofold. First, the IMBH formation efficiency $m_\bullet/M_\star$ has not been extensively studied in the literature along a star cluster mass-size relation. Second, the initially most massive central star clusters in the hierarchical star cluster assembly simulations did not form IMBHs from their stars but rather accreted them from lower-mass star clusters, and we wish to understand the physical background of this phenomenon.

Several studies (e.g. \citealt{Fujii2013}) have pointed out that the merger rate of massive stars can be enhanced only if their mass segregation time-scale is shorter than the life-times of massive stars ($t\sim2$--$3$ Myr). Even for our most massive star cluster models the mass segregation time-scale from Eq. \eqref{eq: tseg} is $t_\mathrm{seg} \lesssim 1$ Myr. Thus, massive stars in massive star clusters should in principle be able to collide with each other. However, based on analytic estimates for the core collapse time-scale, all clusters have not had time to core collapse before $t\sim3$ Myr as shown in Table \ref{table: isolated}.

One can argue that based on virial arguments (e.g. \citealt{Naab2009}) that the mean density of the assembling star cluster keeps decreasing, especially if the accreted systems are small with a fixed total accreted mass. This consequently leads to fewer stellar collisions. While the decreased mean cluster density may contribute to the lack of collisions in the central clusters, we show in this study that VMS growth is suppressed even in isolated, massive star cluster models.

We setup isolated star cluster initial conditions using the same structural parameters as for the individual clusters in the hierarchical assembly runs with ten different masses logarithmically spaced in the particle number range of $1.65\times10^4\lesssim N \lesssim 1.03\times10^6$. We sample ten realizations of each of the models I1-I7 with $N\leq2.53\times10^5$, 20 realizations of the model I8 with $N=4.0\times10^5$ and five realizations of the two more massive setups I9 and I10. The model I8 is examined the most as its mass lies close to the threshold after which the central clusters did not form their own IMBHs in the hierarchical assembly runs. In total, the sample consists of $100$ isolated star cluster models. The basic properties of the star cluster models I1-I10 are listed in Table \ref{table: isolated}. We run each model for $t=5.0$ Myr using the \bifrost{} code at which point stars with initial masses above $m_\mathrm{\star} \gtrsim\msol{42}$ have reached the end of their lives. In the structured star cluster assembly runs no VMS grew from a star with an initial mass below $m_\star\lesssim\msol{80}$, so the most efficient period of VMS growth through stellar collisions is expected to be over at this point.

Stellar collisions occur in almost all isolated simulations. The maximum reached stellar ($\max{m_\star}$) and black hole masses ($\max{m_\bullet}$) in the isolated simulations are presented in Table \ref{table: isolated} and in Fig. \ref{fig: isolated-maxmass}. The maximum VMS mass is $m_\star=\msol{197}$ in the lowest-mass cluster I1, and increases following a linear trend $\max{(m_\star)} \propto M_\star$ until the trend flattens around models I5-I6 with cluster masses $M_\star \sim \msol{6\times10^4}$ -- $\msol{9\times10^4}$. We find that the mass of the most massive star rarely exceeds $\max{(m_\star)} \approx 0.02 \times M_\star$ in the setups I1-I6, corresponding to a maximum IMBH formation efficiency of $2\%$.

Fig. \ref{fig: isolated-maxmass} also shows the maximum stellar mass in the initial models (solid black line). At high cluster masses the initial maximum stellar mass is constant, $m_\star=\msol{150}$. At lower star cluster masses high-mass stars are not initially allowed, following \cite{Yan2023}. The $\max{(m_\star)} \approx 0.02 \times M_\star$ trend continues from model I1 to model I6, even though the suppression of massive stars in the initial conditions only occurs in the models I1--I3. In addition, the relation for the initial cluster mass -- maximum stellar mass is shallower than the linear trend, so the maximum stellar masses in the initial conditions do not alone explain the relation between the cluster masses and the mass of the final collision product. The simple explanation is that the stellar collision rates also depend on the cluster mass, as will be later shown.

The maximum VMS mass of $\max{(m_\star)} \sim \msol{1500}$ is reached in the setup I7 with $M_\star = \msol{1.5\times10^5}$. At higher cluster masses, the suppression of the VMS growth is apparent, and in the highest-mass cluster model I10 ($M_\star = \msol{5.9\times10^5}$) the stars do not grow beyond their initial mass of $m_\star=\msol{150}$.

The maximum BH mass $m_\bullet$ as a function of their host star cluster mass $M_\star$ follows a similar trend as the maximum stellar mass $m_\star$, additionally processed by stellar evolution, especially the (P)PISN mass gap. IMBHs above the (P)PISN mass gap ($m_\bullet\gtrsim \msol{290}$) form in cluster setups from I3 ($M_\star = \msol{2.3\times10^4}$) to I7 ($M_\star = \msol{1.5\times10^5}$), in total $26$ IMBHs. The maximum reached IMBH mass is $m_\bullet = \msol{1672}$. Note that most massive IMBHs are more massive than the most massive VMSs as the IMBHs further grow by disrupting stars after the collapse of the VMS. BHs in the (P)PISN mass gap form in star clusters across a wide range in cluster mass. Out of the $100$ simulated isolated clusters, $25$ form a BH in the mass gap below the IMBH mass limit of $m_\bullet=\msol{100}$. Only two IMBHs are located in the mass gap itself in the IMBH mass range between $\msol{100} \lesssim m_\bullet \lesssim \msol{290}$. The VMS growth and collapse IMBH formation channel is approximately an order of magnitude more efficient in producing IMBH above the mass gap than in the mass gap above $m_\bullet\gtrsim\msol{100}$.

\subsection{Which mergers grow the VMSs?}

\begin{figure}
\includegraphics[width=\columnwidth]{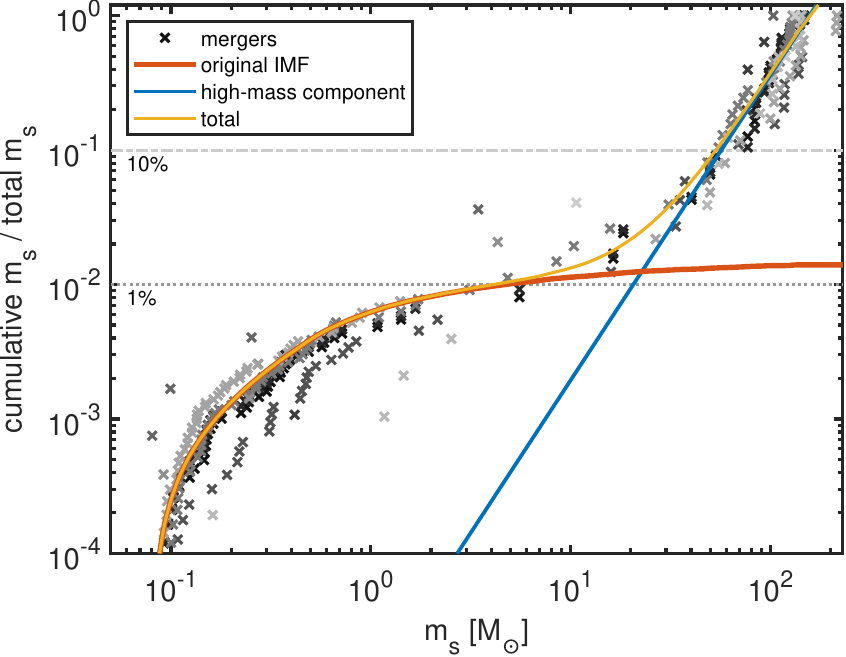}
\caption{The normalized cumulative mass function of the stars merged to the growing VMS in the simulation set I7. The individual mergers are shown in gray and black symbols, the color hue being different in each ten simulations. The mass function (yellow solid line) is comparable to the Kroupa IMF (orange solid line) at low masses below $m_\star\lesssim\msol{2}$. There are relatively few accreted stars in the range of a few tens of Solar masses. At high masses beyond $m_\star\gtrsim\msol{50}$, the mass function has a large overabundance of massive stars (yellow and blue solid lines) compared to the Kroupa IMF. Only $10\%$ of the total accreted mass is provided by stars less massive that $m_\star\lesssim\msol{53}$, and less than $1\%$ by stars with mass less than $m_\star\lesssim\msol{5}$.}
\label{fig: mass-origin}
\end{figure}

Next, we study the merger histories and the mass budget of the VMSs, focusing on the simulation set I7 which produced the most massive VMSs (and IMBHs) in the isolated simulations of this study. As in the structured star cluster assembly runs, typically only one star grows beyond $m_\star=\msol{150}$ through collisions. The progenitor of the star growing via stellar collisions is always a massive star with a mass of at least $m_\star\sim \msol{119}$. The mean mass of initial VMS progenitors in the set I7 is $m_\star\sim\msol{131}$.

We show the cumulative accreted stellar mass from the accreted stars in Fig. \ref{fig: mass-origin}. In each ten simulations of the set I7, we extract the masses ($m_\mathrm{s}$) of the stars merged into the growing VMS, sort them, and calculate the normalized cumulative mass function. We also show for comparison the cumulative mass function of the standard \cite{Kroupa2001} piece-wise power-law IMF. The masses $m_\mathrm{s}$ show a somewhat bimodal distribution, with the lower-mass end being comparable to the Kroupa IMF up to $m_\star\sim\msol{1}$--$\msol{2}$. Beyond this, there are relatively few mergers with stars with masses from $m_\star\sim\msol{1}$--$\msol{2}$ to $m_\star\sim\msol{40}$--$\msol{50}$. At large stellar masses ($m_\star\gtrsim\msol{50}$), the mass distribution of the accreted stars strongly deviates from the Kroupa IMF as a consequence of mass segregation in the simulated star clusters.

While the mergers between the growing VMS and low-mass stars are common, they do not significantly contribute the total mass of the VMS. We find that mergers with stars with masses less than $m_\star \lesssim \msol{5}$ only contribute $\sim 1\%$ to the total mass accreted by the growing VMS, and the contribution of stars with masses $m_\star \lesssim \msol{53}$ is $\sim 10\%$. In total $50\%$ of the mass budget of the VMS is provided by massive stars with masses above $m_\star\gtrsim\msol{114}$. Thus, VMS growth and IMBH formation is mainly fueled by the most massive stars of the IMF in our simulations.

\subsection{Stellar mergers: mass and orbit type classification}

\begin{figure}
\includegraphics[width=\columnwidth]{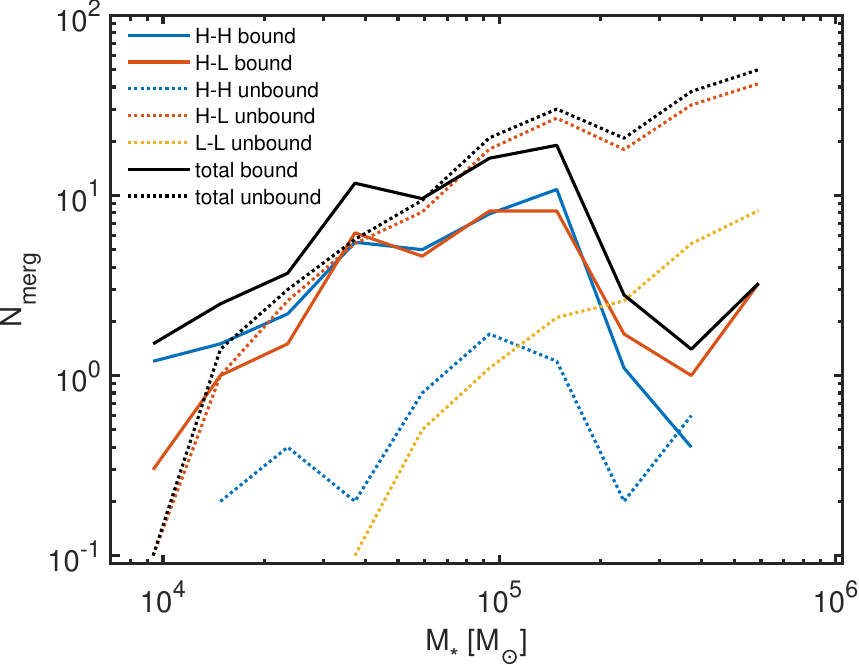}
\caption{Classifying the number of stellar mergers $N_\mathrm{merg}$ (per simulation) in the isolated simulations according to the orbit type and masses of merging stars. Mergers from bound orbits are shown using solid lines and mergers from unbound orbits with dashed lines. The mass threshold between the low-mass (L) and the high-mass stars (H) in this analysis is $m_\star=\msol{10}$. With increasing star cluster mass $M_\star$, mergers from unbound orbits become more common while mergers from any bound orbits (H-L bound and H-H bound, solid orange and blue lines) are suppressed above $M_\star\gtrsim\msol{1.5\times10^5}$. As the numbers of massive stars merging from unbound orbits (H-H unbound, dashed blue line) is in general low, this leads to a suppression of mergers between massive stars (H-H). Most common mergers at high cluster masses are unbound H-L mergers (dashed orange line) while low-mass stars merge with each other exclusively from unbound orbits (L-L unbound, in yellow).
}
\label{fig: isolated-mergers-unbound}
\end{figure}

We classify all the mergers occurred in the isolated star cluster simulations based on the type of their merger orbit (bound or unbound) and the masses of the merged stars. The mass threshold between the low-mass (L) and the high-mass stars (H) in this analysis is $m_\star=\msol{10}$. Most mergers occur from close to parabolic orbits. Approximately $93\%$ of the mergers occur from orbits with eccentricities between $0.9 \lesssim e \lesssim1.1$, and $\sim64\%$ from eccentricities between $0.99 \lesssim e \lesssim1.01$. The semi-major axis distribution of the bound merger orbits is well described by a log-normal distribution with a mean of $\mu_\mathrm{log10(a/pc)} = -2.78$ (corresponding to $a\sim1.7\times10^{-3}$ pc) and $\sigma_\mathrm{log10(a/pc)} = 0.82$, i.e. the typical semi-major axis of a bound mergers orbit is in the milliparsec range. 

The merger mass classification simply divides the stars in to higher-mass (H) and lower-mass (L) stars (H-L threshold mass $m_\star=\msol{10}$), yielding three merger mass categories: H-H between massive stars, mixed H-L mergers, and L-L between low-mass stars. The merger mass and orbit type classification is presented in Fig. \ref{fig: isolated-mergers-unbound}. First, we note that the total number of mergers per simulation $N_\mathrm{merg}$ increases as a function of cluster mass $M_\star$. The trend for $N_\mathrm{merg}$ is not a simple power-law, so the increased number of mergers in massive clusters cannot be purely explained by the fact that more massive clusters simply have more stars. Second, mergers from unbound orbits become increasingly more common compared to mergers from bounds orbits as the cluster mass $M_\star$ increases. At low cluster masses ($M_\star\lesssim \msol{4\times10^4}$) the most common mergers are bound H-H and H-L mergers, and unbound H-L mergers, up to on average $N_\mathrm{merg}\sim 6.2$ of each of such mergers per simulation. High-mass stars can also merge with each other from unbound orbits, but this is more rare, only up to $N_\mathrm{merg}\sim1.7$ unbound H-H mergers per simulation. Bound L-L mergers never occur in the isolated simulation sample. 

Between star cluster masses $M_\star = \msol{4\times10^4}$ and $M_\star = \msol{1.5\times10^5}$, the number of bound H-H and H-L mergers only mildly increases, never exceeding $N_\mathrm{merg}\sim11$ mergers per simulation. Meanwhile, the number of unbound H-L mergers steadily grows, reaching $N_\mathrm{merg}=27$ mergers per simulation, becoming the most common type of mergers at this cluster mass range. The total number of unbound mergers exceed the total number of bound mergers at $M_\star \sim \msol{6\times{10^4}}$. Unbound L-L mergers occur when the mass of the star cluster is $M_\star\gtrsim\msol{4\times10^4}$, and their number monotonically increases as a function of the cluster mass.

In the massive star clusters $M_\star \gtrsim \msol{1.5\times{10^5}}$ the unbound mergers follow their trends from the lower-mass clusters, but the number of bound mergers steeply decreases. The number of bound H-H mergers decreases by over an order of magnitude, while the H-L merger channel is somewhat less affected. In the highest-mass ($M_\star = \msol{5.9\times10^5}$) isolated cluster simulations, $\sim77\%$ of the total mergers are unbound H-L mergers, $17\%$ unbound L-L mergers, and the remaining $6\%$ bound H-L and bound H-H mergers. Together with the results in Fig. \ref{fig: isolated-maxmass} and Fig. \ref{fig: mass-origin}, it appears that the reason for the suppressed VMS growth at high cluster masses is the suppression of the bound massive-massive merger channel at cluster masses above $M_\star \gtrsim \msol{1.5\times{10^5}}$.

\subsection{The time of onset of stellar mergers}

\begin{figure}
\includegraphics[width=\columnwidth]{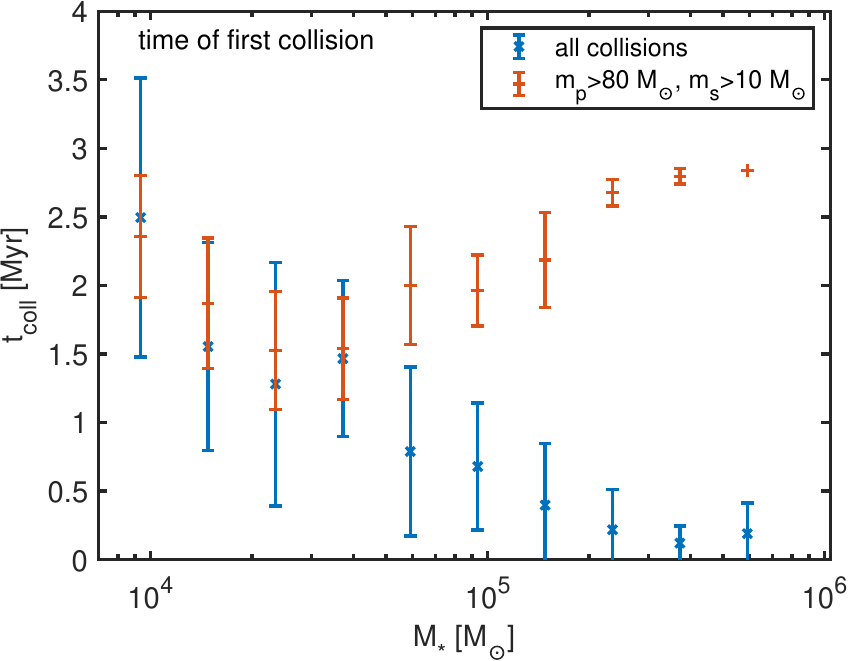}
\caption{The tuning-fork diagram for the stellar merger times, i.e. the time of the first collision $t_\mathrm{coll}$ as a function of the isolated star cluster mass. The first collision times between massive stars ($m_\mathrm{p}>\msol{80}$, $m_\mathrm{s}>\msol{10}$) are shown in orange while we show the first collision times of any stars in blue. Collisions occur on average earlier in more massive clusters due to higher stellar densities. However, $t_\mathrm{coll}$ for the massive stars decouples from this trend at star cluster masses above $M_\star \gtrsim \msol{4\times10^4}$.}
\label{fig: tuning-fork}
\end{figure}

Next, we study the the merger times of stars, specifically the time of the first stellar collision $t_\mathrm{coll}$ of the isolated simulations as a function of the star cluster mass $M_\star$. The results of the first collision time analysis are presented in Fig. \ref{fig: tuning-fork}. 

First, focusing on mergers with any primary ($m_\mathrm{p}$) and secondary ($m_\mathrm{s}$) stellar masses, we find that the mean time for the first merger $t_\mathrm{coll}$ decreases with increasing cluster mass, i.e. massive clusters begin their stellar mergers earlier. The mean $t_\mathrm{coll}$ decreases from $t_\mathrm{coll}\sim2.49\pm1.02$ Myr in the simulation sample I1 ($M_\star=9.4\times10^3$) to $t_\mathrm{coll}\sim0.19\pm0.22$ Myr for the most massive cluster I10 ($M_\star=5.9\times10^5$) in the isolated sample. This can be attributed to the fact that for the chosen initial star cluster mass-size relation the central density $\rho_\mathrm{c}$ monotonically increases with increasing cluster mass $M_\star$.

However, studying the collision times of massive primary stars $(m_\mathrm{p}>\msol{80})$ with massive secondaries $(m_\mathrm{s}>\msol{10})$ we find a qualitatively different trend. Overall, the collision times of massive stars and all stars in Fig. \ref{fig: tuning-fork} form a tuning fork like diagram. At low cluster masses ($M_\star\lesssim\msol{4\times10^4}$) the average time for the first merger of a massive star with a massive secondary corresponds to the average $t_\mathrm{coll}$ for all stars, although with a somewhat smaller scatter. After $M_\star\gtrsim\msol{4\times10^4}$, the first massive star collision times increase, decoupling from the decreasing trend of all stars. The average $t_\mathrm{coll}$ for the massive stars increases from $1.54\pm0.37$ Myr in the sample I4 ($M_\star\lesssim\msol{4\times10^4}$) to the single merger at $t=2.84$ Myr in the most massive isolated cluster sample I10. This is very close to the life-time of the most massive stars in the simulations ($m_\star=\msol{150}$ with $t=2.82$ Myr) at the chosen metallicity $Z=0.01\:Z_\odot$.

The theoretical estimates for the two-body relaxation and mass segregation time-scales (as in Eq. \eqref{eq: rlx} and Eq. \eqref{eq: tseg}) monotonically increase with increasing cluster mass for the cluster mass-size relation adopted for this study. In Fig. \ref{fig: tuning-fork} we see that in our simulations the first collisions occur earlier in more massive clusters. Compared to the core collapse time-scales (as determined from the Lagrangian radii) of the simulation models, the first collisions in the low-mass models occur considerably later than the core collapse time ($t_\mathrm{cc}\gtrsim0.6$ Myr). We note that the core collapse time-scales for the massive star clusters appear to be by up to a factor of few shorter when determined from the Lagrangian radii instead of the analytical estimate of Table \ref{table: isolated} based on the two-body relaxation time-scale. For higher-mass (I6-I10) models, the core collapse time is closer to the time of the first collision involving a high-mass star. In summary, no simple time-scale estimate explains the trends in Fig. \ref{fig: tuning-fork} in a straightforward manner. It is likely that both the central density itself and the mass segregation and the core collapse time-scales all play a role in determining the times for the first collisions.

\subsection{The stellar-dynamical environments of massive stars}

\begin{figure}
\includegraphics[width=\columnwidth]{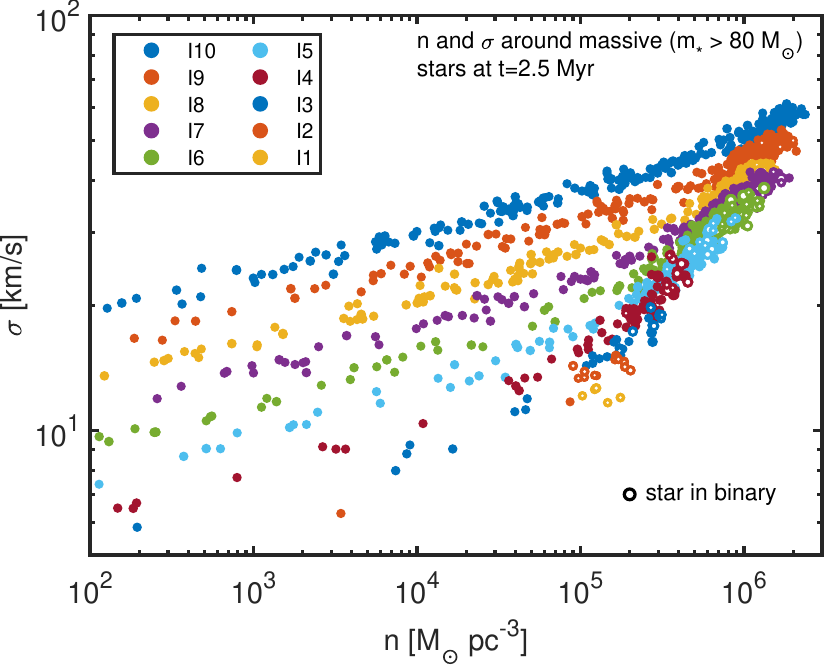}
\caption{The stellar number densities $n$ and velocity dispersions $\sigma$ around massive ($m_\star>\msol{80}$) stars at $t=2.5$ Myr in different isolated star cluster simulations I1-I10 indicated by the color scheme. Massive stars in a binaries ($a<0.01$ pc) with massive secondaries ($m_\mathrm{s}>\msol{10}$) are indicated with open circles with more binaries occurring in the lower-mass setups. The highest number densities and velocity dispersions occur at the cluster centers. Increasing cluster mass from set I1 to I10 results in the increase of the central number density from $n\sim1$--$\nsol{2\times10^5}$ to $n\sim2$--$\nsol{3\times10^6}$ and central velocity dispersion from $\sigma\sim12$ km/s up to $\sigma\sim61$ km/s. The four most massive setups I7-I10 have comparable central number densities $n$, but the central velocity dispersion in the set I7 is only $\sim41$ km/s compared to $\sigma\sim61$ km/s in I10.}
\label{fig: isolated-n-sigma}
\end{figure}

Finally, we examine the stellar-dynamical environments of massive stars characterized by the stellar number density $n$ and the velocity dispersion $\sigma$ in their immediate vicinity at $t=2.5$ Myr when most massive stars evolve away from the MS. The stellar number density $n$ and the velocity dispersion $\sigma$ are calculated from $N_\mathrm{ngb}=100$ closest neighbors of each massive star. The results are displayed in Fig. \ref{fig: isolated-n-sigma}.

The fraction of massive stars in bound binary systems ($a<0.01$ pc) is larger in lower-mass star clusters at $t=2.5$ Myr. The stellar number densities and velocity dispersions around the massive stars are the highest at the cluster centers, as expected. From the lowest mass isolated setup I1 ($M_\star=9.4\times10^3$) to the most massive isolated clusters I10 ($M_\star=5.9\times10^5$) the central stellar number density increases from $n\sim1$--$\nsol{2\times10^5}$ to $n\sim2$--$\nsol{3\times10^6}$ and central velocity dispersion increases from $\sigma\sim12$ km/s up to $\sigma\sim61$ km/s. The four most massive setups (I7--I10) have comparable central stellar number densities. However, their maximum velocity dispersions differ considerably, from $\sim41$ km/s in the set I7 to $\sim61$ km/s in the most massive isolated models I10.

The similar central number densities but differing velocity dispersions indicate that the centers of simulated star clusters I7 ($M_\star=\msol{1.5\times10^5}$) and I10 ($M_\star=\msol{5.9\times10^5}$) are very different dynamical environments. This is due to the steep scaling of few-body process outcomes and collision rates as a function of the velocity dispersion. For example, for equal stellar masses and number densities (but differing velocity dispersions) the three-body binary formation rate estimate of \cite{Goodman1993} in Eq. \eqref{eq: goodman} yields a difference of a factor of $\sim36$ between the sets I7 and I10. In addition, if binaries form dynamically, they contribute to the stellar collisions especially through single-binary interactions \citep{Gaburov2008}. However, the Heggie-Hills \citep{Heggie1975,Hills1975} law states that wide binaries will on average eventually dissolve, and it is more difficult for a binary with a given semi-major axis to be hard in a stellar-dynamical environment with a high velocity dispersion.

\subsection{Suppressing collisions between massive stars in massive star clusters}

\begin{figure}
\includegraphics[width=\columnwidth]{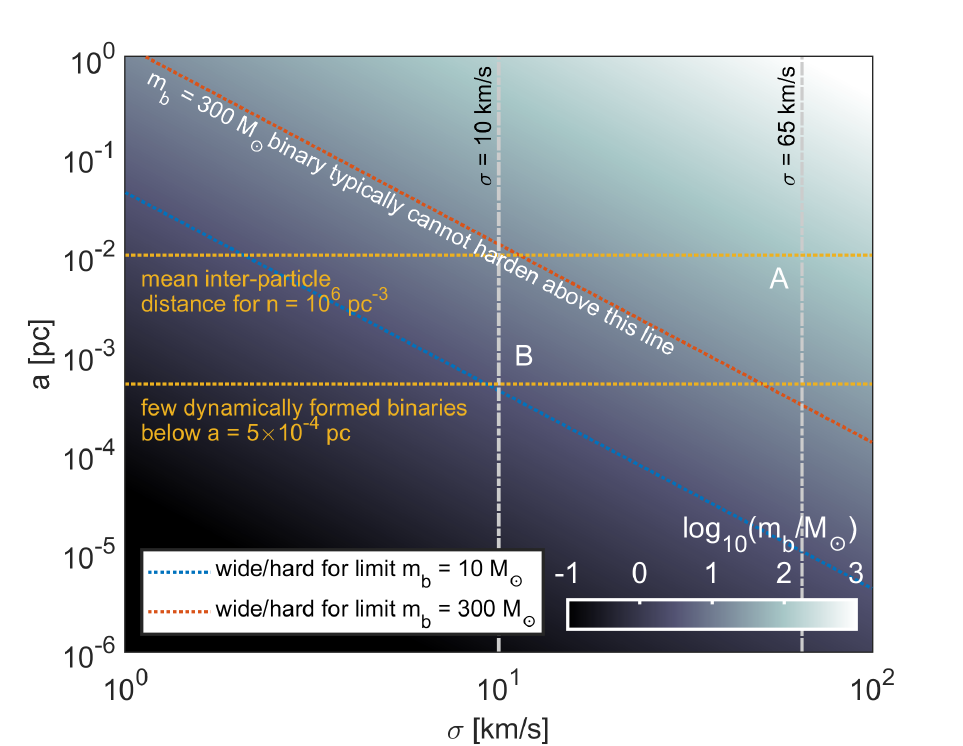}
\caption{The wide-hard binary threshold ($v_\mathrm{orb}=\sigma$) for binaries of different total masses (background color) as a function of their semi-major axis $a$ and the velocity dispersion $\sigma$ of their environment. Dynamically formed binaries typically occur in the range $5\times10^{-4}$ pc $\lesssim a \lesssim 10^{-2}$ pc (yellow dashed horizontal lines). The velocity dispersions $\sigma=10$ km/s and $\sigma=65$ km/s (dash-dotted gray vertical lines) bracket the central velocity dispersions of isolated models I1-I10. The wide-hard threshold is shown for two binaries with $m_\mathrm{b} = m_\mathrm{p} + m_\mathrm{s} = \msol{10}$ (blue line) and $m_\mathrm{b} = \msol{300}$ (orange line). Very massive ($m_\mathrm{b} = \msol{300}$) dynamically formed binaries in the simulations cannot harden in the region A of the parameter space, but will harden in region B. With increasing velocity dispersion $\sigma$ it is increasingly difficult for a binary with a given mass to be located in the region B in which they can harden. Finally, low-mass binaries can never harden, explaining the complete lack of mergers between low-mass star from bound orbits in the isolated simulations.}
\label{fig: heggie}
\end{figure}

We have investigated the origin of the suppressed VMS growth and inhibited IMBH formation initially found in the hierarchical star cluster assembly simulations through analyzing a set of $100$ isolated star cluster simulations. We summarize our findings from the isolated simulations as follows.
\begin{itemize}
    \item The IMBH formation proceeds through a collapse of a VMS formed in a sequence of stellar collisions. IMBHs predominantly form in our simulations in the star cluster mass window between $M_\star\sim\msol{2\times10^4}$ and $M_\star\sim\msol{2\times10^5}$. No more than one star per cluster grows considerably in mass by mergers. 
    \item The vast majority of the VMS mass is acquired through mergers with massive stars. Half of the VMS mass originates from stars more massive than $m_\star\gtrsim\msol{114}$, and only $10\%$ from stars less massive than $m_\star\lesssim\msol{50}$.
    \item $93\%$ of all collisions occur from almost parabolic orbits, $0.9\lesssim e \lesssim 1.1$. The number of collisions between stars of any mass steadily increases as a function of $M_\star$. Collisions between massive stars ($m_\mathrm{p}>\msol{80}$, $m_\mathrm{s}>\msol{10}$) occur predominantly from bound orbits. These bound massive-massive stellar collisions are suppressed in isolated star cluster models with $M_\star\gtrsim\msol{1.5\times10^5}$.   
    \item In more massive star cluster models, the first collision occurs on average earlier with an almost monotonic trend. At low cluster masses, the average first collision time $t_\mathrm{coll}$ between a massive star ($m_\mathrm{p}\gtrsim\msol{80}$) and a massive secondary ($m_\mathrm{s}\gtrsim\msol{10}$) closely follows the $t_\mathrm{coll}$ between any stars. At $M_\star\gtrsim\msol{4\times10^4}$, the first collision time between massive stars decouples from the general trend, and actually increases as a function of $M_\star$. At the highest cluster masses, the average time for the first massive-massive collision is comparable to the life-time of the most massive stars in our isolated simulations.
    \item Finally, the central stellar number densities $n$ are approximately similar in the four most massive isolated cluster setups I7--I10 with $M_\star\gtrsim\msol{1.5\times10^5}$. Only the set I7 of the four setups has sustained VMS growth and frequent IMBH formation. However, given the very different central velocity dispersions from $\sigma\sim41$ km/s (I7) to $\sigma\sim61$ km/s (I10), the centers of the isolated cluster models with different masses are very different stellar-dynamical environments due to typically steep scaling of the outcomes of the dynamical few-body processes as a function of $\sigma$. 
\end{itemize}

The snapshot output interval of the simulations of this study is $\Delta t_\mathrm{output} = 10^{-2}$ Myr. Three-body processes responsible for dynamical binary formation and mergers from binary-single interactions occur on a considerably faster time-scale, and advanced algorithmic techniques such as continuous simulation data output stream \citep{Hausammann2022} would be needed to capture every relevant particle interaction into the output. Considering typical dynamically formed binary semi-major axis $a=2\times10^{-3}$ pc and a high velocity dispersion of $\sigma=50$ km/s, the interaction time-scale for the binary and a third body is of the order of $a/\sigma \sim 4\times10^{-5}$ Myr $\ll\Delta t_\mathrm{output}$. Although our snapshot interval is not frequent enough to resolve interactions responsible for binary formation, we can track the formation, evolution and fate of binaries of massive stars using the output data as the first strong interaction rarely leads to a merger (e.g. \citealt{Atallah2024arxiv}).

We track the most bound binaries ($B = -E \propto m_\mathrm{p} m_\mathrm{s} a^{-1}$) throughout the simulation sets I7 (VMS growth) and I8--I10 (no VMS growth). The primary of the most bound binary is almost immediately a star close to the maximum initial stellar mass $m_\star = \msol{150}$. Binaries with more massive secondaries form through either the 3BBF or exchanges. By $t=1.5$ Myr, binaries with massive secondaries of the order of $m_\mathrm{s}$ have formed. In the lower-mass clusters I7 the most bound binary gradually hardens during the next few $0.1$ Myr--$0.5$ Myr which together with high binary eccentricities enable collisions. However, in the higher-mass (and higher velocity dispersion) clusters I8--I10 some massive binaries still form, but they do not typically harden considerably in the subsequent interactions. 

We argue based on the steep 3BBF velocity dispersion scaling and the Heggie-Hills law that the high velocity dispersion in more massive star clusters both obstructs the formation of massive binaries, and hinders their later hardening, inhibiting the VMS growth and thus IMBH formation. We further elaborate the argument using Fig. \ref{fig: heggie}. Briefly, the binaries that form dynamically in our isolated simulations have typical semi-major axis in the range of $5\times10^{-4}$ pc $\lesssim a \lesssim 10^{-2}$ pc. With the minimum and maximum central velocity dispersions of our isolated models ($10$ km/s $\lesssim a \lesssim 65$ km/s), the semi-major axis range spans a rectangular region in the $(\sigma,a)$ space. We also show the hard-wide binary threshold for a massive equal-mass binary with $m_\mathrm{p}=m_\mathrm{s}=\msol{150}$ in Fig. \ref{fig: heggie}. This line divides the rectangle which the dynamical binaries inhabit into two parts we label A (above the hard-wide threshold) and B (below the threshold). As a consequence of the Heggie-Hills law, binaries in region A will on average further widen and eventually dissolve while binaries in region B can on average further harden in subsequent encounters. Most importantly, for a fixed binary mass, there is a high enough threshold $\sigma$ beyond which a typical dynamically formed binary can only reside in region A, and thus cannot harden. 

The velocity dispersion argument is consistent with the earlier findings in this section, especially the facts that mergers between massive stars from bound orbits are suppressed at sufficiently high cluster masses $M_\star$, and if collisions involving massive stars take place, they occur on average at later times. Finally, the life-times of the most massive stars ($t=2.82$ Myr for $m_\star=\msol{150}$) ensure that the collisions are not just postponed, but completely inhibited as the massive stars reach the ends the their lives before they manage to collide with each other. Our results highlight the importance of the efficiency of the dynamical binary formation channel for the repeated stellar collisions in mass segregated but not yet core collapsed ($t_\mathrm{seg}<t<t_\mathrm{cc}$) massive star clusters.

\section{Discussion}

\subsection{Hierarchical assembly boosts IMBH formation and reduces its stochasticity}

We have shown that in our isolated simulations, massive star clusters with high central velocity dispersions $\sigma$ do not form IMBHs via the VMS collapse channel. In the hierarchical star cluster assembly simulations, this manifests as a fact that the initially most massive star clusters do not form IMBHs from their stars, but rather by inheriting them from less massive star clusters during the star cluster mergers. Isolated comparison simulations with overall density profiles and central densities comparable to the highest and the final assembling cluster densities do not form IMBHs either. Thus, IMBH formation is boosted by the hierarchical star cluster assembly. Intermediate-mass star clusters can better form IMBHs due to their smaller mass segregation $t_\mathrm{seg}$ and core collapse time-scales $t_\mathrm{cc}$ compared to the life-times of massive stars, as well as their lower central velocity dispersions.

Dense, isolated star clusters may still form IMBHs if their central velocity dispersion is not too high. This requires lower star cluster masses $M_\star$.
In our simulations, the threshold for IMBH formation is around $M_\star \sim \msol{2\times10^5}$. However, hierarchical star cluster assembly with this final cluster mass would still most probably produce more than one IMBH in the star cluster. Consequently, the IMBH content of a star cluster with a fixed mass can be drastically different depending on its formation history (monolithic vs hierarchical). In our simulations, young star clusters finishing their assembly had $1$--$5$ IMBHs in total.

Finally, the hierarchical star cluster assembly reduces the stochasticity of IMBH formation. In the isolated simulation sets the cluster models which formed IMBHs did not do so in every tested random realization. In the hierarchical assembly, tens of intermediate-mass star clusters contribute to the final mass of the assembling central cluster. If any of the intermediate-mass clusters end up producing an IMBH, the assembled star cluster will contain an IMBH, at least initially.

\subsection{Surviving the gravitational-wave recoil}

Due to the hierarchical cluster assembly, IMBHs will end up in environments with higher escape velocities than the environment they formed in. This enhances their prospects of being retained in the star clusters in the likely mergers with stellar-mass black holes and the resulting gravitational-wave recoil kicks. As discussed in section \ref{section: gwkick}, an IMBH needs to have a mass of at least $m_\bullet\gtrsim \msol{500}$ to be retained after a merger with most massive stellar BHs below the mass gap ($m_\bullet\sim\msol{63}$) in a star cluster with $M_\star\sim \msol{10^6}$, and $m_\bullet\gtrsim \msol{1000}$ in a star cluster of $M_\star\sim \msol{10^5}$. The most massive IMBHs in each of our cluster assembly simulations have masses above $m_\bullet \gtrsim\msol{1400}$, so they should in principle be resilient against being ejected from their host clusters when merging with stellar-mass black holes.

However, the hierarchical star cluster assembly also poses new challenges for retaining massive IMBHs in their star clusters. In isolated star clusters, typically no more than one IMBH is formed. The hierarchical assembly commonly leads into situations with more than a single IMBH in a single star cluster, enabling IMBH-IMBH mergers in addition to only IMBH-BH mergers. Even though IMBHs formed through the VMS collapse are assumed to be slowly spinning, the gravitational-wave recoil kicks can be strong unless the merging IMBHs have a mass ratio $q=m_\mathrm{s}/m_\mathrm{p}$ close to unity, or $q$ needs to be very small. If the secondary is also in the IMBH mass range, this would require very massive primary IMBHs.

\subsection{Stellar mergers, merger mass loss and primordial binary stars}

In this work we have assumed for simplicity that colliding stars merge without mass loss, unless a common envelope event is triggered instead of full mixing. In reality, stellar collision mass loss may be substantial, especially for stars with low surface gravities, and environments with high collision velocities. Thus, our VMS masses and consequently IMBH masses are likely somewhat overestimated.

However, we do not assume any stellar rejuvenation either. In \sevn, the stellar merger remnant inherits the phase
and percentage of life-time of the most evolved progenitor star, and the interpolation algorithm of \sevn{} finds the new post-merger track
self-consistently. As the durations of the stellar evolutionary phases shorten with increasing stellar mass, the conservative approach does not lead to rejuvenation. It is common in the literature to numerically rejuvenate a star after a stellar merger. The age of the merger product $t(m_\mathrm{p}+m_\mathrm{s})$ can be calculated e.g. as 
\begin{equation}
t(m_\mathrm{p}+m_\mathrm{s}) = \frac{m_\mathrm{p}}{m_\mathrm{p}+m_\mathrm{s}} \frac{t_\mathrm{MS}(m_\mathrm{p}+m_\mathrm{s})}{ t_\mathrm{MS}(m_\mathrm{p}) } t(m_\mathrm{p})
\end{equation}
in which $t_\mathrm{MS}$ is the MS life-time of the star \citep{Meurs1989,PortegiesZwart1999}. In later evolutionary phases the MS life-time $t_\mathrm{MS}$ can be replaced by the duration of the particular phase, and $t(m_\mathrm{p})$ by the time already spent in the particular phase \citep{Mapelli2016}. As extensive rejuvenation will lead to increasingly long life-times for the collision products enabling more future collisions, our choice of no explicit rejuvenation most probably underestimates the final VMS and IMBH masses. Assessing the overall combined effect of the rejuvenation and the lack of stellar merger mass loss is complicated and will require careful further model building and additional simulations beyond this work.

We note that the hierarchical star cluster assembly has implications for the stellar collision mass loss as well. The common reasoning is that more massive stellar systems have higher velocity dispersions $\sigma$, higher average collision velocities thus on average more collision mass loss. At very high collision velocities the stars can even be destroyed. There are a number of potential caveats in the picture. First, a large fraction of collisions may originate from primordial binary stars, in which case the binary population properties and not the host star cluster determines the collision velocities. In the absence of primordial binaries, collisions from interactions between dynamically formed binaries with single stars are important, producing a low-velocity tail in the collision velocity distribution with relatively little expected mass loss \citep{Gaburov2008}. The straightforward effect of the hierarchical star cluster assembly scenario to the stellar collision mass loss is the lowered mean collision velocity. The stellar collisions leading to VMS growth and IMBH formation occur in the lower-mass star clusters with lower velocity dispersions and thus less collision mass loss is expected in the hierarchical assembly process compared to monolithic clusters.

Finally, we emphasize the probable importance of primordial binary stars for the IMBH formation in the hierarchical star cluster assembly scenario. Primordial binaries were not included in the models of this work, but are instead the focus of the next study (FROST-CLUSTERS II) of this project. Primordial binaries enhance the interaction cross-sections ($a$ versus $r_\mathrm{\star}$), potentially increasing the number of stellar collisions. Primordial binaries can ameliorate the suppressed VMS growth at high cluster masses in several ways. First, massive stars tend to reside in binary systems, and a sizable fraction of them are already hard. Second, the 3BBF bottleneck of binary formation with its extreme dependence on the velocity dispersion may vanish if the primordial binary fraction is high enough. Further simulations in the FROST-CLUSTERS project will explore the question in the near future.

\subsection{Gravitational waves}

\begin{figure}
\includegraphics[width=\columnwidth]{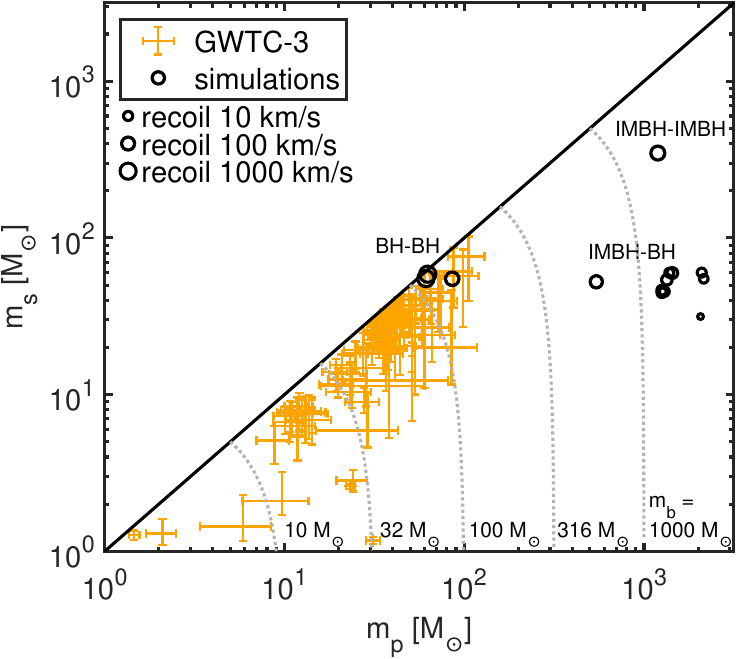}
\caption{The primary and secondary masses in BH-BH mergers of the hierarchical star cluster assembly runs of this study (black, open circles) compared to the primary and secondary masses from the GWTC-3 catalog (in yellow) detected by the Advanced LIGO and Advanced Virgo gravitational-wave observatories up to the end of their third observing run \citep{Abbott2023}. Curves for constant total binary masses $m_\mathrm{b} = m_\mathrm{p} + m_\mathrm{s}$ are displayed from $m_\mathrm{b}=\msol{10}$ to $M_\mathrm{b}=\msol{10^3}$ in gray, dashed lines. Current ground-based GW detectors cannot observe mergers with $m_\mathrm{b} \gtrsim \msol{500}$ very well above $z\gtrsim1$.}
\label{fig: gwmassratio}
\end{figure}

We present the primary $m_\mathrm{p}$ and secondary masses $m_\mathrm{s}$ of merging BH-BH, IMBH-BH and IMBH-IMBH binaries from our hierarchical star cluster assembly simulations H1, H2 and H3 in Fig \ref{fig: gwmassratio}. In addition, Fig. \ref{fig: gwmassratio} shows the primary and secondary masses from the GWTC-3 catalog (in yellow) detected by the Advanced LIGO and Advanced Virgo gravitational-wave observatories up to the end of their third observing run \citep{Abbott2023}. The BH-BH mergers ($\msol{61} \lesssim m_\mathrm{p} \lesssim \msol{86}$, $\msol{52} \lesssim m_\mathrm{s} \lesssim \msol{59}$) during the first $t=50$ Myr of the simulations agree with the GWTC-3 sources very well.

The most numerous gravitational-wave merger category during the first $t=50$ Myr of the cluster assembly simulations in this study are IMBH-BH mergers. Most IMBH-BH mergers have their total masses $m_\mathrm{b}=m_\mathrm{p}+m_\mathrm{s}$ and mass ratios $q=m_\mathrm{s}/m_\mathrm{p}$ ranging from $\msol{1300}\lesssim m_\mathrm{b} \lesssim \msol{2220}$ and $1.5\times10^{-2}\lesssim q \lesssim4.5\times10^{-2}$. One IMBH-BH merger has a lower total mass of $m_\mathrm{b}\sim\msol{600}$ and $q\sim 0.1$. Finally, the only true IMBH-IMBH merger of this study has a total mass of $m_\mathrm{b}\sim \msol{1500}$ and $q\sim0.3$. Based on the literature work (e.g. \citealt{DiCarlo2019,Rodriguez2019,Banerjee2020,Rastello2021,Rizzuto2022,Torniamenti2022,ArcaSedda2023c}), we expect that the inclusion of primordial binary stars in the initial cluster setups in our forthcoming study FROST-CLUSTERS II will populate the BH-BH LIGO/Virgo/KAGRA parameter space in Fig. \ref{fig: gwmassratio}. Beside this regime, our hierarchical star cluster assembly simulations also indicate a distinct second GW population in the IMBH mass range around $m_\mathrm{b}\sim\msol{10^3}$ with typical mass ratios of $1.5\times10^{-2}\lesssim q \lesssim4.5\times10^{-2}$.

Signatures of BH mergers in the IMBH mass range with various mass ratios have been searched from the data of Advanced LIGO and and Advanced Virgo detectors from the first, second and third observing runs \citep{Abbott2017IMBHsearch1stRun,Abbott2019IMBHsearch2ndRun, Abbott2022IMBHsearch3rdRun}. The current ground-based detectors are sensitive to BH-BH mergers at the lower end of the IMBH mass range as demonstrated by the detection of the event GW190521 from the formation of an IMBH with $m_\bullet \sim \msol{150}$ \citep{Abbott2020}. So far, no signatures of gravitational waves from BH-BH mergers in the IMBH mass range have been found, and only upper limits on the IMBH merger rates can have been placed. The searches do not typically extend well beyond total (IM)BH binary masses of $m_\mathrm{b}=m_\mathrm{p}+m_\mathrm{s}\sim \msol{500}$ as the current ground-based detectors become less sensitive at higher total binary masses, corresponding to lower frequencies below $\sim 10$ Hz. In general, more massive merging binaries produce a shorter but louder GW signal with little inspiral but more ringdown signal at the current observable frequency range \citep{Abbott2022IMBHsearch3rdRun}.

In general, IMBH mergers with a total mass of $\msol{500} \lesssim m_\mathrm{b} \lesssim \msol{1000}$ could be observable with the current LIGO/Virgo/KAGRA detectors with a sufficient signal-to-noise ratio of S/N$\sim8$ to cosmological distances up to $\sim$ a few Gpc, corresponding to redshifts $z\lesssim1$ (e.g. \citealt{Mazzolo2014,Haster2016,Mehta2022,Fragione2022b}). For a fixed total mass binary mass $m_\mathrm{b}$, binaries with less massive secondaries $m_\mathrm{s}$ have smaller signal-to-noise ratios and are thus more difficult to detect from large distances than equal-mass binaries. For IMBH-BH mergers ($m_\mathrm{p}\gg m_\mathrm{s}$), the detection horizon distance scales approximately proportional to as $\left(m_\mathrm{s}/m_\mathrm{p}\right)^{1/2}$ when $m_\mathrm{b}<\msol{200}$ (inspiral domination) and with $m_\mathrm{s}/m_\mathrm{p}$ at higher binary masses when the ringdown signal dominates \citep{Haster2016a,Haster2016}, assuming Advanced LIGO design sensitivity. Thus, we expect that most of our IMBH mass scale mergers would not be detectable with the current GW instruments, as our $z=0.01\:\mathrm{Z_\odot}$ simulations mimic the low-metallicity environment at $z\gtrsim6$. However, such mergers from low metallicity pockets at lower redshifts in the Universe are possible, yet likely rare.

Future ground-based (Einstein Telescope, Cosmic Explorer) and space-borne detectors such as LISA will be able to detect IMBH mergers with $m_\mathrm{b}\sim \msol{10^3}$ well up to $z\sim10$ \citep{Jani2020}. For multi-band detections, the Einstein Telescope + LISA network will be the most sensitive for binaries with $m_\mathrm{b}\sim \msol{1.5\times10^3}$--$\msol{2.0\times10^3}$, independent of the mass ratio \citep{Jani2020}. This is exactly the mass range of the most massive IMBHs of \catthree{} category in our hierarchical assembly simulations.

\section{Summary and conclusions}

We have used the updated \bifrost{} N-body simulation code now coupled to stellar evolution to study the formation of VMSs and IMBHs in hierarchically assembling massive star clusters. We assume the universal power-law mass function for the individual young sub-clusters with a slope of $\alpha=-2$ supported both by observations (e.g. \citealt{Elmegreen1996, Zhang1999,Adamo2020}) and state-of-the-art solar-mass-resolution hydrodynamical simulations of star cluster formation \citep{Lahen2020}. 

Besides the observationally motivated shallow power-law slope for the star cluster mass-size relation ($\alpha=0.18$), we choose small sub-pc birth radii $r_\mathrm{h}$ for our initial sub-cluster populations based on the observed sizes of the still embedded nascent star clusters (e.g. \citealt{Marks2012}). An additional motivation for the small sub-cluster sizes are the recent JWST observations of several close, dense, parsec-size, massive ($M_\star\sim\msol{10^6}$) star clusters in the Cosmic Gems arc at $z=10.2$ \citep{Adamo2024arxiv}. The structure of the initial star cluster assembly regions is adopted from representative cluster formation regions of the hydrodynamical low-metallicity dwarf galaxy starburst simulations of \cite{Lahen2020}.

Our main findings are the following. First, IMBHs form in each three hierarchical cluster assembly simulations during the first $t=50$ Myr with maximum IMBH masses $m_\bullet \sim \msol{2200}$. The primary formation channel for IMBH formation is the collapse of a VMS which formed through a sequence of collisions with massive stars, the primary progenitor having an initial mass exceeding $m_\star\gtrsim\msol{80}$. All star clusters forming IMBHs are mass segregated before they merge with the central cluster, but not all of them have reached the core collapse. Tidal disruption of stars, BH-BH, IMBH-BH and IMBH-IMBH (in one case) mergers also contribute to the final mass of the IMBHs, but their contribution is sub-dominant during the first $50$ Myr. Typically only a single IMBH forms in a star cluster, but multiple IMBHs will end up in the most massive star cluster through the cluster assembly process. Multiple IMBHs per final star cluster is essentially an unique prediction of the model as in isolated setups typically no more than a single object can considerably grow by stellar mergers.

Second, no IMBHs originate from stars in the initially most massive star clusters in the hierarchical simulations. Although the initially most massive star clusters have the highest central stellar densities, they also have longest mass segregation timescales and highest central velocity dispersions, impeding the collisions of massive stars and IMBH formation. In addition, when simulating the evolution of isolated massive star clusters with density profiles comparable to the hierarchical assembly end products, no IMBHs form. The key to the presence of IMBHs in assembled massive star clusters with $M_\star>\msol{5\times10^5}$ appears to be the hierarchical assembly process itself. 

The hierarchical star cluster assembly process has additional important implications for the VMS growth and IMBH formation. Mass loss in stellar collisions, though not modeled in this study, may be less severe in hierarchical formation of star clusters due to lower velocity dispersions of the star clusters in which the VMSs formed. Moreover, the IMBHs end up in higher escape velocity environments than their original host star clusters, making the IMBHs more resilient to gravitational-wave recoil kicks when inevitably merging with stellar-mass black holes at later times. However, the fact that massive star clusters in our simulation may host multiple IMBHs poses an additional challenge to their survival in the clusters, as IMBH-IMBH mergers have closer to equal mass ratios compared to IMBH-BH mergers, making strong GW recoil kicks more probable. 

The various GW-driven BH-BH and IMBH-BH mergers and the single IMBH-IMBH merger are detectable with either current LIGO/Virgo/KAGRA GW observatories or the next-generation ground-based and space-borne detectors (e.g. \citealt{Jani2020}). Besides the binary mergers consistent with the current LIGO/Virgo/KAGRA observations, our simulations show another distinct population of GW sources: IMBHs in with masses $m_\bullet \sim \msol{10^3}$--$\msol{2\times10^3}$ merging with stellar-mass BHs yielding mass ratios of $1.5\times10^{-2}\lesssim q \lesssim4.5\times10^{-2}$.

We have further studied the VMS and IMBH growth in dense, isolated, massive star clusters in detail in a set of $100$ star cluster simulations containing up to a million stars. Star clusters with masses $M_\star \gtrsim \msol{2\times10^4}$ form IMBHs above the (P)PISN mass gap. Initially, the maximum reached VMS masses through collisions scale linearly with the cluster mass, $\max{(m_\star)} \approx 0.02\times M_\star$, yielding a BH formation efficiency of $\epsilon_\bullet \sim 0.02$. In low-mass clusters, VMS growth is inhibited by the relatively low central densities and the low number of massive stars. The maximum IMBH mass in the isolated cluster simulations during the first $t=5$ Myr is reached at cluster masses $M_\star \sim \msol{1.5\times10^5}$ with $m_\bullet \sim \msol{1630}$, somewhat lower than in the structured cluster assembly runs. 

The key results of the isolated star cluster simulations are the following. First, more massive star clusters have more stellar mergers, and the first mergers occur on average earlier in more massive clusters. Second, most mergers occur from bound orbits at low cluster masses ($M_\star\lesssim\msol{6\times10^4}$) while unbound mergers make up to over $90\%$ of the total mergers at high cluster masses. Next, most of the VMS mass growth originates from a relatively small number of mergers of VMS primaries with massive secondaries: stars less massive than $m_\star \sim \msol{50}$ only contribute less than $\sim10\%$ to the total mass accreted by the VMSs. VMS-VMS mergers are rare. Mergers between massive stars ($m_\star \gtrsim \msol{10}$) originate mostly from bound orbits, and occur later in more massive clusters. The massive-massive bound mergers become increasingly rare at high cluster masses above $M_\star \gtrsim \msol{2\times10^5}$. This efficiently suppresses the VMS growth and thus IMBH formation in isolated massive star clusters.

The suppression of VMS growth and rapid IMBH formation in massive, isolated star clusters caused by a combination of two factors. When massive stars end their lives, their dense, massive star clusters have mass segregated but not necessarily yet core collapsed. In star clusters with masses in the range of $\msol{2\times10^4}\lesssim M_\star \lesssim \msol{2\times10^5}$ binaries with massive stars can form through the 3BBF channel and subsequent dynamical three-body exchanges. As the velocity dispersion of these clusters is relatively low ($\lesssim40$ km/s), these binaries can harden according to Heggie-Hills law, and combined with their high eccentricities and low pericenter distances also eventually merge with each other. In the case of more massive clusters the velocity dispersion is high and the 3BBF channel is suppressed. If binaries with massive stars form, they are too wide to harden with further interactions with the cluster stars. Thus, massive stellar mergers and the VMS growth is inhibited. Second, the relatively short life-times of massive stars ($t=2.82$ Myr for $m_\star=\msol{150}$) ensure that the onset of massive star mergers is truly prevented and not just postponed as the massive stars reach the ends of their lives. We also note that as the assumed star cluster mass-size relation (an observationally motivated power-law with $\beta = 0.18$) is relatively shallow, so the relative increase in the central densities of the cluster models e.g. from $N=10^5$ to $N=10^6$ is only moderate. For more compact million-body models the core-collapse time-scale would be shorter, potentially allowing for more mergers of massive stars and thus IMBH formation.

Finally, we emphasize that primordial binaries, ignored in this study for simplicity, may help to counter the suppression of VMS growth observed in star cluster setups with high velocity by providing a source of massive binary stars independent of the 3BBF and binary-single exchange channels. We will explore this in the forthcoming second study of the FROST-CLUSTERS project.

\section*{Data availability statement}
The data relevant to this article will be shared on reasonable request to the corresponding author.

\section*{Acknowledgments}
The authors thank the reviewer for a constructive referee report. We also thank Francesco Paolo Rizzuto and Jacob Kosowski from the Theoretical Extragalactic Group of the University of Helsinki, Finland. The numerical simulations were performed using facilities hosted by the Max Planck Computing and Data Facility (MPCDF) in Garching, Germany. TN acknowledges support from the Deutsche Forschungsgemeinschaft (DFG, German Research Foundation) under Germany's Excellence Strategy - EXC-2094 - 390783311 from the DFG Cluster of Excellence "ORIGINS".


\bibliographystyle{mnras}
\interlinepenalty=10000
\bibliography{manuscript}


\appendix
\section{The updated \bifrost{} code: gravitational dynamics}
\subsection{Simplified gradient term calculation for fourth-order forward symplectic integrator}\label{section: hhsfsi-omelyan}

The fourth-order forward symplectic integrator (FSI, \citealt{Chin1997,Chin2005,Chin2007,Dehnen2011,Dehnen2017a}) and its hierarchical version (HHS-FSI, \citealt{Rantala2021}) need to calculate the so-called gradient accelerations $\vect{g}_\mathrm{i}$ for a modified mid-step kick operation in addition to the common Newtonian accelerations $\vect{a}_\mathrm{i} = \vect{a}_\mathrm{i}(\vect{r}_\mathrm{i})$ in order to reach the fourth-order accuracy. We do not derive the integrator here or review its properties, but instead refer the reader to \cite{Rantala2021} and references therein for the details. 

Briefly, the mid-step gradient kick operation requires the calculation of the modified accelerations $\vect{\tilde{a}}_\mathrm{i}$ defined as
\begin{equation}
\begin{split}  
    \vect{\tilde{a}}_\mathrm{i} &= \vect{a}_\mathrm{i} + \frac{\epsilon^2}{48} \vect{g}_\mathrm{i}\\
    \vect{g}_\mathrm{i} &= 2 G \sum_\mathrm{j \neq i}^\mathrm{N} \frac{m_\mathrm{j}}{r^5_\mathrm{ji}} \bigg( r_\mathrm{ji}^2 \vect{a}_\mathrm{ji} - 3 ( \vect{x}_\mathrm{ji} \cdot \vect{a}_\mathrm{ji} ) \vect{x}_\mathrm{ji} \bigg) 
\end{split}
\end{equation}
in which $\epsilon$ is the time-step, $m_\mathrm{j}$ the particle masses, $\vect{x}_\mathrm{ij}$ the relative particle separations and $\vect{a}_\mathrm{ij}$ the relative Newtonian accelerations. However, this step can be simplified by using the so-called Omelyan's approximation \citep{Omelyan2006} as noted by \cite{Chin2007}. The essence of the approximation is the fact that the term $\vect{\tilde{a}}_\mathrm{i}$ can be very well approximated in Keplerian potentials by ordinary Newtonian accelerations calculated using displaced positions
\begin{equation}   
\vect{r}_\mathrm{i}' = \vect{r}_\mathrm{i} + \frac{\epsilon^2}{24} \vect{a}_\mathrm{i}
\end{equation}
so that $\vect{\tilde{a}}_\mathrm{i}(\vect{r}_\mathrm{i}) \approx \vect{a}_\mathrm{i}(\vect{r}_\mathrm{i}')$. Thus, Omelyan's approximation allows for using the same numerical procedures for the gradient acceleration calculation as for the common Newtonian calculations. While the computational expense of the approximation is still comparable to the original method, the algorithmic complexity of the required code is greatly reduced. We find this especially convenient for the CUDA kernels of \bifrost{}.

\subsection{Slow-down algorithmic regularization for subsystem integration}\label{section: sdar}

Instead using the standard logarithmic Hamiltonian (LogH) variant of the algorithmically regularized few-body integration (e.g. \citealt{Mikkola1999,Preto1999,Mikkola2002,Mikkola2006,Mikkola2008a,Hellstrom2010,Rantala2017,Rantala2020,Trani2023}) as in the previous \bifrost{} version \citep{Rantala2023a}, we implement a version of slow-down algorithmic regularization (SDAR, \citealt{Wang2020}) into the novel version of \bifrost. The technical details of the implementation closely follow \cite{Wang2020}. We very briefly outline the main properties and advantages of the method here.

The key idea of slow-down regularization is to identify weakly perturbed binary systems and slow down their relative orbital motion by hand by a factor of $\kappa$ so that the new orbital period of the binary is $\kappa P$ instead of the Keplerian period $P$ \citep{Mikkola1996}. If there are no weakly perturbed binaries in the system, the SDAR reduces to the ordinary algorithmic regularization technique. The increased orbital period obviously makes the orbital integration of the system containing the binary computationally less expensive. The slow-down factor $\kappa>1$ can be determined from the strength of the tidal perturbation onto the binary and the fact that the perturbation is only allowed to change a little during one slow-down period $\kappa P$. From the point of view of the binary, the slow-down orbital motion corresponds to an increased strength of the external perturbation. Formidably, the slow-down orbital evolution of the binary corresponds to its evolution in the secular approximation \citep{Wang2020}. This fact with the lowered computational costs make the SDAR a powerful tool for integrating few-body systems which include weakly perturbed binaries.

In \bifrost, we replace both the LogH algorithmic regularization and the secular integration of three-body systems used in \cite{Rantala2023a} with our implementation of the SDAR method. The main difference of our implementation of SDAR compared to the original SDAR of \cite{Wang2020} is the use of the Gragg-Bulirsch-Stoer (GBS) extrapolation method \citep{Gragg1965,Bulirsch1966} to ensure the user-desired numerical accuracy. The original SDAR implementation instead uses a high-order symplectic integrator. In this study we do not use the slow-down method (that is, we set $\kappa=1$) for subsystems requiring post-Newtonian (PN) equations of motion, i.e. subsystems including black holes. As in \cite{Rantala2023a}, we use PN terms up to the order PN3.5.

Following \cite{Wang2020}, the equations of motion our SDAR implementation originate from a slow-down Hamiltonian $H_\mathrm{sd}$ defined as 
\begin{equation}
    H_\mathrm{sd} = \sum_\mathrm{i}^\mathrm{N_b} \frac{1}{\kappa_\mathrm{i}} H_\mathrm{b,i} + \left( H  - \sum_\mathrm{i}^\mathrm{N_b} H_\mathrm{b,i} \right),
\end{equation}
in which $H=T+U$ is the Hamiltonian of the original system, $N_\mathrm{b}$ the number of binary systems, $H_\mathrm{b,i}$ their two-body Hamiltonians and $\kappa_\mathrm{i}$ their individual slow-down factors. The slow-down Hamiltonian can be separated into kinetic and potential terms as $H_\mathrm{sd} = T_\mathrm{sd} + U_\mathrm{sd}$ with $T_\mathrm{sd}$ and $U_\mathrm{sd}$ defined as
\begin{equation}
\begin{split}
T_\mathrm{sd} &= \sum_\mathrm{i}^\mathrm{N_b} \frac{1}{\kappa_\mathrm{i}} T_\mathrm{b,i} + \sum_\mathrm{i}^\mathrm{N_b} T_\mathrm{b,cm,i} + \sum_\mathrm{i=2N_\mathrm{b}+1}^\mathrm{N} T_\mathrm{i} + B_\mathrm{sd}\\
U_\mathrm{sd} &= \sum_\mathrm{i}^\mathrm{N_b} \frac{1}{\kappa_\mathrm{i}} U_\mathrm{b,i} + \left[ U - \sum_\mathrm{i}^\mathrm{N_b} \frac{1}{\kappa_\mathrm{i}} U_\mathrm{b,i}\right].
\end{split}
\end{equation}
In the definition of $T_\mathrm{sd}$ the first term is the kinetic energy of the slow-down binaries, while the second and the third terms are the unaltered kinetic energies of binary center-of-masses and single particles. The indexes are ordered so that the binaries are first in the arrays, and then the singles. The last term is the so-called momentum of time term $p_\mathrm{t}=-H_\mathrm{sd}(t=0)=B_\mathrm{sd}$ in the extended phase-space of algorithmic regularization, i.e. the initial binding energy. The potential term $U_\mathrm{sd}$ is somewhat more simple, consisting of the internal potential energy in the slow-down binaries (first term) and the potential energy of the rest of the system (the term in the brackets).

Now, the time-transformed equations of motion can be constructed. First, the equation of motion for time $t$ as a function of the new independent variable $s$ follows directly from the chosen time transformation $\mathrm{ds} = U_\mathrm{sd} \mathrm{dt}$ as 
\begin{equation}
    \derfrac{t}{s} = \frac{1}{U_\mathrm{sd}} = \frac{1}{T_\mathrm{sd}}.
\end{equation}
Note that here $T_\mathrm{sd}$ includes the momentum-of-time binding energy term $B_\mathrm{sd}$, which is typically displayed on its own when discussing the time transformation (e.g. \citealt{Rantala2017,Rantala2020}). Next, the equations of motion can be written for the single particles as
\begin{equation}
\begin{split}
    \derfrac{\vect{r}_\mathrm{i}}{s} &= \frac{1}{T_\mathrm{sd}} \vect{v}_\mathrm{i}\\
    \derfrac{\vect{v}_\mathrm{i}}{s} &= \frac{1}{U_\mathrm{sd}} \vect{a}_\mathrm{i}
\end{split}
\end{equation}
and for the particles in binary systems as
\begin{equation}
\begin{split}
    \derfrac{\vect{r}_\mathrm{k,i}}{s} &= \frac{1}{T_\mathrm{sd}} \left[\frac{1}{\kappa_\mathrm{i}} \left( \vect{v}_\mathrm{k,i} - \vect{v}_\mathrm{cm,i}\right) + \vect{v}_\mathrm{cm,i} \right]\\
    \derfrac{\vect{v}_\mathrm{k,i}}{s} &= \frac{1}{U_\mathrm{sd}} \left[\frac{1}{\kappa_\mathrm{i}} \vect{a}_\mathrm{bin,i} + \left( \vect{a}_\mathrm{k,i} - \vect{a}_\mathrm{bin,i}\right)\right].
\end{split}
\end{equation}
Here $\vect{a}_\mathrm{bin,i}$ refer to the acceleration within each individual binary.

We use the so-called jumping-$\kappa$ method closely following \cite{Wang2020} to change the slow-down factors of individual binaries. The slow-down factor is determined using the relative strength at the apocenter approach already present in a simple form in \cite{Mikkola1996}. For binary with a label $i$ and $N_\mathrm{p}$ perturbers the slow-down factor $\kappa_\mathrm{i}$ is
\begin{equation}
    \kappa_\mathrm{i} = k_\mathrm{ref} \frac{m_\mathrm{i,1} m_\mathrm{i,2}}{(m_\mathrm{i,1} + m_\mathrm{i,2}) r_\mathrm{i,apo}^3} \sum_\mathrm{j}^\mathrm{N_\mathrm{p}} \frac{\norm{\vect{r}_\mathrm{j} - \vect{r}_\mathrm{i,cm}}^3}{m_\mathrm{j}}
\end{equation}
in which $k_\mathrm{ref} = 10^{-6}$. The maximum slow-down factor is calculated by estimating the how rapidly the perturbation itself is changing. Denoting $\langle X \rangle_\mathrm{m}$ as the mass-weighted average of quantity $X_\mathrm{j}$ over the perturbers $j$, we have
\begin{equation}
\kappa_\mathrm{i,max} = \frac{c}{P} \frac{\norm{\langle \vect{r}_\mathrm{j} - \vect{r}_\mathrm{i,com} \rangle_\mathrm{m}}}{\norm{\langle \vect{v}_\mathrm{j} - \vect{v}_\mathrm{i,com} \rangle_\mathrm{m}}},
\end{equation}
in which $c=0.1$.

We enforce the step-size control and ensure the numerical accuracy of the SDAR integration by using the Gragg-Bulirsch-Stoer (GBS) extrapolation method \citep{Gragg1965,Bulirsch1966}. We use at maximum $k_\mathrm{max}=8$ sub-step divisions, and the user given tolerance parameters are $\eta_\mathrm{gbs} = 5\times10^{-8}$ for the GBS algorithm and $10^{-2}$ for the end-time iteration procedure. For further details on practical implementation of algorithmically regularized integrators and the GBS method in regularized integration see \cite{Rantala2017,Rantala2020,Rantala2023a}. For further details of the SDAR method we refer the reader to \cite{Wang2020}.

\section{The updated \bifrost{} code: stellar evolution, collisions and mergers}

\subsection{\sevn{} and this study}

Fast stellar population synthesis codes for single and binary star evolution, pioneered by \texttt{SSE} and \texttt{BSE} by \cite{Hurley2000,Hurley2002}, with a straightforward direct N-body or cluster Monte Carlo interface are essential for realistic simulations of star cluster evolution. \texttt{SSE} uses computationally efficient fitting polynomials to the stellar evolution tracks of \cite{Pols1998} to follow the evolution of stars. Instead of the pre-computed fitting polynomials, recent fast population synthesis codes, such as \textsc{combine} \citep{Kruckow2018}, \textsc{metisse} \citep{Agrawal2020}, \textsc{posydon} \citep{Fragos2023} and \sevn{} \citep{Iorio2023} rely on multi-dimensional interpolation of the stellar tracks. A major advantage of the multidimensional interpolation compared to the fitting formulas is the ability to change the stellar tracks once new ones become available.

\sevn{} \citep{Iorio2023, Mapelli2020, Spera2015, Spera2017} is a fast stellar population synthesis code, written in C\texttt{++}. Binary evolution prescriptions are based on analytic and semi-analytic formulas following \citep{Hurley2002} plus updates for common envelope treatments and stellar mergers \citep{Spera2019, Iorio2023}. The reason for us choosing \sevn{} to be coupled to \bifrost{} over the other options are the following. First, \sevn{} is written in C\texttt{++} making the C/CUDA coupling to \bifrost{} more straightforward. Second, \sevn{} is easily available, and the code framework enables importing improved stellar tracks in the future. In this work (FROST-CLUSTERS I), we model systems with no primordial binary stars, so we focus here in single stellar evolution, and stars in dynamically formed binaries evolve as they were single stars. The effect of primordial binary dynamics and binary stellar evolution will be a topic for a forthcoming follow-up study (FROST-CLUSTERS II).

\sevn{} can rapidly evolve stars with a wide range of masses and metallicities by interpolating pre-calculated H-rich and pure-He stellar tracks selected by the user. At the moment, \sevn{} has two \texttt{PARSEC}-based \citep{Bressan2012, Chen2015, Costa2021, Nguyen2022} track sets and one set of \texttt{MIST} \citep{Choi2016} tracks available. As we are interested in the evolution of massive stars and their collisions, for this work we choose the \texttt{PARSEC}-based \texttt{SEVNtracks\_parsec\_ov04\_AGB} tracks ranging in stellar mass from $\msol{2.2} \leq m_\star \leq \msol{600}$ as they reach higher stellar masses than the other available stellar tracks. The tracks assume an overshooting parameter of $\lambda_\mathrm{ov} = 0.4$, and we use a fixed metallicity of $Z=0.0002=0.01\:Z_\odot$ in all of our runs. The other H-rich \texttt{PARSEC} tracks not used in this work, \texttt{SEVNtracks\_parsec\_ov05\_AGB} tracks have an overshooting parameter of $\lambda_\mathrm{ov} = 0.5$, extend up to $m_\star=\msol{450}$ in stellar mass, but only for specific metallicities the tracks extend up to $m_\star=\msol{600}$. The available \texttt{MIST} \citep{Choi2016} \texttt{SEVNtracks\_MIST\_AGB} only extend up to $m_\star=\msol{150}$.

In practice, the main difference between the two \texttt{PARSEC} tracks sets for our simulations would be the stellar life-times from the beginning of the main sequence (MS) until the beginning of the remnant phase, and the remnant masses. The life-times from the beginning of the MS to the beginning of the remnant phase at the selected metallicity for the $\lambda_\mathrm{ov} = 0.5$ tracks would be typically $4\%$--$8\%$ shorter compared to the $\lambda_\mathrm{ov} = 0.4$ tracks. The use of $\lambda_\mathrm{ov} = 0.5$ tracks would lead to increased remnant BH masses by up to $\sim 20\%$--$25\%$ for stars with initial masses of $m_\star \sim \msol{15}$--$\msol{20}$, and decreased remnant masses near the lower end of the (P)PNS gap by $\sim 17\%$--$23\%$. The (P)PISN mass gap is somewhat narrower for $\lambda_\mathrm{ov} = 0.5$ tracks at our metallicity of $Z=0.01\:Z_\odot$ compared to the $\lambda_\mathrm{ov} = 0.4$ tracks.

\begin{table}
\begin{tabular}{l l l}
\hline
\sevn{} stellar process & model & reference\\
\hline
Core-collapse SN & delayed & \cite{Fryer2012}\\
Neutrino mass loss & lattimer89 & \cite{Lattimer1989}\\
(P)PISN processes & mapelli20 & \cite{Woosley2017}\\
&  & \cite{Spera2017}\\
& & \cite{Mapelli2020}\\
SN kicks & unified & \cite{Giacobbo2020}\\
\end{tabular}
\caption{SN-related stellar evolution models used in this work.}
\label{table: sevn-processes}
\end{table}

In addition to the track interpolation, \sevn{} models a variety of single stellar evolution processes, which include supernova-related aspects of stellar evolution not included in the pre-computed evolutionary tracks. The most relevant such processes to this work are the core collapse supernovae (SN), SN neutrino mass loss \citep{Zevin2020, Lattimer1989}, (P)PISN processes \citep{Woosley2007, Yoshida2016, Woosley2017} and SN kicks. The models used in this work are listed in Table \ref{table: sevn-processes}. For in-depth descriptions of the individual physical processes, see \cite{Iorio2023} and references therein.

\subsection{Interfacing \sevn{} with \bifrost}\label{section: sevn}

\subsubsection{Code interface}
We use \sevn{} as a library and have written a library interface in \bifrost{} using C\texttt{++} and standard C. The basic building blocks of the interface are the various wrapper functions to access evolving stellar properties from \sevn{} in \bifrost, functions for memory allocation, initialization of stars, evolving the stars, and finally performing stellar collisions and mergers. As a fast population synthesis code, \sevn{} is efficient enough to be used in a serial manner in \bifrost. As the slow-fast hierarchical Hamiltonian splitting approach for time-stepping in \bifrost{} requires regular pivot sorting of the data \citep{Rantala2021}, we store the \sevn{} stellar evolution data into a separate structure in \bifrost{}, and only sort pointer indexes of the particles to the stellar evolution data alongside particle masses, positions and velocities.

The stellar evolution is performed after the FSI phase in the \bifrost{} HHS-FSI. Every star in the \sevn{} mass range is not evolved each time they enter the FSI. Instead, each star is assigned a time stamp for their next evolution time $t_\mathrm{next}$ based on the simulation time and their stellar evolution time-step $\epsilon_\mathrm{se}$ from \sevn. If the simulation time when entering the FSI exceeds the next evolution time $t_\mathrm{next}$, the star is evolved using \sevn, and $t_\mathrm{next}$ is updated. In addition, at the end of each \bifrost{} integration interval, each star is evolved up to that time.

\subsubsection{Time-step assignment}
The \bifrost{} time-step assignment for simulation particles is modified if stellar evolution is enabled. For a star, given the dynamical time-step $\epsilon_\mathrm{dyn} = \min{(\epsilon_\mathrm{ff}, \epsilon_\mathrm{fb}, \epsilon_\mathrm{\nabla})}$ and the stellar evolution time-step $\epsilon_\mathrm{se}$ obtained from \sevn{}, the time-step is $\epsilon = \min{(\epsilon_\mathrm{dyn},\epsilon_\mathrm{se})}$. For additional details of the adaptive time-step algorithm in \sevn{}, see \cite{Iorio2023}. In addition, stars very close to the end of their lives (\texttt{phase\_star=6}) may have their time-steps reduced due to the possible SN kick received at the beginning of the remnant phase \texttt{phase\_star=7} so that the dynamical interactions with other stars after the SN kick are accurately treated. Namely, we use $\epsilon = \min{(\epsilon_\mathrm{dyn}, \epsilon_\mathrm{se}, \epsilon_\mathrm{SNkick})}$ with $\epsilon_\mathrm{SNkick} = \eta_\mathrm{fb} r_\mathrm{ngb} / \tilde{v}_\mathrm{kick}$ in which $\eta_\mathrm{fb}\sim0.1$ is the typical fly-by time-step factor of \bifrost{}, $r_\mathrm{ngb}\sim10^{-3}$ pc the typical spatial size of the subsystem regions and $\tilde{v}_\mathrm{kick} \sim 10^3$ km/s a typical large SN kick velocity. These order-of-magnitude estimates yield a SN kick time-step of the order of $\epsilon_\mathrm{SNkick}\sim10^{-7}$ Myr.

\subsubsection{Low-mass stars and very massive stars}

The stellar tracks \texttt{SEVNtracks\_parsec\_ov04\_AGB} used in this work range from $\msol{2.2} \leq m_* \leq \msol{600}$ in initial stellar mass. The longest simulations times in this work are $50$ Myr, and stars with masses of $\lesssim \msol{2.2}$ remain in the main sequence experience little evolution during this time. Thus, in this work we do not evolve such low-mass stars and treat them as they were in the beginning of their main sequence phase. Low-mass stars may become evolving and enter the \sevn{} mass range through stellar mergers with other low-mass stars.

As massive stars grow by mergers, they may exceed the maximum mass of $\msol{600}$. These very massive stars follow simplified stellar evolution by scaling the $m_\star=\msol{600}$ \texttt{SEVNtracks\_parsec\_ov04\_AGB} stellar track. The VMS inherits its life-time and spin magnitude from the $m_\star=\msol{600}$ track while the wind mass loss and the VMS radius are scaled from the $m_\star=\msol{600}$ using power-law scaling. The mass loss rate scales proportional to $m_\mathrm{VMS}/\msol{600}$ while the VMS radius scales as $(m_\mathrm{VMS}/\msol{600})^{1/3}$ to maintain a constant stellar density. We note that while our VMS treatment is simplified, detailed VMS and SMS models have been developed (e.g. \citealt{Gieles2018}). Including such a model in \bifrost{} remains a topic for future work.

\subsection{Collisions and mergers}

\subsubsection{The collision detection procedure}

We perform the procedures for checking for particle collisions within the regularized subsystem integration routine. The current \bifrost{} code version uses a two-fold procedure for particle collision detection based on the separation of two particles $r_\mathrm{sep} = \norm{\vect{r}_\mathrm{j}-\vect{r}_\mathrm{i}}$ and their osculating pericenter distance $r_\mathrm{p}$ calculated from their osculating two-body orbital elements. A two-body collision criterion based on osculating two-body orbital elements alone may suffer from a boosted collision rate due to false positive collisions: while the osculating elements indicate a collision, it is not guaranteed to occur in a system with more than two bodies.

In the current code version there are three main types of collisions: BH-BH mergers, BH-star tidal disruption events, and star-star collisions. Each type of collision is associated with a characteristic radial separation $r_\mathrm{coll}$ below which a collision certainly occurs, i.e. $r_\mathrm{sep}<r_\mathrm{coll}$. We term these true collisions. We also nevertheless check the pericenter criterion ($r_\mathrm{p}<r_\mathrm{coll}$) using the osculating two-body orbital elements. The particle pairs fulfilling the osculating pericenter collision criterion while not fulfilling the separation criterion ($r_\mathrm{sep}>r_\mathrm{coll}$) are treated as collision candidates. 

If a collision candidate is detected within the SDAR subsystem integration, we first disable the slow-down treatment, i.e. set $\kappa=1.0$. Next, we evaluate the time to the next pericenter (bound orbits) or to the only pericenter passage (hyperbolic orbits) of the collision candidate particle pair using their osculating two-body orbital elements. Now instead of letting the GBS routine choose the regularized time-step, we force the regularized integrator to take (typically less than a few) steps until the actual pericenter is reached. At the pericenter we re-evaluate the collision criteria to actually resolve whether there is a collision or not.

\subsubsection{BH-BH mergers}
A BH-BH occurs in the current version of \bifrost{} based on the radius of the innermost stable circular orbit of a zero-spin BH defined as $r_\mathrm{ISCO} = 6Gm_\bullet/c^2$. The BH pair collision criterion is thus $r_\mathrm{sep}<\max{(r_\mathrm{ISCO,p},r_\mathrm{ISCO,s})}$. The mass, spin and recoil velocity of the remnant BH is calculated using the fitting formulas to numerical relativity simulations by \cite{Zlochower2015}. The natal BH spins are calculated using the Geneva model \citep{BelczynskiKlencki2020} as in their Eq. (3). Examples for recoil kick velocities for various BH binary mass ratios and spins are elaborated in section \ref{section: gwkick}. The gravitational-wave mass loss is typically of the order of $\sim 5\%$ of the original total mass of the system (e.g. \citealt{JimenezForteza2017}).

\subsubsection{Tidal disruption of stars by BHs}

We use a simple order-of-magnitude estimation for the radius below which a tidal disruption event occurs (TDE) as $r_\mathrm{sep}< r_\mathrm{TDE} = 1.3 (m_\bullet/m_\star)^{1/3} R_\star$ based on the standard cube root scaling of \cite{Kochanek1992}. Most TDEs in the simulations in our runs occur from almost parabolic orbits with low pericenters. We assume an accretion factor of $f_\mathrm{c}=0.5$ following \cite{Rizzuto2023}, so half of the stellar material is instantly accreted into the BH while the remaining half is removed from the simulation. Most tidal disruption events in this study occur between IMBHs and stars with masses less than $m_\star\lesssim\msol{100}$, and TDEs are a subdominant channel in IMBH growth in the setups we have examined. Our simple approach in this work does not include partial or repeated tidal disruption events.

\subsubsection{Stellar collisions and mergers}

In the earlier \bifrost{} version of \cite{Rantala2023a} no distinction was made between a stellar collision (the radii of the two stars overlap) and a stellar merger (two stars are mixed into one). If two non-remnant stars had their pericenter distance smaller than the sum of their radii, i.e. $r_\mathrm{p}<r_\mathrm{\star,p}+r_\mathrm{\star,s}$, the stars were immediately merged.

In the current \bifrost{} version, we use \sevn{} to evaluate the outcome of each stellar collision ($r_\mathrm{p}<r_\mathrm{\star,p}+r_\mathrm{\star,s}$) within the \sevn{} mass range. Instead of merging and complete mixing of the two stars, a stellar collision can lead to a common envelope (CE) phase from which the two stars may survive without actually merging. This occurs especially in the late life stages of stars. Binary systems surviving the CE phase have reduced orbital separations and low eccentricities. Stars colliding from hyperbolic orbits are in practice merged from almost parabolic bound orbits as the current version of \sevn{} assumes bound binary orbits for colliding stars. 

As in \sevn, in this study we assume no rejuvenation and no mass loss in stellar mergers. In \sevn, when two stars merge, their total, He core and CO core masses are simply summed. The merger product inherits the evolutionary phase and percentage of life of the more evolved progenitor star, and the \sevn{} interpolation algorithm finds the new post-merger track self-consistently. For the somewhat more complex post-CE merger procedure, see \cite{Iorio2023}. Low-mass stellar mergers ($m_\mathrm{p}<\msol{2.2}$, $m_\mathrm{s}<\msol{2.2}$) and any stellar mergers involving stars above $m_\star>\msol{600}$ are treated without \sevn{} simply by summing up the stellar masses and updating the stellar radii. 


\bsp	
\label{lastpage}
\end{document}